\pdfoutput=1
\documentclass[12pt,a4paper,titlepage]{article}
\usepackage[utf8]{inputenc}
\usepackage[top=50pt,bottom=50pt,left=68pt,right=66pt]{geometry}
\usepackage{amsmath}
\usepackage{booktabs}
\usepackage{amssymb}
\usepackage[title]{appendix}
\usepackage{mathrsfs}
\usepackage{float}
\usepackage{multirow}
\usepackage{graphicx,caption,subcaption}
\usepackage[space]{grffile}
\newcommand{\overbar}[1]{\mkern1.5mu\overline{\mkern-1.5mu#1\mkern-1.5mu}\mkern 1.5mu}

\begin{document}
\renewcommand{\arraystretch}{1.3}

\makeatletter
\def\@hangfrom#1{\setbox\@tempboxa\hbox{{#1}}%
      \hangindent 0pt
      \noindent\box\@tempboxa}
\makeatother


\def\un#1{\relax\ifmmode\@@underline#1\else
        $\@@underline{\hbox{#1}}$\relax\fi}


\let\under=\unt                 
\let\ced=\ce                    
\let\du=\du                     
\let\um=\Hu                     
\let\sll=\lp                    
\let\Sll=\Lp                    
\let\slo=\os                    
\let\Slo=\Os                    
\let\tie=\ta                    
\let\br=\ub                     


\def\a{\alpha}
\def\b{\beta}
\def\c{\chi}
\def\d{\delta}
\def\e{\epsilon}
\def\f{\phi}
\def\g{\gamma}
\def\h{\eta}
\def\i{\iota}
\def\j{\psi}
\def\k{\kappa}
\def\l{\lambda}
\def\m{\mu}
\def\n{\nu}
\def\o{\omega}
\def\p{\pi}
\def\q{\theta}
\def\r{\rho}
\def\s{\sigma}
\def\t{\tau}
\def\u{\upsilon}
\def\x{\xi}
\def\z{\zeta}
\def\D{\Delta}
\def\F{\Phi}
\def\G{\Gamma}
\def\J{\Psi}
\def\L{\Lambda}
\def\O{\Omega}
\def\P{\Pi}
\def\Q{\Theta}
\def\S{\Sigma}
\def\U{\Upsilon}
\def\X{\Xi}


\def\ve{\varepsilon}
\def\vf{\varphi}
\def\vr{\varrho}
\def\vs{\varsigma}
\def\vq{\vartheta}


\def\ca{{\cal A}}
\def\cb{{\cal B}}
\def\cc{{\cal C}}
\def\cd{{\cal D}}
\def\ce{{\cal E}}
\def\cf{{\cal F}}
\def\cg{{\cal G}}
\def\ch{{\cal H}}
\def\ci{{\cal I}}
\def\cj{{\cal J}}
\def\ck{{\cal K}}
\def\cl{{\cal L}}
\def\cm{{\cal M}}
\def\cn{{\cal N}}
\def\co{{\cal O}}
\def\cp{{\cal P}}
\def\cq{{\cal Q}}
\def\car{{\cal R}}
\def\cs{{\cal S}}
\def\ct{{\cal T}}
\def\cu{{\cal U}}
\def\cv{{\cal V}}
\def\cw{{\cal W}}
\def\cx{{\cal X}}
\def\cy{{\cal Y}}
\def\cz{{\cal Z}}


\def\Sc#1{{\hbox{\sc #1}}}      
\def\Sf#1{{\hbox{\sf #1}}}      



\def\slpa{\slash{\pa}}                            
\def\slin{\SLLash{\in}}                                   
\def\bo{{\raise-.3ex\hbox{\large$\Box$}}}               
\def\cbo{\Sc [}                                         
\def\pa{\partial}                                       
\def\de{\nabla}                                         
\def\dell{\bigtriangledown}                             
\def\su{\sum}                                           
\def\pr{\prod}                                          
\def\iff{\leftrightarrow}                               
\def\conj{{\hbox{\large *}}}                            
\def\ltap{\raisebox{-.4ex}{\rlap{$\sim$}} \raisebox{.4ex}{$<$}}   
\def\gtap{\raisebox{-.4ex}{\rlap{$\sim$}} \raisebox{.4ex}{$>$}}   
\def\TH{{\raise.2ex\hbox{$\displaystyle \bigodot$}\mskip-4.7mu \llap H \;}}
\def\face{{\raise.2ex\hbox{$\displaystyle \bigodot$}\mskip-2.2mu \llap {$\ddot
        \smile$}}}                                      
\def\dg{\sp\dagger}                                     
\def\ddg{\sp\ddagger}                                   

\font\tenex=cmex10 scaled 1200


\def\sp#1{{}^{#1}}                              
\def\sb#1{{}_{#1}}                              
\def\oldsl#1{\rlap/#1}                          
\def\slash#1{\rlap{\hbox{$\mskip 1 mu /$}}#1}      
\def\Slash#1{\rlap{\hbox{$\mskip 3 mu /$}}#1}      
\def\SLash#1{\rlap{\hbox{$\mskip 4.5 mu /$}}#1}    
\def\SLLash#1{\rlap{\hbox{$\mskip 6 mu /$}}#1}      
\def\PMMM#1{\rlap{\hbox{$\mskip 2 mu | $}}#1}   %
\def\PMM#1{\rlap{\hbox{$\mskip 4 mu ~ \mid $}}#1}       %
\def\Tilde#1{\widetilde{#1}}                    
\def\Hat#1{\widehat{#1}}                        
\def\Bar#1{\overline{#1}}                       
\def\sbar#1{\stackrel{*}{\Bar{#1}}}             
\def\bra#1{\left\langle #1\right|}              
\def\ket#1{\left| #1\right\rangle}              
\def\VEV#1{\left\langle #1\right\rangle}        
\def\abs#1{\left| #1\right|}                    
\def\leftrightarrowfill{$\mathsurround=0pt \mathord\leftarrow \mkern-6mu
        \cleaders\hbox{$\mkern-2mu \mathord- \mkern-2mu$}\hfill
        \mkern-6mu \mathord\rightarrow$}
\def\dvec#1{\vbox{\ialign{##\crcr
        \leftrightarrowfill\crcr\noalign{\kern-1pt\nointerlineskip}
        $\hfil\displaystyle{#1}\hfil$\crcr}}}           
\def\dt#1{{\buildrel {\hbox{\LARGE .}} \over {#1}}}     
\def\dtt#1{{\buildrel \bullet \over {#1}}}              
\def\der#1{{\pa \over \pa {#1}}}                
\def\fder#1{{\d \over \d {#1}}}                 


\def\frac#1#2{{\textstyle{#1\over\vphantom2\smash{\raise.20ex
        \hbox{$\scriptstyle{#2}$}}}}}                   
\def\half{\frac12}                                        
\def\sfrac#1#2{{\vphantom1\smash{\lower.5ex\hbox{\small$#1$}}\over
        \vphantom1\smash{\raise.4ex\hbox{\small$#2$}}}} 
\def\bfrac#1#2{{\vphantom1\smash{\lower.5ex\hbox{$#1$}}\over
        \vphantom1\smash{\raise.3ex\hbox{$#2$}}}}       
\def\afrac#1#2{{\vphantom1\smash{\lower.5ex\hbox{$#1$}}\over#2}}    
\def\partder#1#2{{\partial #1\over\partial #2}}   
\def\parvar#1#2{{\d #1\over \d #2}}               
\def\secder#1#2#3{{\partial^2 #1\over\partial #2 \partial #3}}  
\def\on#1#2{\mathop{\null#2}\limits^{#1}}               
\def\bvec#1{\on\leftarrow{#1}}                  
\def\oover#1{\on\circ{#1}}                              

\def\[{\lfloor{\hskip 0.35pt}\!\!\!\lceil}
\def\]{\rfloor{\hskip 0.35pt}\!\!\!\rceil}
\def\Lag{{\cal L}}
\def\du#1#2{_{#1}{}^{#2}}
\def\ud#1#2{^{#1}{}_{#2}}
\def\dud#1#2#3{_{#1}{}^{#2}{}_{#3}}
\def\udu#1#2#3{^{#1}{}_{#2}{}^{#3}}
\def\calD{{\cal D}}
\def\calM{{\cal M}}

\def\szet{{${\scriptstyle \b}$}}
\def\ulA{{\un A}}
\def\ulM{{\underline M}}
\def\cdm{{\Sc D}_{--}}
\def\cdp{{\Sc D}_{++}}
\def\vTheta{\check\Theta}
\def\fracm#1#2{\hbox{\large{${\frac{{#1}}{{#2}}}$}}}
\def\ha{{\fracmm12}}
\def\tr{{\rm tr}}
\def\Tr{{\rm Tr}}
\def\itrema{$\ddot{\scriptstyle 1}$}
\def\ula{{\underline a}} \def\ulb{{\underline b}} \def\ulc{{\underline c}}
\def\uld{{\underline d}} \def\ule{{\underline e}} \def\ulf{{\underline f}}
\def\ulg{{\underline g}}
\def\items#1{\\ \item{[#1]}}
\def\ul{\underline}
\def\un{\underline}
\def\fracmm#1#2{{{#1}\over{#2}}}
\def\footnotew#1{\footnote{\hsize=6.5in {#1}}}
\def\low#1{{\raise -3pt\hbox{${\hskip 0.75pt}\!_{#1}$}}}

\def\Dot#1{\buildrel{_{_{\hskip 0.01in}\bullet}}\over{#1}}
\def\dt#1{\Dot{#1}}

\def\DDot#1{\buildrel{_{_{\hskip 0.01in}\bullet\bullet}}\over{#1}}
\def\ddt#1{\DDot{#1}}

\def\DDDot#1{\buildrel{_{_{\hskip 0.01in}\bullet\bullet\bullet}}\over{#1}}
\def\dddt#1{\DDDot{#1}}

\def\DDDDot#1{\buildrel{_{_{\hskip 
0.01in}\bullet\bullet\bullet\bullet}}\over{#1}}
\def\ddddt#1{\DDDDot{#1}}

\def\Tilde#1{{\widetilde{#1}}\hskip 0.015in}
\def\Hat#1{\widehat{#1}}


\newskip\humongous \humongous=0pt plus 1000pt minus 1000pt
\def\caja{\mathsurround=0pt}
\def\eqalign#1{\,\vcenter{\openup2\jot \caja
        \ialign{\strut \hfil$\displaystyle{##}$&$
        \displaystyle{{}##}$\hfil\crcr#1\crcr}}\,}
\newif\ifdtup
\def\panorama{\global\dtuptrue \openup2\jot \caja
        \everycr{\noalign{\ifdtup \global\dtupfalse
        \vskip-\lineskiplimit \vskip\normallineskiplimit
        \else \penalty\interdisplaylinepenalty \fi}}}
\def\li#1{\panorama \tabskip=\humongous                         
        \halign to\displaywidth{\hfil$\displaystyle{##}$
        \tabskip=0pt&$\displaystyle{{}##}$\hfil
        \tabskip=\humongous&\llap{$##$}\tabskip=0pt
        \crcr#1\crcr}}
\def\eqalignnotwo#1{\panorama \tabskip=\humongous
        \halign to\displaywidth{\hfil$\displaystyle{##}$
        \tabskip=0pt&$\displaystyle{{}##}$
        \tabskip=0pt&$\displaystyle{{}##}$\hfil
        \tabskip=\humongous&\llap{$##$}\tabskip=0pt
        \crcr#1\crcr}}


\def\eV{\,{\rm eV}}
\def\keV{\,{\rm keV}}
\def\MeV{\,{\rm MeV}}
\def\GeV{\,{\rm GeV}}
\def\TeV{\,{\rm TeV}}
\def\sv{\left<\sigma v\right>}
\def\({\left(}
\def\){\right)}
\def\cm{{\,\rm cm}}
\def\K{{\,\rm K}}
\def\kpc{{\,\rm kpc}}
\def\beq{\begin{equation}}
\def\eeq{\end{equation}}
\def\bea{\begin{eqnarray}}
\def\eea{\end{eqnarray}}


\newcommand{\be}{\begin{equation}}
\newcommand{\ee}{\end{equation}}
\newcommand{\nbe}{\begin{equation*}}
\newcommand{\nee}{\end{equation*}}

\newcommand{\fr}{\frac}
\newcommand{\lb}{\label}

\thispagestyle{empty}


{\hbox to\hsize{
\vbox{\noindent September 2020 \hfill IPMU20-0074 }}
\noindent ~revised version \hfill }

\noindent
\vskip2.0cm
\begin{center}

{\Large\bf Primordial black holes from modified supergravity}

\vglue.3in

Yermek Aldabergenov~${}^{a,b}$, Andrea Addazi~${}^{c,d}$, and Sergei V. Ketov~${}^{e,f,g}$
\vglue.3in

${}^a$~Department of Physics, Faculty of Science, Chulalongkorn University\\
Thanon Phayathai, Pathumwan, Bangkok 10330, Thailand\\
${}^b$~Institute of Experimental and Theoretical Physics, Al-Farabi Kazakh National University,
71 Al-Farabi Avenue, Almaty 050040, Kazakhstan\\
${}^c$~Center for Theoretical Physics, College of Physics, Science and Technology \\
Sichuan University,  610065 Chengdu,  China\\
${}^d$~INFN, Sezione Roma Tor Vergata, I-00133 Rome, Italy \\
${}^e$~Department of Physics, Tokyo Metropolitan University\\
1-1 Minami-ohsawa, Hachioji-shi, Tokyo 192-0397, Japan \\
${}^f$~Research School of High-Energy Physics, Tomsk Polytechnic University\\
2a Lenin Avenue, Tomsk 634028, Russian Federation\\
${}^g$~Kavli Institute for the Physics and Mathematics of the Universe (WPI)
\\The University of Tokyo Institutes for Advanced Study, Kashiwa 277-8583, Japan\\
\vglue.1in

 yermek.a@chula.ac.th, addazi@scu.edu.cn, ketov@tmu.ac.jp
\end{center}

\vglue.3in

\begin{center}
{\Large\bf Abstract}
\end{center}

The modified supergravity approach is applied to describe a formation of Primordial Black Holes (PBHs) after Starobinsky inflation. Our approach naturally leads to the two-(scalar)-field attractor-type double inflation, whose first stage is driven by scalaron and whose second stage is driven by another scalar field which belongs to a supergravity multiplet. The scalar potential and the kinetic terms are derived, the vacua are studied, and the inflationary dynamics of those two scalars is investigated. We numerically compute the power spectra and we find  the ultra-slow-roll regime leading to an enhancement (peak) in the scalar power spectrum. This leads to an efficient formation of PBHs. We  estimate the masses of PBHs and we find their density fraction (as part of Dark Matter). We show that our modified supergravity models are in agreement with inflationary observables, while they predict the PBH masses in a range between $10^{16}$ g and $10^{20}$ g. In this sense, modified supergravity provides a natural top-down approach for explaining and unifying the origin of inflation and the PBHs Dark Matter.  

\vglue.1in
\noindent

\newpage

\section {Introduction}

The prospect that Dark Matter (DM) is composed of Primordial Black Holes (PBHs) is an intriguing and highly motivated alternative to any particle physics explanations such as Weak Interacting Massive Particles (WIMPs), gravitino or axion dark matter. Indeed, such a possibility reverses the strategy for DM phenomenology: DM signals may appear in cosmological data rather than colliders, direct detection searches or indirect detection in astroparticle physics. The idea of PBHs was proposed by {\it Zeldovich} and {\it Novikov} \cite{Novikov:1967tw}, and then by {\it Hawking} \cite{Hawking:1971ei} who realized that some primordial density fluctuations may lead to PBH seeds in the early Universe. There are several mechanisms that may catalyze the formation of PBHs: (i) gravitational instabilities induced from scalar fields \cite{M1} such as axion-like particles or multi-field inflation, (ii) bubble-bubble collisions from first order phase transitions (see Refs.~\cite{M4,M5,Addazi:2018nzm} for recent discussions), and (iii) formation of critical topological defects such as cosmic strings \cite{Vilenkin:2018zol} and domain walls 
\cite{Belotsky:2018wph,Liu:2019lul} in the early Universe.

After accretion, some PBHs may survive in the Universe today and provide candidates for (non-particle) Dark Matter (DM) \cite{Barrow:1992hq}. More recently, PBHs attracted considerable attention in the literature, related 
to observational progress in lensing, cosmic rays, and Cosmic Microwave Background (CMB) radiation, see Refs.~\cite{Sasaki:2018dmp,Ketov:2019mfc} for a review of observational constraints on PBHs and their prospects for being a fraction of or a whole DM. 

On the theoretical side, PBHs are considered as a probe of very high energy physics and quantum gravity "even if they never formed" \cite{Carr:2003bj}. Numerous phenomenological scenarios were proposed for PBH formation and, especially, for PBH generation after inflation in the early Universe, under the assumption that PBHs contribute to DM
(see e.g., Refs.~\cite{Pi:2017gih,Germani:2018jgr,Fumagalli:2020adf,Palma:2020ejf,Cai:2019bmk,Cai:2018dig,Deng:2018wmy} and the references therein).  Indeed, the whole PBH DM case leaves only two limited windows for allowed PBH masses around either $10^{-15}$ or $10^{-12}$ of the Solar mass.

Therefore, it is of interest to study a possible theoretical origin of PBHs at a more fundamental level than General
Relativity (GR) by using string theory, as a candidate for quantum gravity, and supergravity as the first step in that direction. Moreover, because of the constraints imposed by local supersymmetry on possible couplings, a viable description of PBHs in supergravity may lead to significant discrimination of phenomenological models of inflation and PBHs. 

Due to the absence of large non-Gaussianities and isocurvature perturbations in the current observational CMB data \cite{Akrami:2018odb}, {\it single-field} inflationary models  were distinguished and discriminated within the large landscape of inflation mechanisms. The Starobinsky $R^2$ inflation \cite{Starobinsky:1980te} seems to be favored as the best phenomenological fit. Then a required growth (by a factor of $10^{7}$ compared to the CMB amplitude) of the amplitude of fluctuations to be responsible for PBH seeds can be achieved by modifying the inflaton scalar potential with a nearly inflection point \cite{Garcia-Bellido:2017mdw, Motohashi:2017kbs,Passaglia:2018ixg}. Details of the PBH production after single-field inflation are very much dependent upon a choice of inflaton potential. This requires a significant fine-tuning for PBHs as a candidate of DM. Standard (Einstein) supergravity can accommodate single-field inflationary models in the (new) minimal setup, with the only restriction to the inflaton potential as a real function squared \cite{Farakos:2013cqa,Ferrara:2013rsa,Aldabergenov:2016dcu,Aldabergenov:2017bjt}.~\footnote{See e.g., Ref.~\cite{Addazi:2018pbg} for a specific example of the inflaton potential with an inflection point in supergravity.} 

Since there are no fundamental reasons for the absence of non-Gaussianities and iso-curvature perturbations (they just have to be below the observational limits), {\it multi-field} inflationary models were also extensively studied. The required growth of primordial fluctuations can be achieved by tachyonic instabilities, say, in the waterfall phase of hybrid inflation 
\cite{GarciaBellido:1996qt,Kawasaki:2015ppx}.  Moreover, the PBH production may be a generic feature of {\it two-field} 
inflation \cite{Braglia:2020eai}. On the other side, multi-field inflation considerably extends a number of physical
degrees of freedom and possible interactions, which reduce predictive power.

Thus, supersymmetry is expected to be even more important in multi-field inflation by limiting the number of fields involved (in the minimal setup) and severely restricting their interactions.

As a guiding principle, in this paper we elaborate on a possible "supergravitational" origin of both inflation and PBHs, by using only supergravity fields and their locally supersymmetric interactions, without adding extra matter fields. Only the {\it minimal} number of the physical degrees of freedom associated with an $N=1$ full supergravity multiplet is used. In its spirit, our approach is similar to Starobinsky inflation based on gravitational interactions only (see Ref.~\cite{Ketov:2019toi} for a recent review of Starobinsky inflation in gravity and supergravity). The Starobinsky inflation is based on the modified $(R+\zeta R^2)$ gravity, which can be further extended to modified supergravity in the minimal setup \cite{Ketov:2012jt,Ketov:2013dfa,Addazi:2017rkc} leading to the effective {\it two-field} double inflation. We will show that the emerging double-field inflationary model is suitable for a formation of PBH seeds after the first inflation.
In this sense, Starobinsky supergravity naturally relates inflation with the dark matter genesis. 

Our paper is organized as follows. In Sec.~2 we introduce the general {\it modified} supergravity setup and give a specific example of the bosonic terms arising in the simplest non-trivial model. In Sec.~3 we introduce the duality transformations between the modified supergravity and the standard supergravity (in Jordan and Einstein frames) in terms of the field components (of the bosonic part) and in terms of the superfields. In Sec.~4 we study the vacuum structure of our basic model and the effective inflationary dynamics of its two scalars. Section 5 is devoted to an investigation of two-field inflation in our basic model defined by keeping only the leading terms in a generic modified supergravity action.  We demonstrate consistency of the basic model with CMB observations but also find the necessity of extreme fine-tuning of initial conditions for PBH generation. In Sec.~6 we extend our basic model by two subleading terms within  {\it the same} modified supergravity framework, and study in detail the two modifications of our basic model, corresponding to activation of only one of the two subleading terms. We numerically compute the power spectra, and estimate PBH masses and their density fraction, in both cases. We find that our extended models are capable to {\it simultaneously} describe viable (Starobinsky-type) inflation and PBH production after inflation, with limited fine-tuning of the parameters, and an attractor-type behavior in one of our models. In Sec.~7, we give our conclusions and comments. Some technical details are summarized in Appendices A and B. 

\section{Modified supergravity setup}

Let us consider a modified supergravity theory with the general Lagrangian (in curved superspace of the old-minimal supergravity in four spacetime dimensions, with $M_{\rm Pl}=1$) \cite{Cecotti:1987sa,Ketov:2013dfa}
\begin{equation}
    {\cal L}=\int d^2\Theta 2{\cal E}\left[-\frac{1}{8}(\overbar{\cal D}^2-8{\cal R})N({\cal R},\overbar{\cal R})+{\cal F}({\cal R})\right]+{\rm h.c.}~,\label{L_master}
\end{equation}
which is parametrized by two arbitrary functions, a non-holomorphic real potential $N$ and a holomorphic potential 
$\cal F$, of the covariantly chiral scalar curvature superfield $\cal R$ of the old-minimal supergravity.~\footnote{We use
the standard (Wess-Bagger) notation \cite{Wess:1992cp} for supergravity in superspace with a few adjustments mentioned in Appendix \ref{App_superspace}.}  Some relevant details about supergravity in superspace are collected in Appendix A. It should be mentioned that the master Eq.~(\ref{L_master}) goes beyond the supergravity textbooks and describes a {\it modified} supergravity because the standard (Einstein) supergravity
actions  are the extensions of Einstein-Hilbert term, whereas Eq.~(\ref{L_master}) is more general and reduces to the
pure Einstein supergravity action only in the very special case of $N=0$ and ${\cal F}=-3{\cal R}$. In other words, 
Eq.~(\ref{L_master}) can be considered as a {\it generic} modified supergravity extension of $(R+R^2)$ gravity (see below).  

Let us expand the functions $N$ and $\cal F$ in Taylor series and keep only the leading terms, as our first probe of modified supergravity. Then our simplest non-trivial ansatz reads
\begin{gather}
    N=\fracmm{12}{M^2}{\cal R\overbar{R}}-\fracmm{\xi}{2}({\cal R\overbar{R}})^2~,\quad
    {\cal F}=\alpha+3\beta{\cal R}~,\label{N_F_choice}
\end{gather}
where we have introduced the real parameters  $M$ and $\xi$, and the complex parameters  $\alpha$ and $\beta$.
The ansatz in Eq.~(\ref{N_F_choice}) was already proposed in Ref.~\cite{Addazi:2017rkc}, and it also
appeared in the dual scalar-tensor supergravity (see  Sec. 3) in Ref.~\cite{Kallosh:2013xya} where it was shown that the 
$\xi$-term is essential for curing a tachyonic instability of inflation.

After expanding the Lagrangian above in terms of the field components (see Appendix \ref{App_superspace} for the definitions of the field components), we obtain the bosonic part as follows:
\begin{align}
    e^{-1}{\cal L}=&-\fracmm{1}{12}\left[3(\beta+\bar{\beta})-\fracmm{24}{M^2}|X|^2+11\xi|X|^4-\fracmm{2}{9}\left(\fracmm{6}{M^2}-\xi|X|^2\right)b_mb^m\right]\left(R+\frac{2}{3}b_mb^m\right)+\nonumber\\
    &+\left(\fracmm{6}{M^2}-\xi|X|^2\right)\left(\fracmm{1}{72}R^2-2\partial_mX\partial^m\overbar{X}+\fracmm{1}{18}(\nabla_mb^m)^2-\fracmm{1}{162}(b_mb^m)^2\right)+\nonumber\\
    &+\fracmm{i}{2}(\beta-\bar{\beta})\nabla_mb^m-\fracmm{i}{3}\left(\fracmm{12}{M^2}-\xi|X|^2\right)b^m(\overbar{X}\partial_mX-X\partial_m\overbar{X})-U(X,\overbar{X})~,\label{L_master_comp}
\end{align}
where the scalar potential $U(X,\overbar{X})$ reads
\begin{equation}
    U=-6(\alpha\overbar{X}+\bar{\alpha}X)-6(\beta+\bar{\beta})|X|^2-\fracmm{48}{M^2}|X|^4+18\xi|X|^6~,\label{U(X)}
\end{equation}
and we demand ${\rm Re}\beta<0$ for the correct sign of the Einstein--Hilbert term.

The scalar potential \eqref{U(X)} has an anti-de-Sitter (AdS) minimum unless $\alpha$ vanishes, so we set $\alpha=0$. This uplifts the minimum at $X=0$ to a Minkowski vacuum provided that the parameters are chosen appropriately (see the next Sections). Then (at $X=0$) the canonical normalization of the Einstein--Hilbert term fixes $\beta=-1$ (or ${\rm Re}\beta=-1$ in general). Next, as will be shown below, the parameter $M$ will be the mass of Starobinsky scalaron, so that it can be fixed by identifying scalaron with inflaton via CMB measurements. Hence, we are left with a {\it single} free parameter $\xi$ that will determine the shape of the scalar potential.

\section{Dual supergravity}

It is remarkable that the higher-derivative modified supergravity (\ref{L_master}) can be transformed to the standard
supergravity (in Jordan frame, without higher derivatives) as was first demonstrated by {\it Cecotti} in 1987 \cite{Cecotti:1987sa}, similarly to the well known duality between a modified $f(R)$ gravity and a scalar-tensor gravity.  Moreover, a duality transformation can be done in the manifestly supersymmetric way, when using  superspace \cite{Gates:2009hu,Ketov:2013dfa}. In this Section, we first apply the duality transformation to the Lagrangian  \eqref{L_master_comp} in the familiar field components and then dualize the whole superfield action in 
Eq.~\eqref{L_master}. Of course, both approaches lead to the same physics and the Lagrangians coincide after some  field redefinitions, but only the superspace approach is manifestly supersymmetric.

\subsection{Dual bosonic part in field components}\label{Ss_dual_comp}

Let us introduce the notation 
\begin{equation}
    \fracmm{M^4\xi}{144}\equiv\zeta \quad {\rm and} \quad |X|\equiv\fracmm{M}{2\sqrt{6}}\,\sigma~,
\end{equation}
where $\sigma$ is the radial part of the complex scalar $X$. Its angular part (let us call it $\theta$) does not appear in the potential because we set 
$\alpha=0$.~\footnote{With $\alpha=0$ our model has the global R-symmetry under which $X$ is rotated by a phase.} We also set $\theta=b_m=0$ 
for simplicity.

Then the action \eqref{L_master_comp}  takes the form
\begin{equation}
    e^{-1}{\cal L}=\frac{1}{2}f(R,\sigma)-\frac{1}{2}(1-\zeta\sigma^2)(\partial\sigma)^2-U~,\label{L_f(R)}
\end{equation}
where we find
\begin{gather}
    f(R,\sigma)=\left(1+\frac{1}{6}\sigma^2-\frac{11}{24}\zeta\sigma^4\right)R+\fracmm{1}{6M^2}(1-\zeta\sigma^2)R^2~,\label{f(R,sigma)}\\
    U=\frac{1}{2}M^2\sigma^2\left(1-\frac{1}{6}\sigma^2+\frac{3}{8}\zeta\sigma^4\right)~.\label{U(sigma)}
\end{gather}

By following the standard procedure, we introduce the auxiliary field $\chi$ and rewrite the action as
\begin{equation}
    e^{-1}{\cal L}=\frac{1}{2}\left[f_\chi(R-\chi)+f\right]-\frac{1}{2}(1-\zeta\sigma^2)(\partial\sigma)^2-U~,\label{L_f(chi)}
\end{equation}
where $f_\chi\equiv\frac{\partial f}{\partial\chi}$, and $f\equiv f(\chi,\sigma)$ is the function \eqref{f(R,sigma)} with $R$ replaced by $\chi$. Varying with respect to $\chi$ leads to the action \eqref{L_f(R)}. A transfer to Einstein frame is obtained via Weyl rescaling,
\begin{gather}
    g_{mn}\rightarrow f_{\chi}^{-1}g_{mn}~,~~~e\rightarrow f_{\chi}^{-2}e~,\nonumber\\
    ef_\chi R\rightarrow eR-\frac{3}{2}ef^{-2}_{\chi}(\partial f_\chi)^2~.\label{Weyl_resc}
\end{gather}
Therefore, the function
\begin{gather}
    f_\chi=A+B\chi  \quad {\rm with}\nonumber\\
    A\equiv 1+\frac{1}{6}\sigma^2-\frac{11}{24}\zeta\sigma^4 \quad {\rm and} \quad B\equiv \fracmm{1}{3M^2}(1-\zeta\sigma^2)~,
\end{gather}
can be identified with Starobinsky scalaron that can be brought to the canonically normalized field $\varphi$ via the identification
\begin{equation}
    f_\chi=\exp{ \left[ \sqrt{\frac{2}{3}}\varphi\right] }~,
\end{equation}
so that
\begin{equation}
    \chi=\fracmm{1}{B}\left(e^{\sqrt{\frac{2}{3}}\varphi}-A\right) \quad {\rm and} \quad f=\fracmm{1}{2B}\left(e^{2\sqrt{\frac{2}{3}}\varphi}-A^2\right)~,
\end{equation}
which is essentially the change of variables from $\chi$ to $\varphi$.

After the Weyl rescaling \eqref{Weyl_resc} the Lagrangian \eqref{L_f(chi)} takes the following form in terms of the canonical scalaron $\varphi$:
\begin{equation}
    e^{-1}{\cal L}=\frac{1}{2}R-\frac{1}{2}(\partial\varphi)^2-\frac{1}{2}(1-\zeta\sigma^2)e^{-\sqrt{\frac{2}{3}}\varphi}(\partial\sigma)^2-V~,\label{L_varphi}
\end{equation}
where the two-field scalar potential reads
\begin{align}
    V&=\fracmm{1}{4B}\left(1-Ae^{-\sqrt{\frac{2}{3}}\varphi}\right)^2+e^{-2\sqrt{\frac{2}{3}}\varphi}U=\nonumber\\
    &=\fracmm{3M^2}{4(1-\zeta\sigma^2)}\left[1-e^{-\sqrt{\frac{2}{3}}\varphi}-\frac{1}{6}\sigma^2\left(1-\frac{11}{4}\zeta\sigma^2\right)e^{-\sqrt{\frac{2}{3}}\varphi}\right]^2+\fracmm{M^2}{2}e^{-2\sqrt{\frac{2}{3}}\varphi}\sigma^2\left(1-\frac{1}{6}\sigma^2+\frac{3}{8}\zeta\sigma^4\right).\label{V_varphi}
\end{align}

As is clear from the Lagrangian \eqref{L_varphi}, when $\sigma^2>1/\zeta$, the scalar $\sigma$ becomes a ghost. However, when approaching 
$\sigma^2=1/\zeta$, the potential \eqref{V_varphi} becomes singular, so that it would take the infinite amount of energy to turn $\sigma$ into a ghost (assuming  its starting value in the region $\sigma^2<1/\zeta$). It is also worth noticing that the inflaton mass $M$ enters the potential as the overall factor, so that it does 
not affect the shape of the potential.

\subsection{Superfield dual version}

As was demonstrated in Ref.~\cite{Ketov:2013dfa}, the dual superfield Lagrangian of Eq.~\eqref{L_master} is obtained by introducing the Lagrange multiplier (chiral) superfield $\bf T$ as~\footnote{We use the bold font for the chiral superfields $\bf S$ and $\bf T$, and the regular font for their leading field components.}
\begin{equation}
    {\cal L}=\int d^2\Theta 2{\cal E}\left\{-\frac{1}{8}(\overbar{\cal D}^2-8{\cal R})N({\bf S},\overbar{\bf S})+{\cal F}({\bf S})+6{\bf T}({\bf S}-{\cal R})\right\}+{\rm h.c.} \label{L_Lambda}
\end{equation}
Varying it with respect to $\bf T$ gives back the original Lagrangian \eqref{L_master} by identifying the chiral superfield $\bf S$ with $\cal R$.

When using instead the superspace identity
\begin{equation}
    \int d^2\Theta 2{\cal E}(\overbar{\cal D}^2-8{\cal R})({\bf T}+\overbar{\bf T})+{\rm h.c.}=-16\int d^2\Theta 2{\cal E}{\cal R}{\bf T}+{\rm h.c.}~,
\end{equation}
the Lagrangian \eqref{L_Lambda} can be rewritten to
\begin{equation}
    {\cal L}=\int d^2\Theta 2{\cal E}\left\{\frac{3}{8}(\overbar{\cal D}^2-8{\cal R})\left[{\bf T}+\overbar{\bf T}-\frac{1}{3}N({\bf S},\overbar{\bf S})\right]+{\cal F}({\bf S})+6{\bf T}{\bf S}\right\}+{\rm h.c.} \label{L_Lambda2}
\end{equation}
Given the functions $N({\cal R},\overbar{\cal R})$ and $\cal F({\cal R})$ according to Eq.~\eqref{N_F_choice}, the Lagrangian \eqref{L_Lambda2} can be rewritten to the standard form,
\begin{equation}
    {\cal L}=\int d^2\Theta 2{\cal E}\left[\frac{3}{8}(\overbar{\cal D}^2-8{\cal R})e^{-K/3}+W\right]+{\rm h.c.}~,\label{L_superfield_dual}
\end{equation}
where the K\"ahler potential $K$ and the superpotential $W$ are given by ({\it cf.} Ref.~\cite{Kallosh:2013xya}) 
\begin{gather}
    K=-3\log({\bf T}+\overbar{\bf T}-\tilde{N})~,~~~\tilde{N}\equiv\frac{1}{3}N={\bf S}\overbar{\bf S}-\frac{3}{2}\zeta({\bf S}\overbar{\bf S})^2~,\label{Kael}\\
    W=3M{\bf S}\left({\bf T}-\frac{1}{2}\right)~, \label{supV}
\end{gather}
after rescaling ${\bf S}\rightarrow M{\bf S}/2$ and using the parameter $\zeta\equiv M^4\xi/144$.

It is straightforward to derive the corresponding bosonic terms in field components. We find
\begin{equation}
    e^{-1}{\cal L}=\frac{1}{2}R-K_{i\bar{j}}\partial_m\Phi^i\partial^m\overbar{\Phi}^j-e^K\left(K^{i\bar{j}}D_iW D_{\bar{j}}\overbar{W}-3|W|^2\right)~,\label{L_Phi}
\end{equation}
where $\Phi^i=(T,S)$, $i=1,2$, and the K\"ahler metric reads
\begin{equation}
K_{i\bar{j}}=
\begin{pmatrix}
K_{T\overbar{T}} & K_{T\overbar{S}}\\
K_{S\overbar{T}} & K_{S\overbar{S}} 
\end{pmatrix}=
\fracmm{3}{P^2}\begin{pmatrix}
1 & -\tilde{N}_{\bar{S}}\\
-\tilde{N}_{S} & \tilde{N}_{S}\tilde{N}_{\bar{S}}+P\tilde{N}_{S\bar{S}}
\end{pmatrix}
\end{equation}
with $P\equiv T+\overbar{T}-\tilde{N}$. The inverse K\"ahler metric is given by
\begin{equation}
K^{i\bar{j}}=
\begin{pmatrix}
K^{T\overbar{T}} & K^{T\overbar{S}}\\
K^{S\overbar{T}} & K^{S\overbar{S}} 
\end{pmatrix}=
\fracmm{P}{3}\begin{pmatrix}
P+\tilde{N}^{S\overbar{S}}\tilde{N}_S\tilde{N}_{\overbar{S}} & \tilde{N}^{S\overbar{S}}\tilde{N}_S\\
\tilde{N}^{S\overbar{S}}\tilde{N}_{\overbar{S}} & \tilde{N}^{S\overbar{S}}
\end{pmatrix}~.
\end{equation}

Because of the non-vanishing non-diagonal elements of the K\"ahler metric, the kinetic part of the Lagrangian mixes the derivatives of $S$ and $T$,
\begin{equation}
    e^{-1}{\cal L}_{\rm kin}=-\fracmm{3}{P^2}\left[\partial T\partial\overbar{T}-\tilde{N}_S\partial S\partial\overbar{T}-\tilde{N}_{\overbar{S}}\partial T\partial\overbar{S}+(\tilde{N}_S\tilde{N}_{\overbar{S}}+P\tilde{N}_{S\overbar{S}})\partial S\partial\overbar{S}\right]~.
\end{equation}

In order to bring the Lagrangian to the form \eqref{L_varphi}, where the contributions of $b_m$~\footnote{The vector field $b_m$ in higher-derivative supergravity contributes a physical \textit{scalar} $\nabla_m b^m$, as was shown in Ref. \cite{Ketov:2013dfa}. In the dual matter-coupled supergravity, this scalar can be associated with the axionic field ${\rm Im}T$.} and the angular part of $X$ are ignored, we set ${\rm Im}T=0$, $S=|S|$, and denote $|S|=\sigma/\sqrt{6}$. The kinetic mixing between $\partial S$ and $\partial T$ can be eliminated by using $P=T+\overbar{T}-\tilde{N}$ as the independent (real) scalar instead of ${\rm Re}T$. We get the canonical normalization of its kinetic term in the parametrization $P=\exp{\left[\sqrt{\frac{2}{3}}\varphi
\right]}$ as follows:
\begin{equation}
    e^{-1}{\cal L}_{\rm kin}=-\frac{1}{2}(\partial\varphi)^2-\frac{1}{2}(1-\zeta\sigma^2)e^{-\sqrt{\frac{2}{3}}\varphi}(\partial\sigma)^2~,
\end{equation}
that exactly matches the kinetic part of Eq.~\eqref{L_varphi}. 

It is also straightforward (albeit tedious) to check that the scalar potential of Eq.~\eqref{L_Phi} also coincides with that of Eq.~\eqref{V_varphi} after using the field redefinitions diagonalizing the scalar kinetic matrix above.

\section{Critical points}

To study vacuum equations in our basic model, we denote $e^{-\sqrt{\frac{2}{3}}\varphi}\equiv x$ and rewrite the scalar potential \eqref{V_varphi} as
\begin{equation}
    V=\fracmm{1}{4B}\left(1-Ax\right)^2+x^2U~, {\rm where} 
\begin{cases}
A=1+\fracmm{1}{6}\sigma^2-\fracmm{11}{24}\zeta\sigma^4~,\\
B=\fracmm{1}{3M^2}(1-\zeta\sigma^2)~,\\
U=\fracmm{M^2}{2}\sigma^2\left(1-\fracmm{1}{6}\sigma^2+\fracmm{3}{8}\zeta\sigma^4\right)~.
\end{cases}\label{V_vac}
\end{equation}

The equations for critical points read
\begin{align}
    \partial_x V&=\fracmm{A}{2B}(Ax-1)+2xU=0~,\label{V_x}\\
    \partial_\sigma V&=\fracmm{2xA'B+(1-Ax)B'}{4B^2}(Ax-1)+x^2U'=0~,\label{V_sigma}
\end{align}
where the primes denote the derivatives with respect to $\sigma$. A simple solution to these equations is
\begin{equation}
    Ax=1~,\quad U=U'=0~.\label{AxUUp}
\end{equation}
It gives rise to the vanishing potential \eqref{V_vac} for $\sigma_0=\varphi_0=0$. There is another solution by taking $U'=0$ and obtaining
\begin{equation}
    \sigma^2=\fracmm{2}{27\zeta}(2\pm\sqrt{4-162\zeta})~.\label{sigma_11}
\end{equation}
On the other hand, the condition $U=0$ is solved by
\begin{equation}
    \sigma^2=\fracmm{2}{9\zeta}(1\pm\sqrt{1-54\zeta})~.\label{sigma_12}
\end{equation}
Equating Eqs.~\eqref{sigma_11} and \eqref{sigma_12} leads to an equation on the parameter $\zeta$ with a solution $\zeta=1/54\approx 0.019$ provided that the "plus" branch is chosen in Eq.~\eqref{sigma_11}. It means, when $\zeta=1/54$, we have three Minkowski minima: $\sigma_0$ and $\pm|\sigma_1|$ where $\sigma_1^2$ is given by Eqs.~\eqref{sigma_11} or \eqref{sigma_12}.

\begin{figure}
\centering
\begin{subfigure}{.49\textwidth}
  \centering
  \includegraphics[width=1\linewidth]{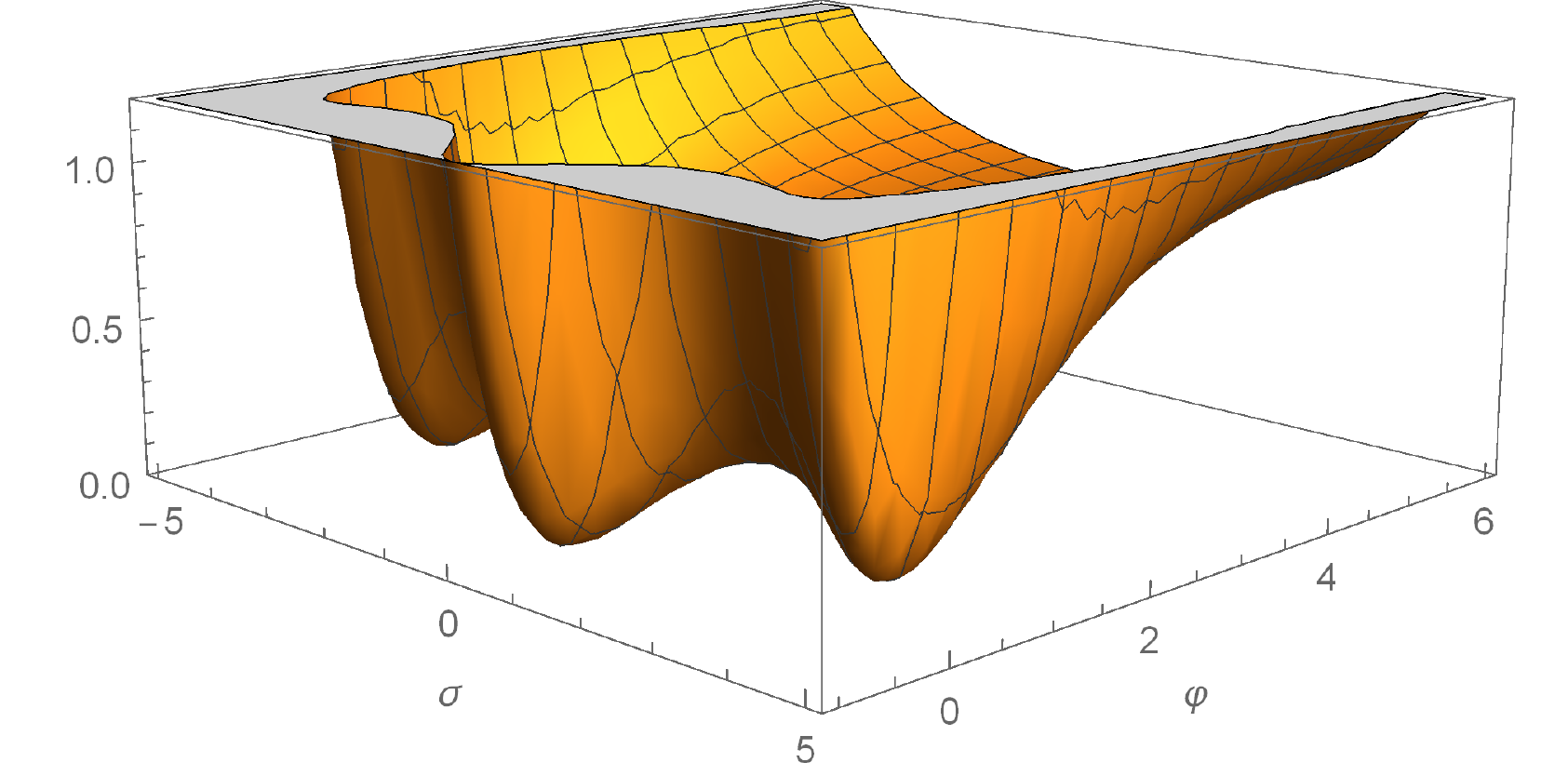}
  \caption{}
  \label{V_3d1}
\end{subfigure}
\begin{subfigure}{.49\textwidth}
  \centering
  \includegraphics[width=1\linewidth]{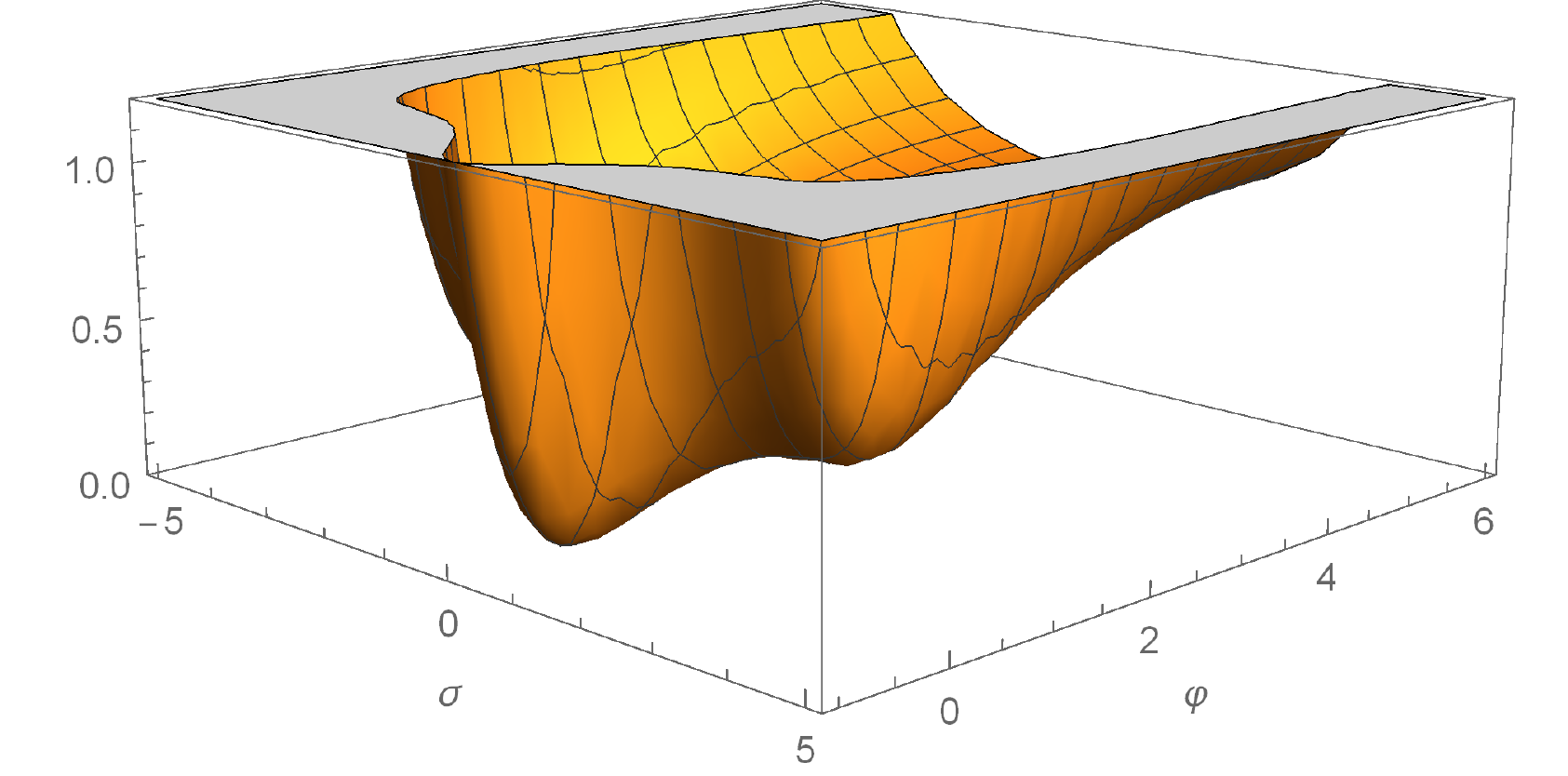}
  \caption{}
  \label{V_3d2}
\end{subfigure}
\captionsetup{width=.9\linewidth}
\caption{The scalar potential $V/M^2$ of Eq.~\eqref{V_vac}. The plot (a): $\zeta=1/54\approx 0.019$ with three Minkowski minima. The plot (b): $\zeta=0.027$ with a single Minkowski minimum at $\sigma=0$ and two inflection points.}
\label{V_3d}
\end{figure}

When $\zeta\neq 1/54$, the two minima at $\pm\sigma_1$ are not given by Eqs.~\eqref{sigma_11} and \eqref{sigma_12}, being more general solutions to the vacuum equations \eqref{V_x} and \eqref{V_sigma}. In particular, when  $0<\zeta<1/54$, the minima at $\pm\sigma_1$ are AdS, while for $1/54<\zeta<0.027$ the minima are uplifted to metastable de-Sitter (dS). When $\zeta\approx 0.027$, there are two inflection points, whereas for $\zeta>0.027$ all the critical points, except of $\sigma=0$, disappear. The scalar potential $V/M^2$ is shown in Fig.~\ref{V_3d1} (at $\zeta=1/54$) and Fig.~\ref{V_3d2} (at $\zeta=0.027$).

Let us comment on the scalar masses for the model \eqref{Kael}\eqref{supV}. Expanding around the Minkowski vacuum at $\varphi=\sigma=0$, we find $M_\varphi=M_\sigma=M$, where $M_\varphi$ and $M_\sigma$ are the masses of $\varphi$ and $\sigma$, respectively. As for the axion ${\rm Im}T$, after its proper normalization we find that it also has the mass $M$. However, the last scalar $\theta$ has vanishing mass around the minimum, so it must be generated by additional means. This, together with the more detailed analysis of the dynamics of ${\rm Im}T$ and $\theta$ deserves a separate investigation that we leave to future works. Here we will focus on the two scalars $\varphi$ and $\sigma$.

\section{Two-field inflationary dynamics}

Having derived the Lagrangian with the two-field scalar potential from the modified supergravity, in this Section we investigate its suitability for describing cosmological inflation in agreement with CMB observations.

\subsection{Field equations}\label{field_eqns}

The Lagrangian in Eqs.~\eqref{L_varphi} and \eqref{V_varphi} takes the form of a Non-Linear 
Sigma-Model (NLSM) minimally coupled to gravity,
\begin{equation}
    e^{-1}{\cal L}=\frac{1}{2}R-\frac{1}{2}G_{AB}\partial\phi^A\partial\phi^B-V~,\label{L_NLSM}
\end{equation}
where $\phi^A=\{\varphi,\sigma\}$, $A=1,2$, and the NLSM metric is given by
\begin{equation}
    G_{AB}=
\begin{pmatrix}
1 & 0\\
0 & (1-\zeta\sigma^2)e^{-\sqrt{\frac{2}{3}}\varphi} 
\end{pmatrix}~.
\end{equation}

Varying the Lagrangian \eqref{L_NLSM} with respect to the scalar fields yields equations of motion in the form
\begin{equation}
    \Box\phi^C+\Gamma^C_{AB}\partial\phi^A\partial\phi^B=G^{AC}\partial_A V~,
\end{equation}
where $\Box\equiv\nabla_m\nabla^m$ is the spacetime Laplace-Beltrami operator, and $\Gamma^C_{AB}$ are the Christoffel symbols of the NLSM target space. The non-vanishing Christoffel symbols are \begin{equation}
    \Gamma^\sigma_{\sigma\varphi}=-\fracmm{1}{\sqrt{6}}~,\quad \Gamma^\varphi_{\sigma\sigma}=\fracmm{1}{\sqrt{6}}(1-\zeta\sigma^2)e^{-\sqrt{\frac{2}{3}}\varphi}~,\quad \Gamma^\sigma_{\sigma\sigma}=-\fracmm{\zeta\sigma}{1-\zeta\sigma^2}~~~.
\end{equation}
After using these results, and the Friedmann--Lemaitre--Robertson--Walker (FLRW) spacetime metric $g_{mn}={\rm diag}(-1,a^2,a^2,a^2)$ with the time-dependent scale factor $a(t)$, the equations of motion take the form
\begin{gather}
    \ddot\varphi+3H\dot{\varphi}+\fracmm{1}{\sqrt{6}}(1-\zeta\sigma^2)e^{-\sqrt{\frac{2}{3}}\varphi}\dot{\sigma}^2+\partial_\varphi V=0~,\label{KG1}\\
    \ddot\sigma+3H\dot{\sigma}-\fracmm{\zeta\sigma\dot{\sigma}^2}{1-\zeta\sigma^2}-\sqrt{\fracmm{2}{3}}\dot{\varphi}\dot{\sigma}+\fracmm{e^{\sqrt{\frac{2}{3}}\varphi}}{1-\zeta\sigma^2}\partial_\sigma V=0~,\label{KG2}
\end{gather}
where the dots stand for the time derivatives.

The Friedmann equations for the system \eqref{L_NLSM} read
\begin{align}
    3H^2&=\frac{1}{2}\dot{\varphi}^2+\frac{1}{2}(1-\zeta\sigma^2)e^{-\sqrt{\frac{2}{3}}\varphi}\dot{\sigma}^2+V~,\label{Fried1}\\
    \dot{H}&=-\frac{1}{2}\dot{\varphi}^2-\frac{1}{2}(1-\zeta\sigma^2)e^{-\sqrt{\frac{2}{3}}\varphi}\dot{\sigma}^2~,\label{Fried2}
\end{align}
where the Hubble function has been introduced, $H\equiv \dot{a}/a$.

For numerical computations it is useful to rescale time as $\tilde{t}\equiv Mt$ (when using $\tilde{t}$, the dots will denote the derivatives with respect to $\tilde{t}$) with the rescaled Hubble function $\tilde{H}=H/M$.

\subsection{Inflationary parameters}

In this Subsection we employ the covariant formalism that is well known in the literature, see e.g., 
Refs.~\cite{Schutz:2013fua,Gundhi:2018wyz,Canko:2019mud} and the references therein, with 
the slow-roll parameter
\begin{equation}
    \epsilon \equiv -\fracmm{\dot{H}}{H^2}=-\fracmm{\dot{\tilde{H}}}{\tilde{H}^2}~~~.\label{epsilon}
\end{equation}

In a two-field analysis, it is useful to define the field-space velocity and acceleration (turn rate) unit vectors as
\begin{equation}
    \Sigma^A\equiv\fracmm{\dot{\phi}^A}{|\dot{\phi}|}~~,\quad \Omega^A\equiv\fracmm{\omega^A}{|\omega|}~~,\label{sigma_Omega}
\end{equation}
respectively, where the absolute value of a field-space vector $a^A$ is defined by $|a|\equiv\sqrt{G_{AB}a^Aa^B}$, and 
the acceleration vector $\omega^A$ is defined by
\begin{equation}
    \omega^A\equiv \dot{\Sigma}^A+\Gamma^A_{BC}\Sigma^B\dot{\phi}^C~~~
    \begin{cases}
    \omega^\varphi=\dot{\Sigma}^\varphi+\fracmm{1}{\sqrt{6}}(1-\zeta\sigma^2)e^{-\sqrt{\frac{2}{3}}\varphi}\Sigma^\sigma\dot{\sigma}~,\\
    \omega^\sigma=\dot{\Sigma}^\sigma-\fracmm{1}{\sqrt{6}}(\Sigma^\varphi\dot{\sigma}+\Sigma^\sigma\dot{\varphi})-\fracmm{\zeta\sigma}{1-\zeta\sigma^2}\Sigma^\sigma\dot{\sigma}~.
    \end{cases}
\end{equation}

Another useful quantity is the effective mass matrix,
\begin{equation}
    {\cal M}^A_B\equiv G^{AC}\nabla_B\partial_C V-R^A_{CDB}\dot{\phi}^C\dot{\phi}^D~,
\end{equation}
where $R^A_{CDB}$ is the Riemann tensor of the NLSM scalar manifold, with the non-vanishing components
\begin{equation}
    {R^{\varphi}}_{\sigma\sigma\varphi}=\fracm{1}{6}(1-\zeta\sigma^2)e^{-\sqrt{\frac{2}{3}}\varphi}~,\quad {R^{\sigma}}_{\varphi\varphi\sigma}=\fracm{1}{6}~~.
\end{equation}

With the above definitions we can introduce the adiabatic and isocurvature parameters
\begin{equation}
    \eta_{\Sigma\Sigma}\equiv\fracmm{{\cal M}^A_B\Sigma_A\Sigma^B}{V}~,\quad \eta_{\Omega\Omega}\equiv\fracmm{{\cal M}^A_B\Omega_A\Omega^B}{V}~,
\end{equation}
respectively, where $\eta_{\Sigma\Sigma}$ plays the role of the second slow-roll parameter, while $\eta_{\Omega\Omega}$ is proportional to the effective isocurvature mass.

The transfer functions are defined as follows:
\begin{align}
    T_{SS}(t_1,t_2)&\equiv\exp\left[\int^{t_2}_{t_1}dt'\beta(t')H(t')\right]~,\\
    T_{RS}(t_1,t_2)&\equiv 2\int^{t_2}_{t_1}dt'|\omega(t')|T_{SS}(t_1,t_2)~,\label{Transfer_fns}
\end{align}
where
\begin{equation}
    \beta(t)\equiv -2\epsilon+\eta_{\Sigma\Sigma}-\eta_{\Omega\Omega}-\fracmm{4|\omega|^2}{3H^2}~.
\end{equation}
The transfer functions describe the evolution of perturbations on superhorizon scales, i.e. from the moment of horizon exit $t_1$ (of the $k$-mode of interest) until some later time $t_2$.

The inflationary observables (CMB tilts) can be computed as (by assuming that isocurvature modes are suppressed)
\begin{equation}
    n_s=1-6\epsilon+2\eta_{\Sigma\Sigma} \quad {\rm and }\quad r=\fracmm{16\epsilon}{1+T_{RS}^2}~.\label{nsr_def}
\end{equation}
As $T_{RS}$ is real, the maximum value of the tensor-to-scalar ratio is $r_{\rm max}=16\epsilon$, while it can be computed without the transfer functions. In Appendix \ref{App_transfer} we estimate both $T_{\rm RS}$ and $T_{\rm SS}$ 
and find them negligible. Therefore, isocurvature effects can be ignored at CMB scales indeed.

According to the latest PLANCK data \cite{Akrami:2018odb}, the observed values of $n_s$ and $r$ are
\begin{equation}
    n_s=0.9649\pm 0.0042~{\rm (1\sigma~CL)}\quad {\rm and} \quad r <0.064~{\rm (2\sigma~CL)}~.\label{nsr_obs}
\end{equation}

\subsection{Inflationary solutions}

Let us first consider the case of $\zeta=1/54\approx 0.019$ with three Minkowski minima in Fig.~\ref{V_3d1}.

\begin{figure}
\centering
\begin{subfigure}{.49\textwidth}
  \centering
  \includegraphics[width=0.85\linewidth]{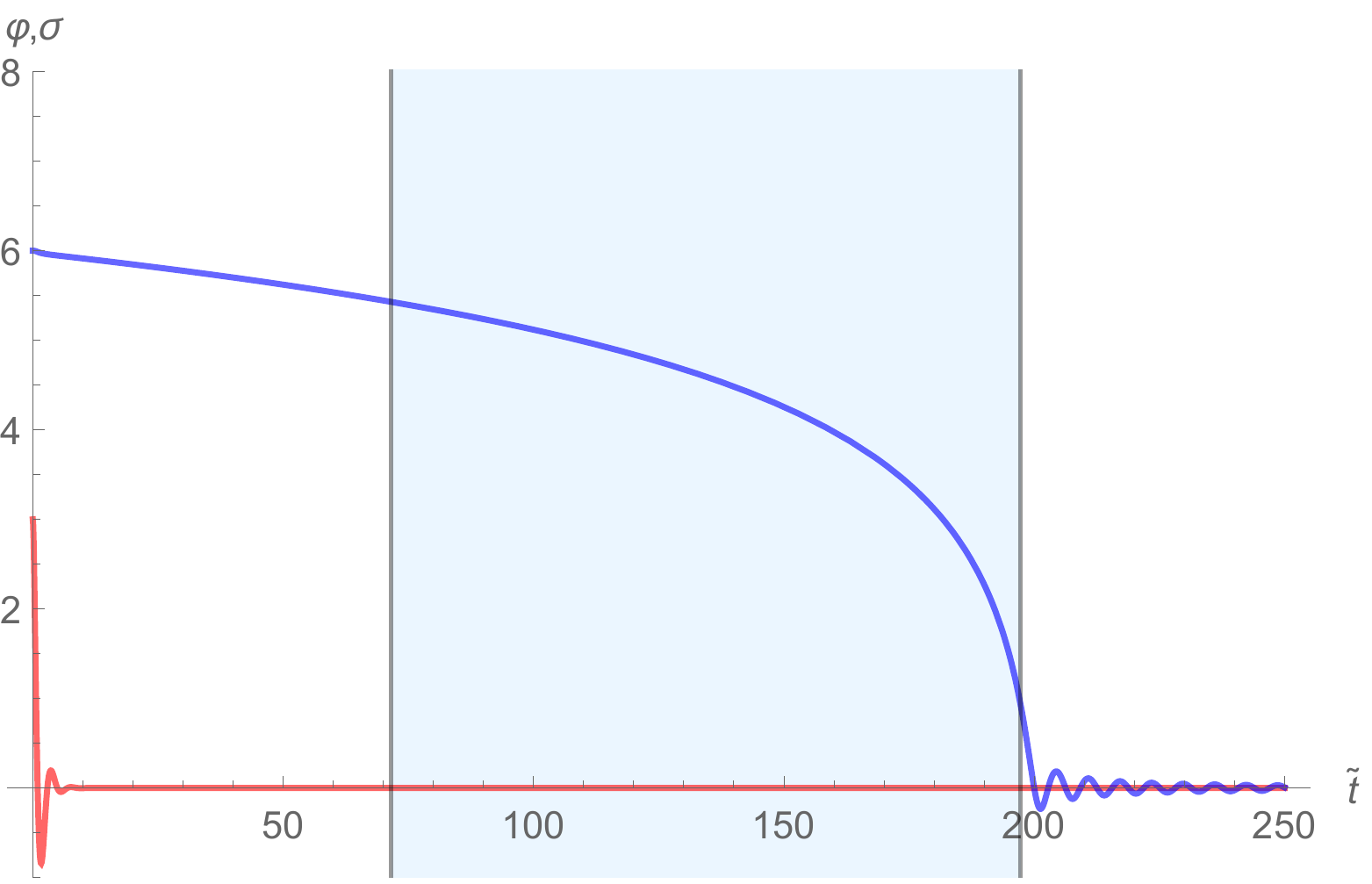}
  \caption{The solutions $\varphi(\tilde{t})$ (blue) and $\sigma(\tilde{t})$ (red).}
  \label{fs_sol_1}
\end{subfigure}
\begin{subfigure}{.49\textwidth}
  \centering
  \includegraphics[width=.75\linewidth]{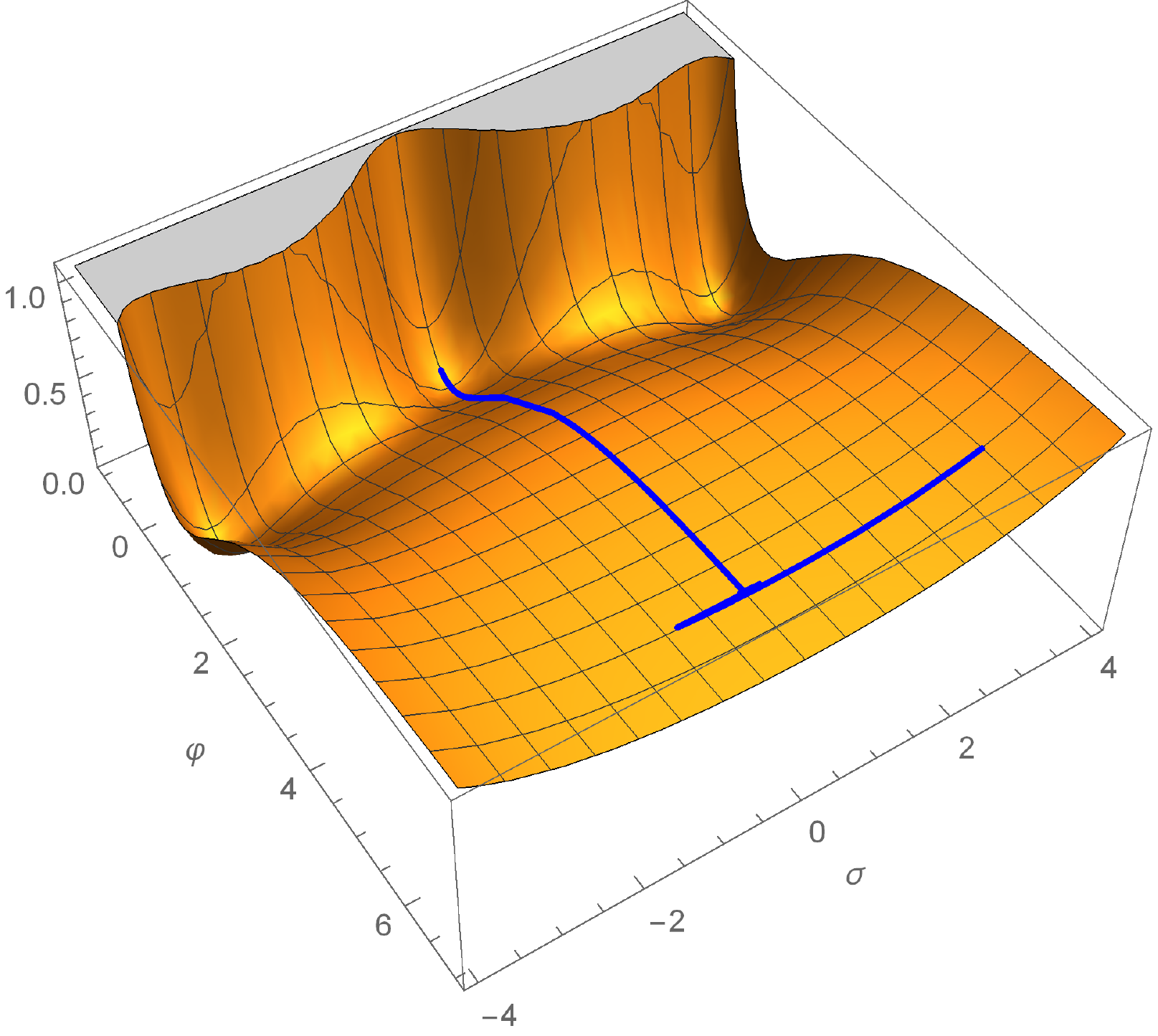}
  \caption{The field-space trajectory of the solutions.}
  \label{V_sol_1}
\end{subfigure}
\captionsetup{width=.9\linewidth}
\caption{The solutions (I) to the field equations \eqref{KG1} and \eqref{KG2} with the initial conditions 
$\varphi(0)=6,\sigma(0)=3$ and the vanishing initial velocities. The blue shaded region represents the time period of the last 60 
e-foldings.}
\label{fsV_sol_1}
\end{figure}

We numerically solve the field equations \eqref{KG1}, \eqref{KG2} and \eqref{Fried1} with the initial conditions 
$\varphi(0)=6,\sigma(0)=3,$ and the vanishing initial velocities, so let us call it the solution (I). The scalar field solutions are plotted in Figure \ref{fs_sol_1}, and their trajectories in the scalar potential are plotted in Figure \ref{V_sol_1}. It can be seen that $\sigma$ quickly drops to its minimum $\sigma=0$, so that the trajectory becomes similar to that in the single-field Starobinsky inflation. In fact, this is a {\it generic} feature when the initial velocities are zero (or almost zero), $\varphi(0)\gtrsim 6$ and $|\sigma(0)|\lesssim\sigma_{\rm max}$, where $\sigma_{\rm max}=1/\sqrt{\zeta}$ is the upper bound on $\sigma$ where the potential is infinite. When $\zeta=1/54$ we find $\sigma_{\rm max}\approx 7.35$.

The solution (I) leads to the spectral tilt and the tensor-to-scalar ratio as $n_s\approx 0.9624$ and 
$r_{\rm max}\approx 0.004$,~\footnote{We evaluate $n_s$ and $r$ at the CMB pivot scale $k=0.05~{\rm Mpc}^{-1}$   identified with the scale exiting the horizon around $54$ e-folds before the end of inflation, and assume the standard reheating temperature of the order $10^9$ GeV \cite{Liddle:1994dx} that is also expected in the modified supergravity setup \cite{Terada:2014uia}.} which are consistent with the observed values and the theoretical (Starobinsky) predictions of chaotic single-field inflation.

As the initial value $\sigma(0)$ approaches $\sigma_{\rm max}$ and/or as the initial velocities become non-negligible, the trajectory starts to curve. Also a smaller value of $\varphi(0)$ makes it easier to curve the trajectory.

\begin{figure}
\centering
\begin{subfigure}{.49\textwidth}
  \centering
  \includegraphics[width=0.85\linewidth]{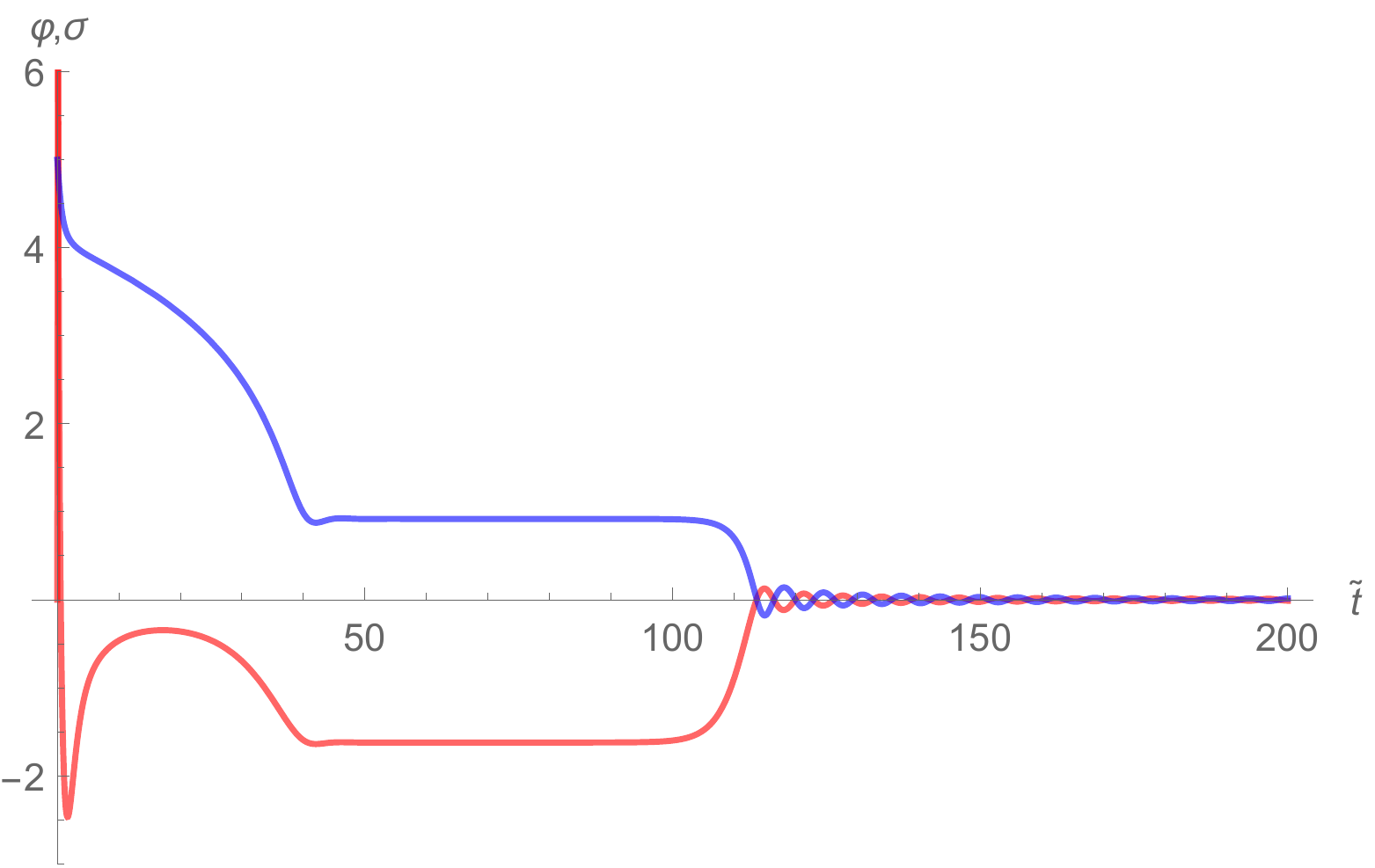}
  \caption{The solutions $\varphi(\tilde{t})$ (blue) and $\sigma(\tilde{t})$ (red).}
  \label{fs_sol_2}
\end{subfigure}
\begin{subfigure}{.49\textwidth}
  \centering
  \includegraphics[width=.75\linewidth]{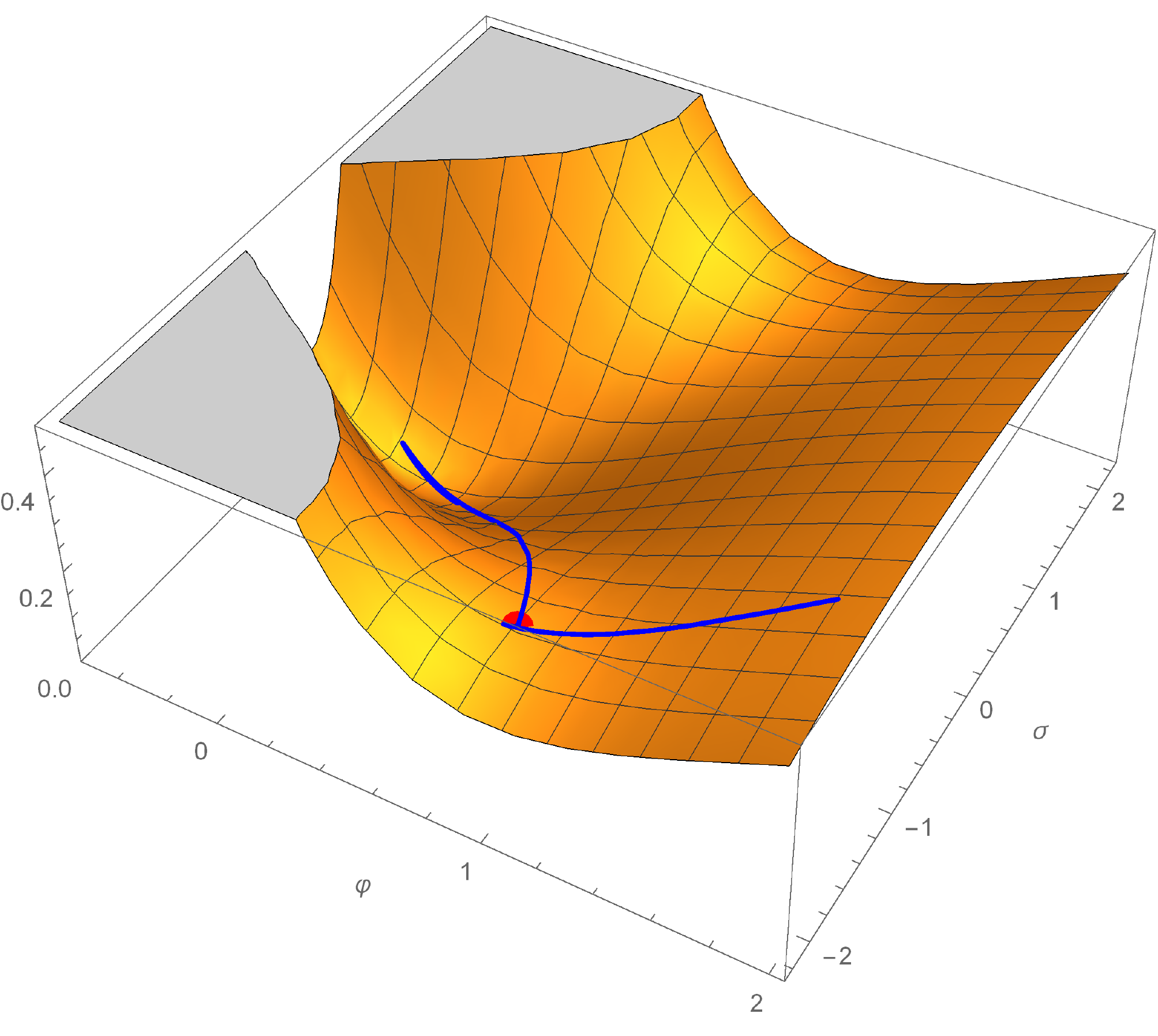}
  \caption{The field-space trajectory of the solutions.}
  \label{V_sol_2}
\end{subfigure}
\begin{subfigure}{.49\textwidth}
  \centering
  \includegraphics[width=.85\linewidth]{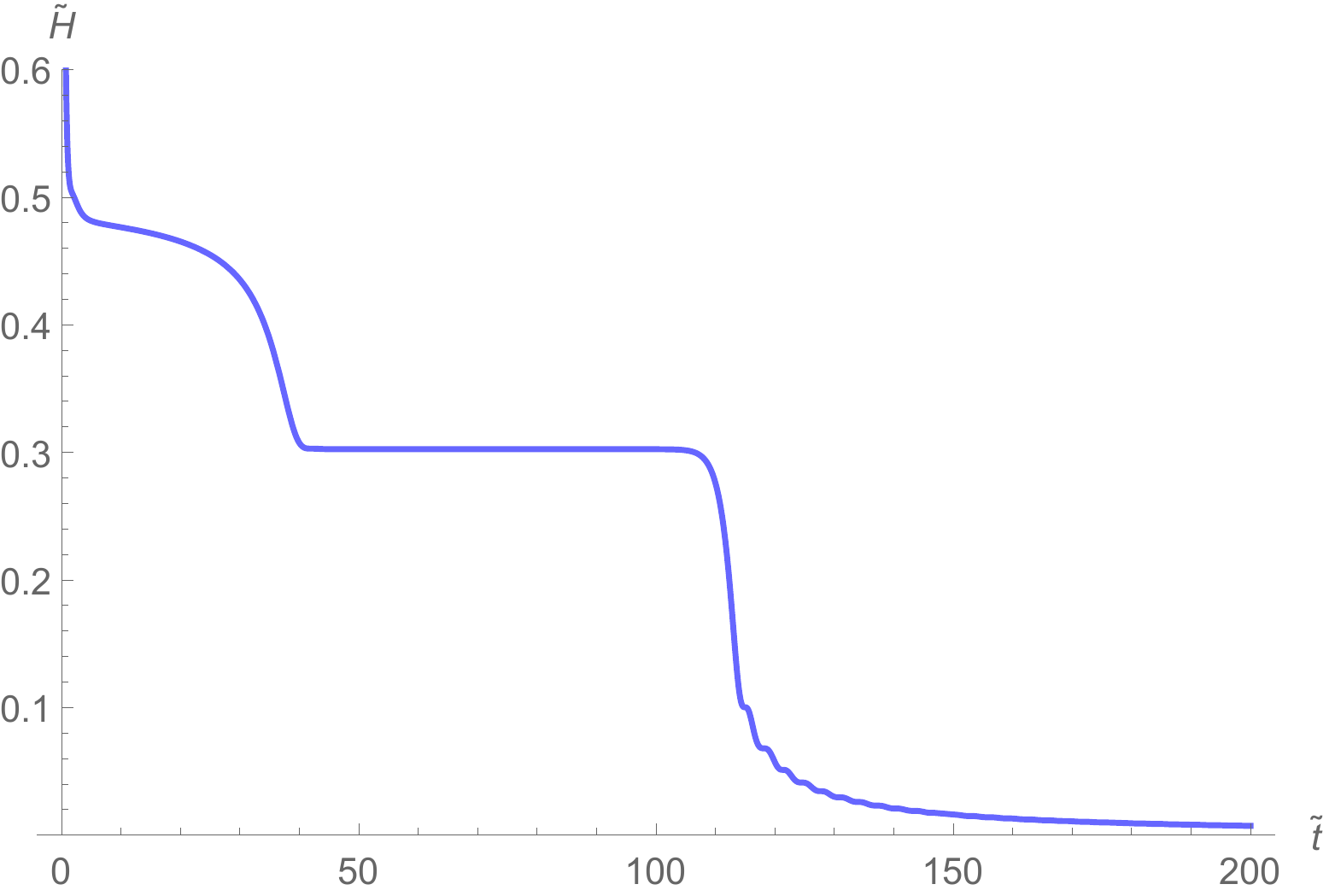}
  \caption{The Hubble function  $\tilde{H}(\tilde{t})$.}
  \label{H_sol_2}
\end{subfigure}
\begin{subfigure}{.49\textwidth}
  \centering
  \includegraphics[width=.85\linewidth]{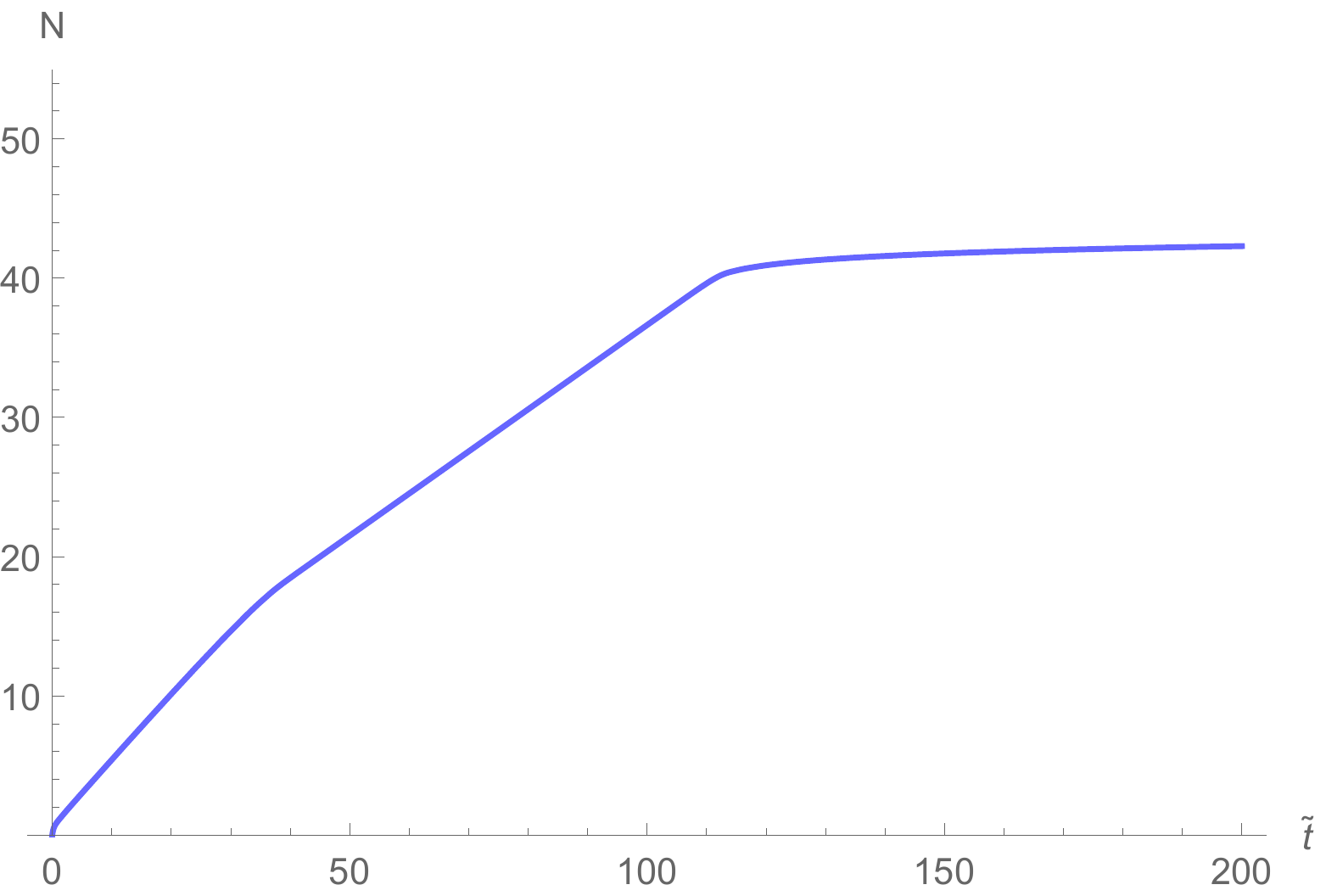}
  \caption{The e-foldings number $N(\tilde{t})$.}
  \label{N_sol_2}
\end{subfigure}
\captionsetup{width=.9\linewidth}
\caption{The solutions (II) to the field equations \eqref{KG1} and \eqref{KG2} with the initial conditions \eqref{IC2}. The red spot in (\ref{V_sol_2}) represents one of the two saddle points of the potential.}
\label{fsV_sol_2}
\end{figure}

As regards PBH production after inflation, let us consider the field-space trajectory going through the {\it saddle} point of the potential, that is a maximum in the $\sigma$-direction and a (local) minimum in the $\varphi$-direction. Then the saddle point divides inflation into two stages. We found a set of initial conditions that leads to such trajectory with
\begin{equation}
    \varphi(0)=5~,\quad \dot{\sigma}(0)=79.784527415607~, \quad \sigma(0)=\dot{\varphi}(0)=0~.\label{IC2}
\end{equation}
Let us call the corresponding solution as the solution (II).  We include its plots in Figures \ref{fs_sol_2} and \ref{V_sol_2}. The time-dependence of the Hubble function and the e-foldings number  defined by $\dot{N}=H$  are shown in Figures \ref{H_sol_2} and \ref{N_sol_2}. The total number of e-foldings is around $40$, though it can be larger for  larger values of $\varphi(0)$ with more fine-tuning of the initial velocities.

Thus, in order to achieve the two-stage inflation, where the field-space trajectory passes through the saddle point, we have to fine-tune the initial conditions as in Eq.~\eqref{IC2}, though the last choice is not unique. The reason is, when $\varphi$ is large, the potential takes the shape of a valley with the minima at $\sigma=0$, so a generic behavior of $\sigma$ is to quickly relax at $\sigma=0$, and let $\varphi$ drive the entire inflationary period. The same remains true if we change the shape of the potential as in Figure \ref{V_3d2} by changing $\zeta$ (the only difference is the saddle point to be replaced by an inflection point).

\section{Generalized attractor-type models}

Having learned the lessons in the previous Sections, we conclude that our basic ansatz in Eq.~\eqref{N_F_choice} 
for functions $N({\cal R},\overbar{\cal R})$ and ${\cal F}({\cal R})$ is too restrictive because it requires extreme fine tuning of the initial conditions for PBHs production. Therefore, we generalize our ansatz by adding the next-order corrections as
\begin{gather}
    N=\fracmm{12}{M^2}|{\cal R}|^2-\fracmm{72}{M^4}\zeta|{\cal R}|^4-\fracmm{768}{M^6}\gamma|{\cal R}|^6~~,
    \label{N_choice2}\\
    {\cal F}=-3{\cal R}+\fracmm{3\sqrt{6}}{M}\delta {\cal R}^2~~,\label{N_F_choice2}
\end{gather}
where we have introduced two new parameters $\gamma$ and $\delta$ with their normalization chosen for later convenience. We keep $\alpha=0$, $\beta=-3$ and $\zeta\equiv M^4\xi/144$,  ignore $b_m$ and the angular mode of 
${\cal R}|=X$, and set $X=M\sigma/\sqrt{24}$, as in the previous Sections. In the framework of the dual matter-coupled supergravity \eqref{L_superfield_dual}, the $\gamma$-term resides in the K\"ahler potential that can be affected by quantum corrections, whereas the $\delta$-term resides in the superpotential that does not receive
 (perturbative) quantum corrections.

After repeating the procedure outlined in Subsection \ref{Ss_dual_comp}, we obtain the Einstein frame Lagrangian as follows:
\begin{equation}
    e^{-1}{\cal L}=\fracmm{1}{2}R-\fracmm{1}{2}(\partial\varphi)^2-\fracmm{3M^2}{2}Be^{-\sqrt{\frac{2}{3}}\varphi}(\partial\sigma)^2-\fracmm{1}{4B}\left(1-Ae^{-\sqrt{\frac{2}{3}}\varphi}\right)^2-e^{-2\sqrt{\frac{2}{3}}\varphi}U~,\label{L_varphi2}
\end{equation}
where the functions $A,B,U$ are given by
\begin{align}
    A&=1-\delta\sigma+\frac{1}{6}\sigma^2-\frac{11}{24}\zeta\sigma^4-\frac{29}{54}\gamma\sigma^6~,\nonumber\\
    B&=\fracmm{1}{3M^2}(1-\zeta\sigma^2-\gamma\sigma^4)~,\label{ABU_tilde}\\
    U&=\fracmm{M^2}{2}\sigma^2\left(1+\frac{1}{2}\delta\sigma-\frac{1}{6}\sigma^2+\frac{3}{8}\zeta\sigma^4+\frac{25}{54}\gamma\sigma^6\right)~.\nonumber
\end{align}
When $\gamma=\delta=0$, all that reduces to Eqs.~\eqref{L_varphi} and \eqref{V_varphi}, as it should. Similary to the basic model, there is the infinite wall in the scalar potential, which prevents $\sigma$ from obtaining values leading to the wrong sign of its kinetic term.

The relevant field equations of the generalized model are
\begin{align}
    0&=\ddot\varphi+3H\dot{\varphi}+\fracmm{1}{\sqrt{6}}(1-\zeta\sigma^2-\gamma\sigma^4)e^{-\sqrt{\frac{2}{3}}\varphi}\dot{\sigma}^2+\partial_\varphi V~,\label{KG1_gamma}\\
    0&=\ddot\sigma+3H\dot{\sigma}-\fracmm{\zeta\sigma+2\gamma\sigma^3}{1-\zeta\sigma^2-\gamma\sigma^4}\dot{\sigma}^2-\sqrt{\fracmm{2}{3}}\dot{\varphi}\dot{\sigma}+\fracmm{e^{\sqrt{\frac{2}{3}}\varphi}}{1-\zeta\sigma^2-\gamma\sigma^4}\partial_\sigma V~,\label{KG2_gamma}\\
    0&=\frac{1}{2}\dot{\varphi}^2+\frac{1}{2}(1-\zeta\sigma^2-\gamma\sigma^4)e^{-\sqrt{\frac{2}{3}}\varphi}
    \dot{\sigma}^2+\dot{H}~,\label{Fried1_gamma}\\
    0&=V-3H^2 -\dot{H}~.\label{Fried2_gamma}
\end{align}

The generalized model defined by Eqs.~(\ref{N_choice2}) and (\ref{N_F_choice2}) appears to be rather complicated for a  detailed numerical analysis, so we study only two special cases, the one with $\delta=0$ (dubbed the $\gamma$-extension) and the one with $\gamma=0$ (dubbed the $\delta$-extension), in what follows.

\subsection{The $\gamma$-extension}

\begin{figure}
\centering
\begin{subfigure}{.49\textwidth}
  \centering
  \includegraphics[width=1\linewidth]{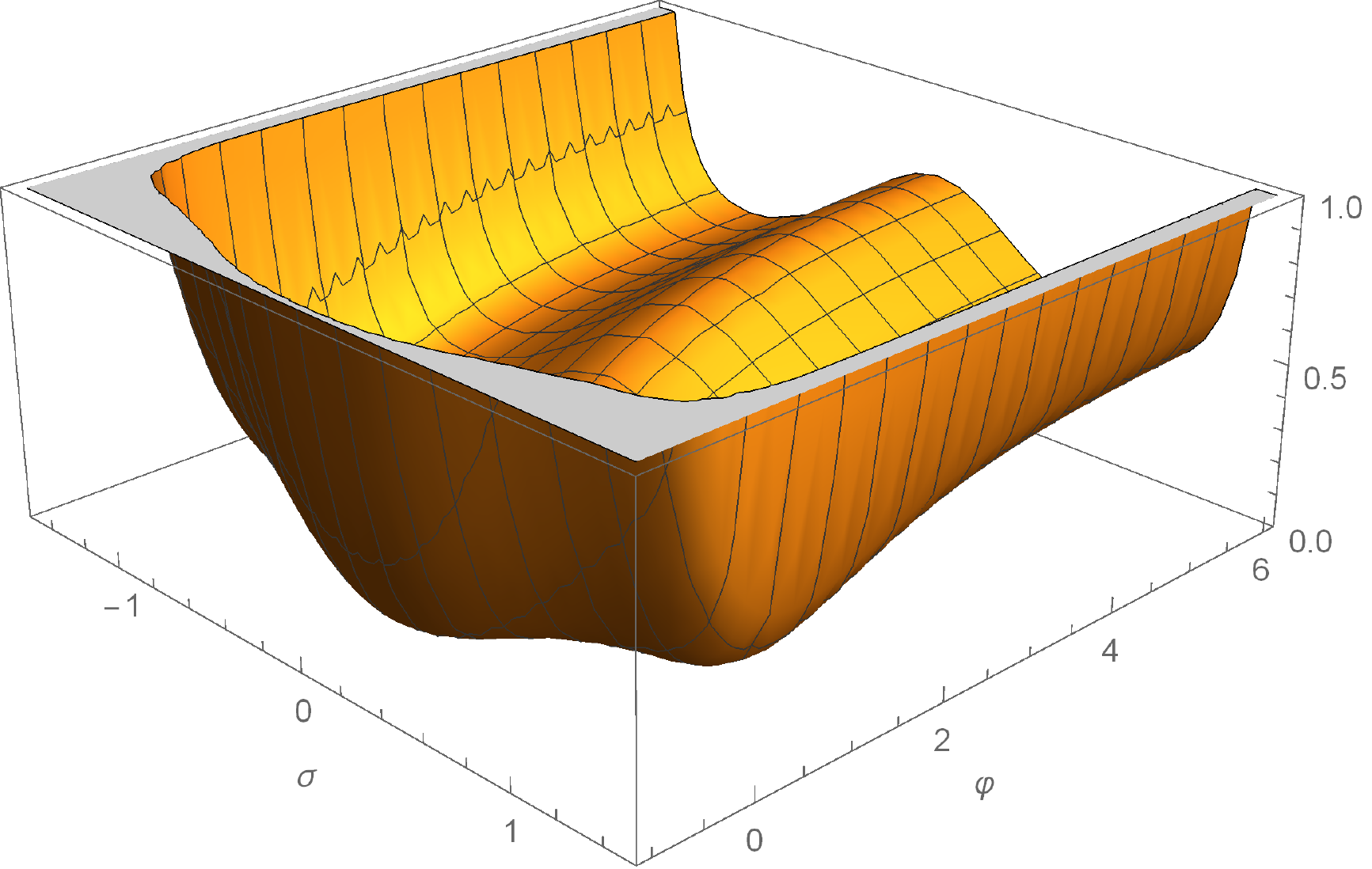}
  \label{V_3d1_gamma}
\end{subfigure}
\begin{subfigure}{.49\textwidth}
  \centering
  \includegraphics[width=.85\linewidth]{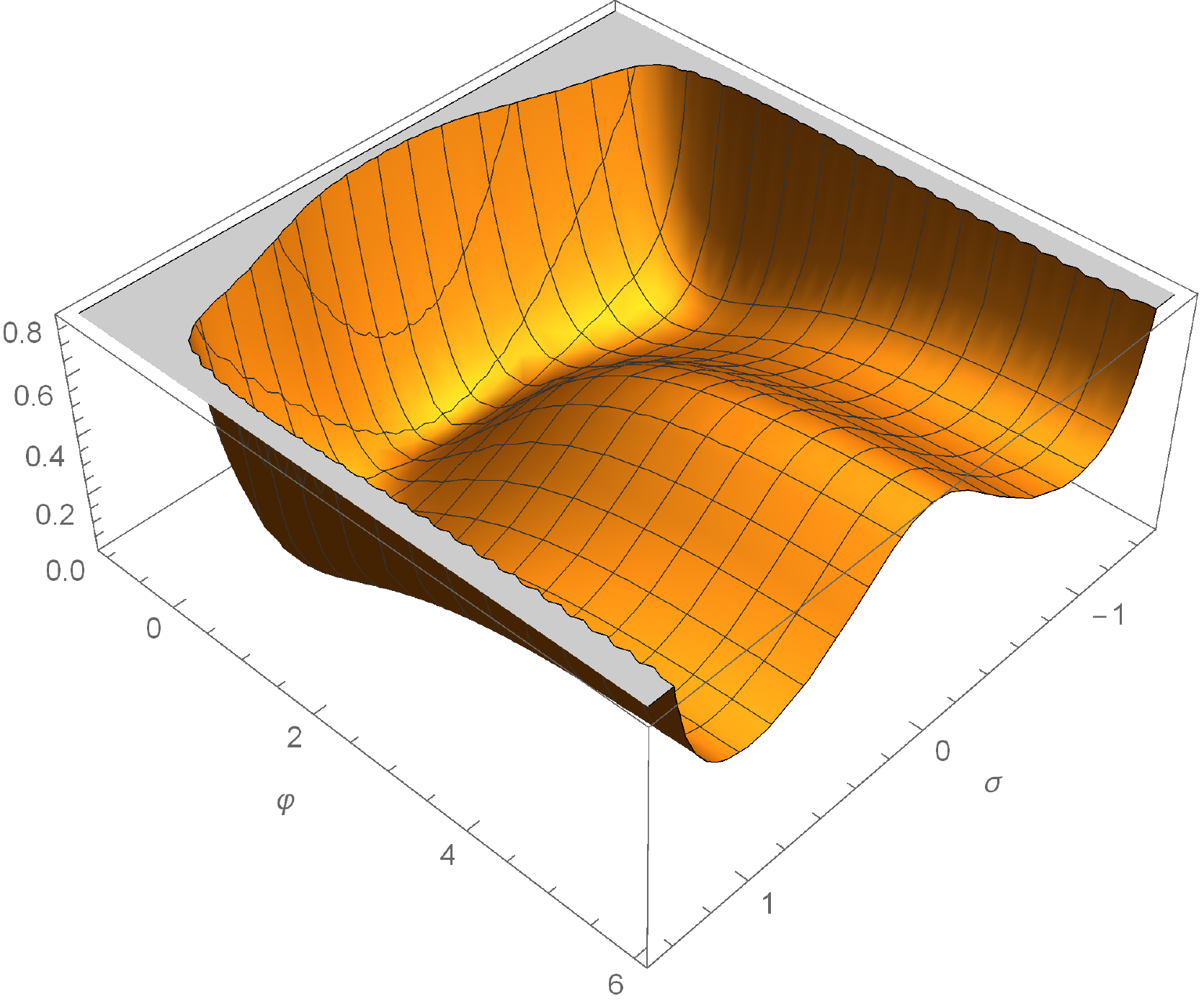}
  \label{V_3d2_gamma}
\end{subfigure}
\captionsetup{width=.9\linewidth}
\caption{The scalar potential $V/M^2$ in the Lagrangian \eqref{L_varphi2} for $\delta=0$, $\gamma=1$, $\zeta=-1.7774$.}
\label{V_3d_gamma}
\end{figure}

As a representative of the $\g$-extension ($\delta=0$), we  choose the parameters $\gamma=1$ and $\zeta=-1.7774$, see Fig.~\ref{V_3d_gamma}. This choice is interesting because the scalar potential (for $\varphi\gg 1$) has two valleys where $\sigma\neq 0$, and a single Minkowski minimum at $\sigma=\varphi=0$. The first Slow-Roll (SR) inflation is possible along either of the valleys.  The valleys merge into the Minkowski minimum by passing through inflection points (or near-inflection points) followed by the second, Ultra-Slow-Roll (USR), inflationary stage.~\footnote{Actually, despite the name, during an ultra-slow-roll regime, the scalar field(s) roll down the potential faster than during a slow-roll regime (see e.g. Ref. \cite{Motohashi:2014ppa}).}

\begin{figure}
\centering
\begin{subfigure}{.49\textwidth}
  \centering
  \includegraphics[width=0.85\linewidth]{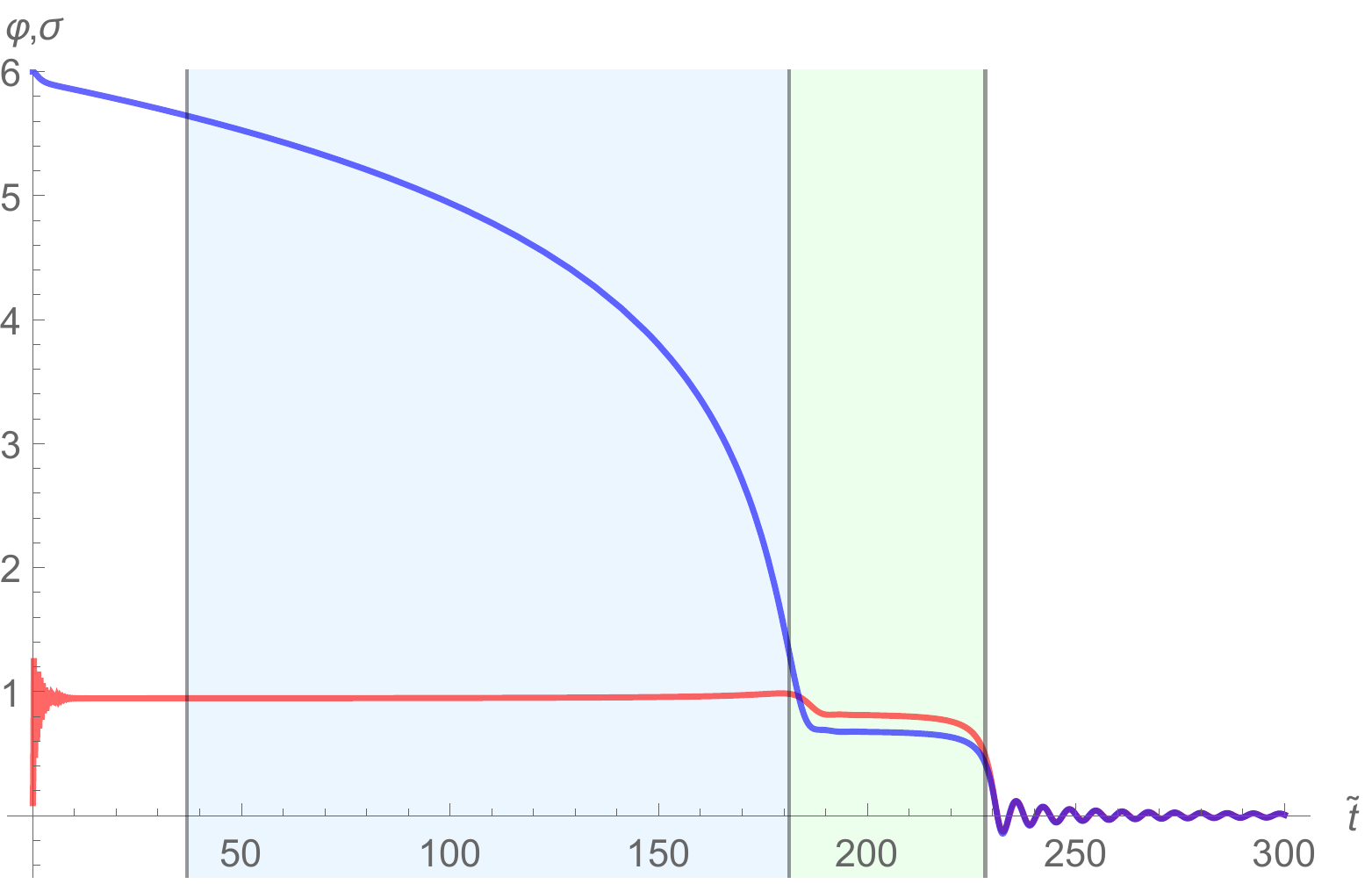}
  \caption{}
  \label{fs_sol_gamma}
\end{subfigure}
\begin{subfigure}{.49\textwidth}
  \centering
  \includegraphics[width=.75\linewidth]{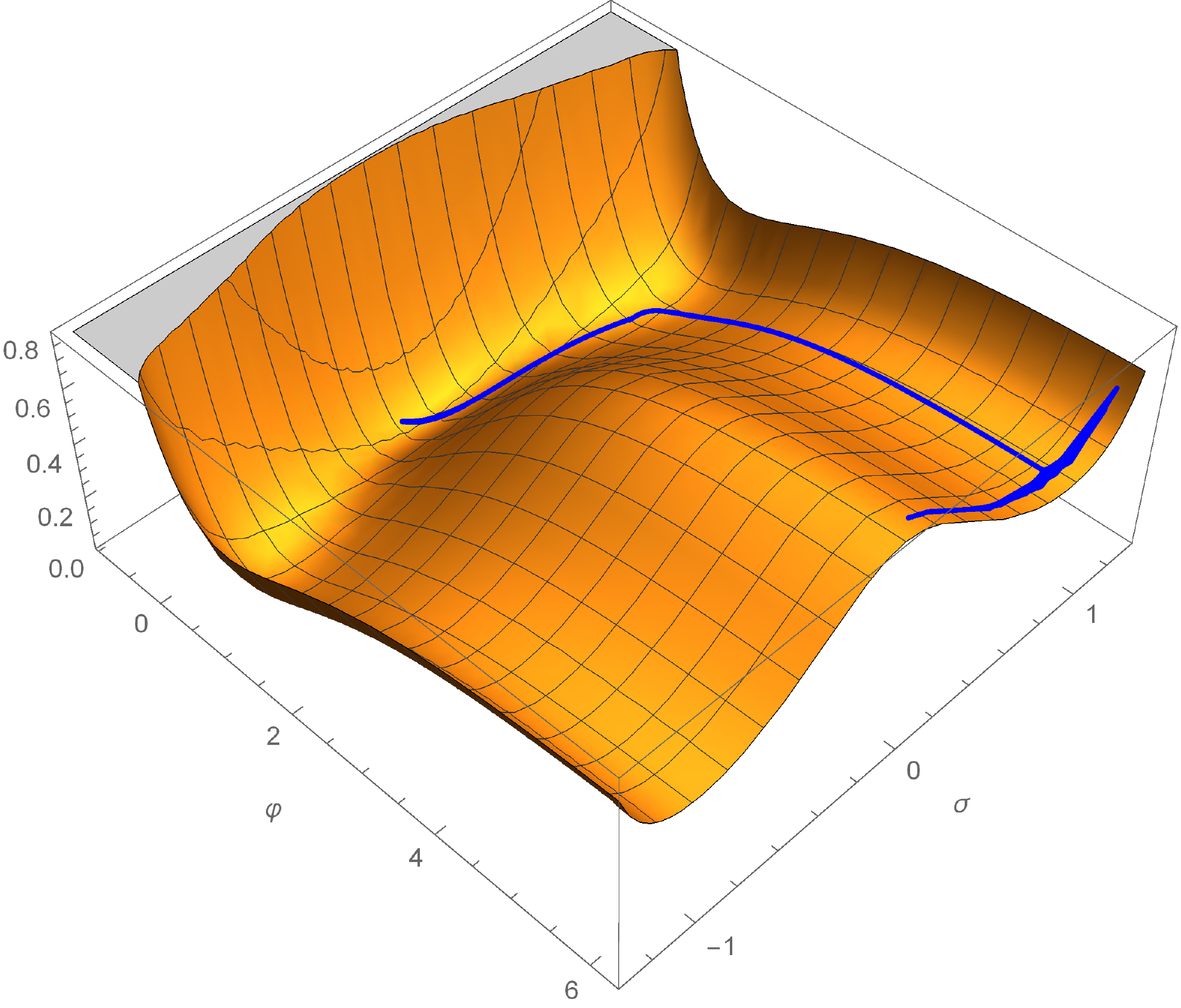}
  \caption{}
  \label{V_sol_gamma}
\end{subfigure}
\begin{subfigure}{.32\textwidth}
  \centering
  \includegraphics[width=.9\linewidth]{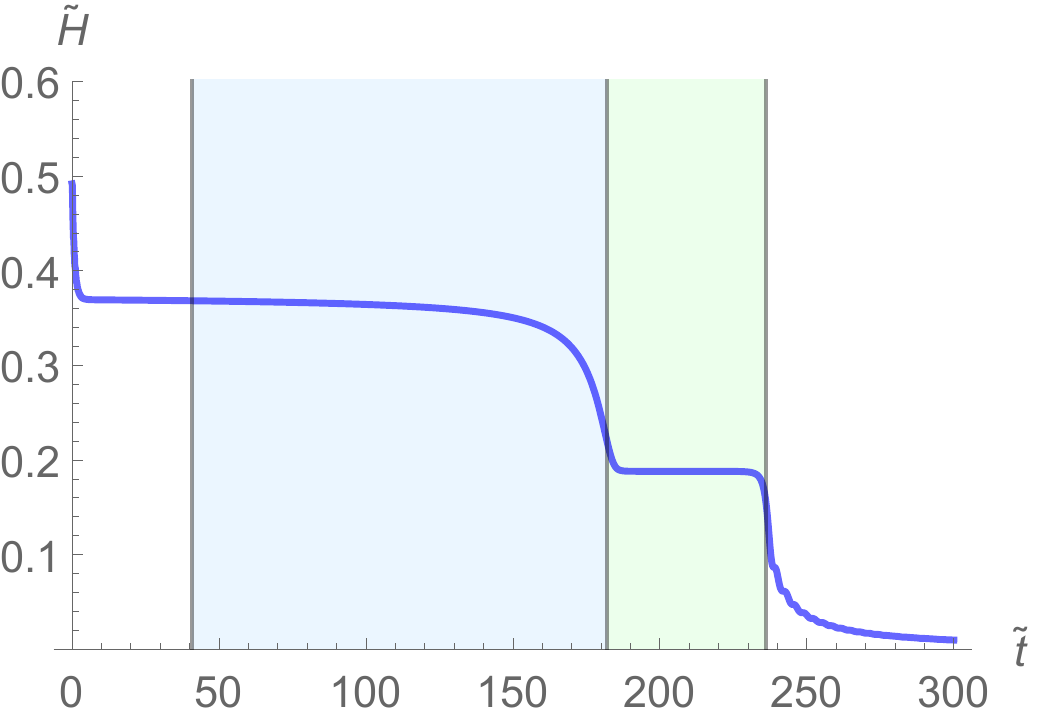}
  \caption{}
  \label{H_sol_gamma}
\end{subfigure}
\begin{subfigure}{.32\textwidth}
  \centering
  \includegraphics[width=.9\linewidth]{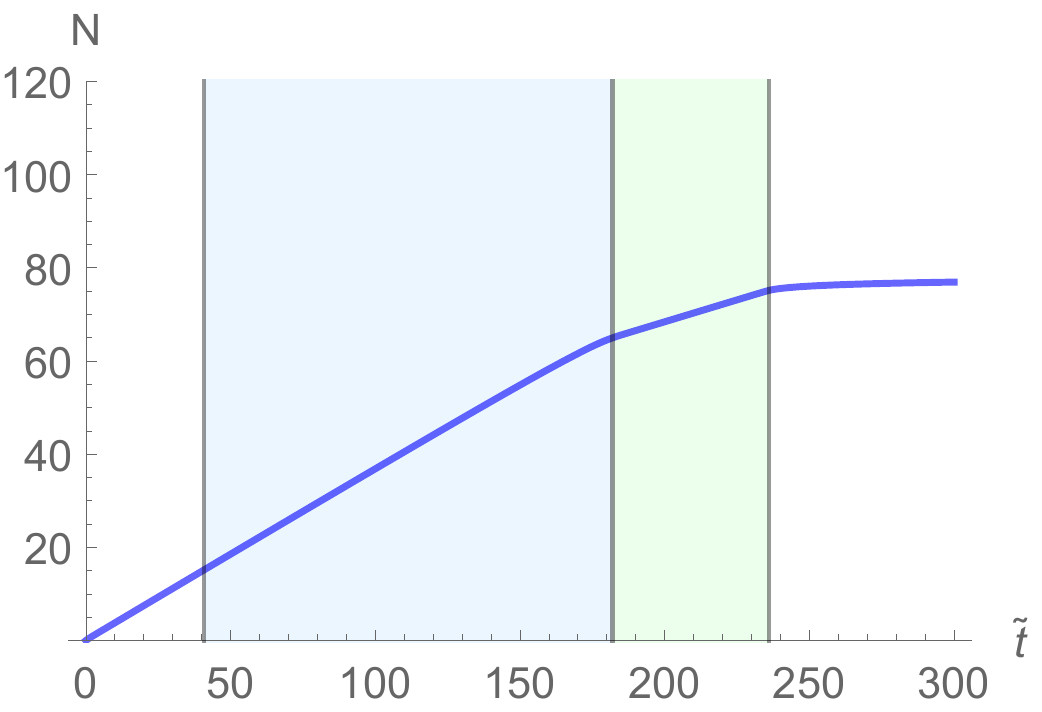}
  \caption{}
  \label{N_sol_gamma}
\end{subfigure}
\begin{subfigure}{.32\textwidth}
\centering
\includegraphics[width=.95\linewidth]{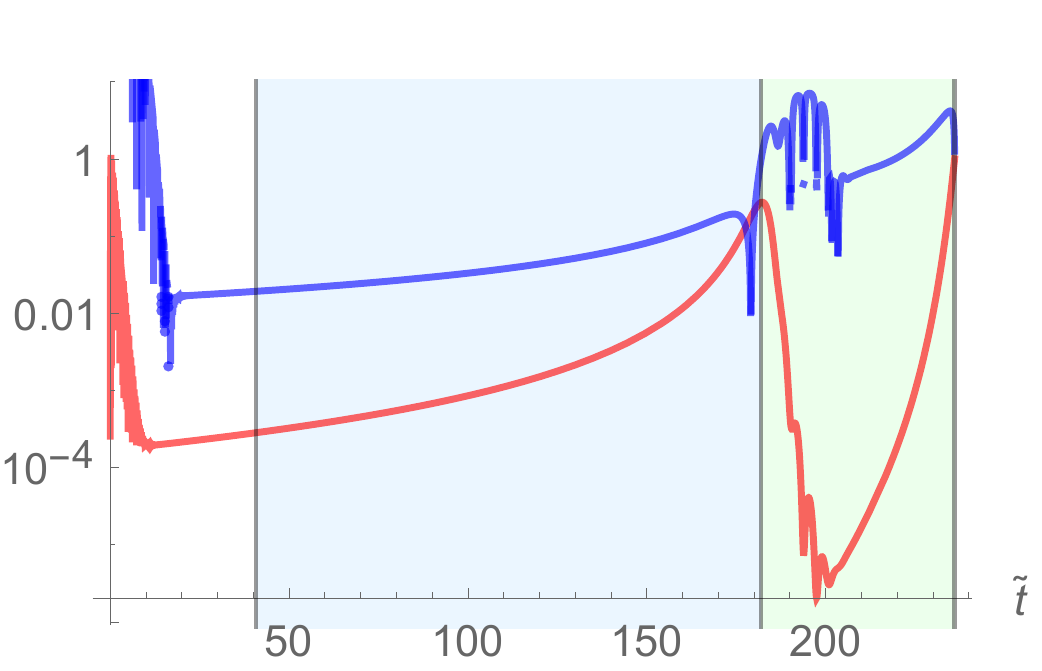}
\captionsetup{width=.9\linewidth}
\caption{}
\label{en_sol_gamma}
\end{subfigure}
\captionsetup{width=.9\linewidth}
\caption{(a) The solution to the field equations \eqref{KG1_gamma} and \eqref{KG2_gamma} with the initial conditions $\varphi(0)=6,\sigma(0)=0.1$, the vanishing initial velocities, and the choice of the parameters as $\delta=0$, $\gamma=1$, $\zeta=-1.7774$. The blue shaded region represents the first stage of inflation, and the green shaded region represents the second stage. (b) The trajectory of the solution. (c) The corresponding Hubble function. (d) The e-foldings number. 
(e) The slow-roll parameters $\epsilon$ (red) and $\eta_{\Sigma\Sigma}$ (blue).}
\label{fsV_sol_gamma}
\end{figure}

After numerically solving the equations of motion \eqref{KG1_gamma}--\eqref{Fried2_gamma} we plot the solutions in Fig.~\ref{fsV_sol_gamma}. The total number of (observable) e-foldings is set to $\Delta N=60$, and the end of the first stage of inflation is defined by the time when $\eta_{\Sigma\Sigma}$ first crosses unity (see Fig.~\ref{en_sol_gamma}). We could also define the end of the first stage from the local maximum of $\epsilon$, which nearly coincides with the former definition. As may be expected from the USR period $\epsilon_{\rm USR}\ll\epsilon_{\rm SR}$ and can be seen in Fig.~\ref{en_sol_gamma}, it leads to an {\it enhancement} in the scalar power spectrum indeed. Inflation ends when $\epsilon=1$, as usual. With the chosen parameters, the first stage lasts $\Delta N_1=50$ e-foldings, whereas the second stage lasts for  $\Delta N_2=10$: in the subsequent Figures the first stage of inflation is represented by the blue shaded region, whereas the second stage is marked by the green shaded region, whenever is relevant. The length of the second stage is controlled by the parameter $\zeta$ for a given $\gamma$.

By using Eq.~\eqref{nsr_def}, we find the observables at the CMB scale as
\begin{equation}
    n_s\approx 0.9545 \quad {\rm and}\quad r_{\rm max}\approx 0.006~.\label{nsr_gamma1_dn10}
\end{equation}

\textbf{The parameter space}. The parameter choice leading to a scalar potential with the suitable properties (as described above) is not unique, and for any $\gamma$ greater than $\sim 0.004$ there is a value of $\zeta$ that leads to a similar shape of the potential (with two inflection points, unique Minkowski minimum, etc.). For a given $\gamma$, one can solve the system of equations
\begin{equation}
    \partial_{\varphi}V=\partial_{\sigma}V={\bf H}=0~,\label{Hessian}
\end{equation}
where ${\bf H}$ is the Hessian determinant of the potential, in order to obtain the value of $\zeta$ leading to the desired inflection points. Then, by fine-tuning $\zeta$ around that value, one can change a duration of the USR stage $\Delta N_2$.

\begin{figure}
\centering
  \includegraphics[width=.5\linewidth]{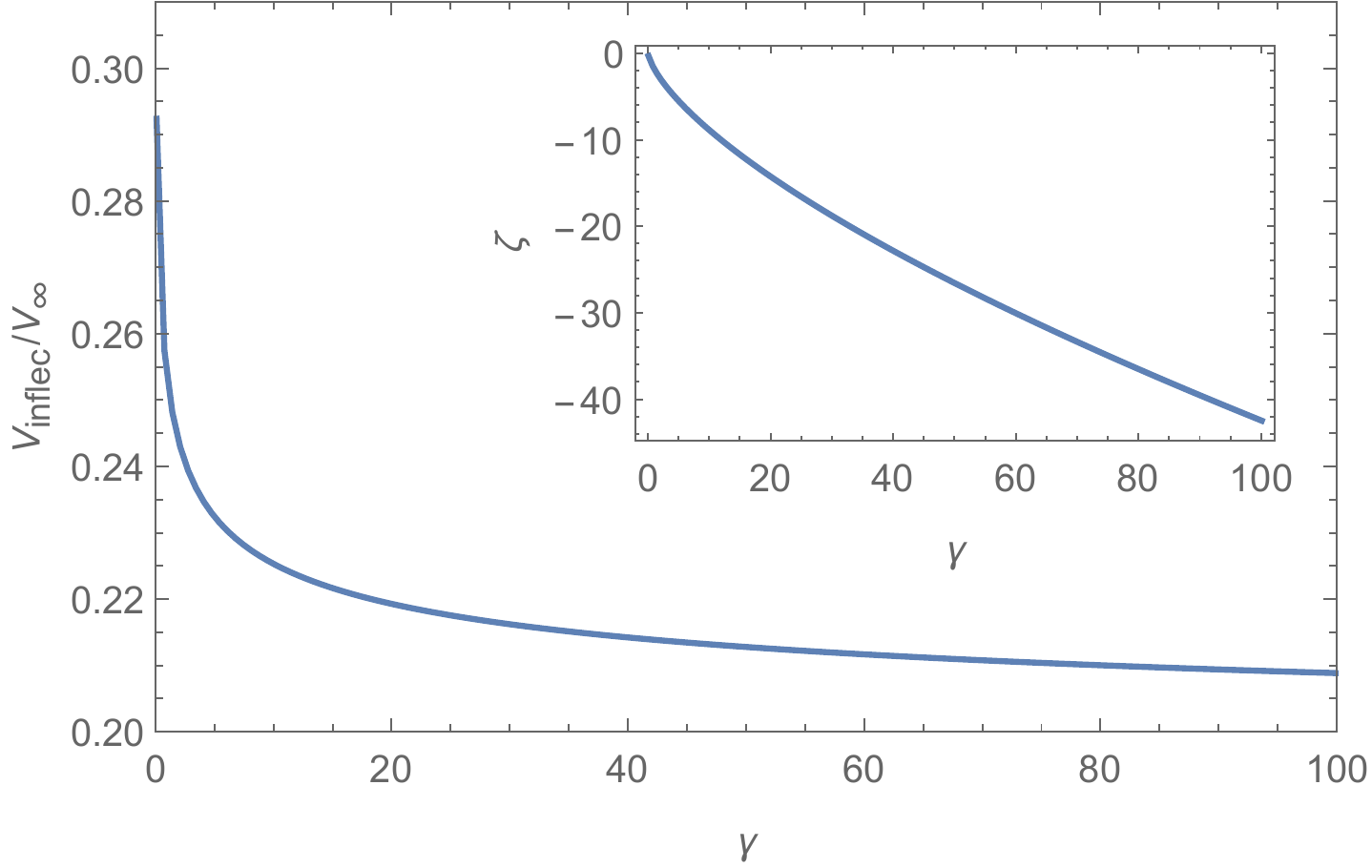}
\captionsetup{width=.9\linewidth}
\caption{The ratio of the scalar potential at the inflection point to its asymptotic value when $\varphi\rightarrow\infty$, as a function of $\gamma$. The embedded plot represents the solution to Eq.~\eqref{Hessian} in terms of $\zeta(\gamma)$.}
\label{V_inflection_gamma}
\end{figure}

In order to see how $\gamma$ changes the shape of the scalar potential, let us evaluate the ratio $V_{\rm inflec.}/V_{\infty}$ as a function of $\gamma$, where $V_{\rm inflec.}$ is the value of the potential at an inflection point, and $V_{\infty}$ is the asymptotic value of the potential when $\varphi\rightarrow\infty$ and $\sigma$ is at its local minimum, which corresponds  to the SR stage. This ratio represents the depth of the inflection points relative to the SR valleys, and it does not significantly change the curvature of the inflationary path in the $\varphi-\sigma$ plane. The plot of $V_{\rm inflec.}/V_{\infty}$ versus $\gamma$, as well as the trajectory in the $\gamma-\zeta$ plane, which solves Eqs.~\eqref{Hessian}, are shown in Fig.~\ref{V_inflection_gamma}. After taking all that into account, we conclude the control over the overall shape of the potential is limited due to the attractor-type behavior of $V_{\rm inflec.}/V_\infty$ at large $\gamma$.

When $\gamma=1$, a solution to Eq.~\eqref{Hessian} gives $\zeta\approx -1.774$ that, in turn, leads to $\Delta N_2\approx 6.3$. However, in our example we slightly departed from that value of $\zeta$ and set $\zeta=-1.7774$ in order to obtain $\Delta N_2=10$. Our strategy is to compute power spectra for different choices of $\gamma$ while keeping $\Delta N_2=10$. The latter condition fixes the value of $\zeta$. Next, we examine the impact of a variation of $\Delta N_2$.

\textbf{The power spectrum at fixed $\Delta N_2$}. We numerically compute the power spectrum of curvature perturbations by using the {\it transport method} introduced in Refs.~\cite{Mulryne:2009kh,Mulryne:2010rp} with the \textit{Mathematica} package described in Ref.~\cite{Dias:2015rca}. We compute the spectrum around the pivot scale $k_*$ that leaves the horizon at the end of the first stage, i.e. $\Delta N_2$ e-folds before the end of inflation (let us call this scale $k_{\Delta N_2}$). The inflaton mass is adjusted in each case around $0.6\times 10^{-5}M_{\rm Pl}$ by requiring $P_\zeta\approx 2\times 10^{-9}$ at the CMB scale.

\begin{figure}
\centering
  \includegraphics[width=.55\linewidth]{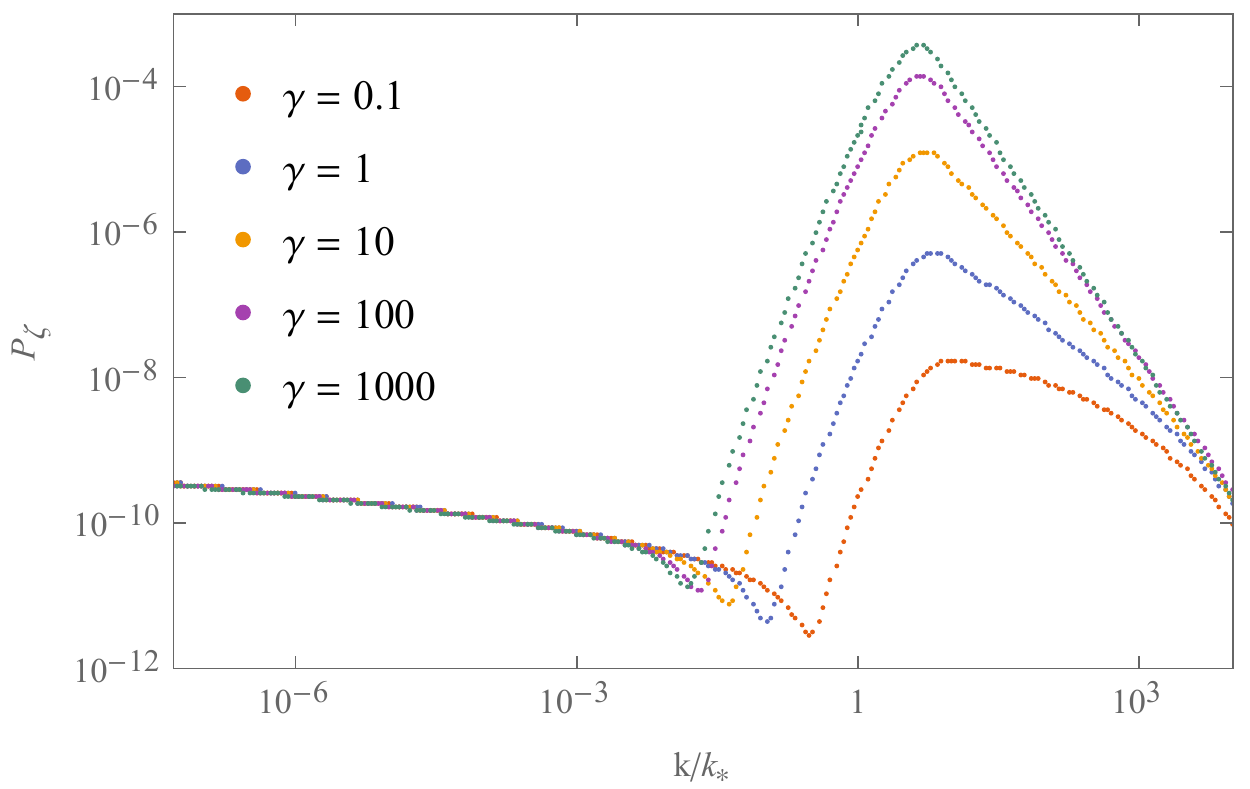}
\captionsetup{width=.9\linewidth}
\caption{The power spectrum $P_{\zeta}$ around the pivot scale $k_*=k_{\Delta N_2}$ at $\Delta N_2=10$ for several values of $\gamma$.}
\label{Pk_int_gamma}
\end{figure}

The power spectrum for various values of $\gamma$ is shown in Fig.~\ref{Pk_int_gamma}. The parameters considered are collected in Table \ref{tab_gamma}, where $\zeta$ is tuned to satisfy $\Delta N_2=10$. A change of $n_s$ and $r_{\rm max}$ (still given by Eq.~\eqref{nsr_gamma1_dn10}) is negligible for those parameters.

\begin{table}[ht]
\centering
\begin{tabular}{l r r r r r}
\toprule
$\gamma$ & $0.1$ & $1$ & $10$ & $100$ & $1000$\\
$\zeta$ & $-0.31165$ & $-1.7774$ & $-8.91495$ & $-42.7976$ & $-201.722$\\\bottomrule
\hline
\end{tabular}
\captionsetup{width=.9\linewidth}
\caption{The parameters used in a computation of the power spectrum in Fig.~\ref{Pk_int_gamma} with $\Delta N_2=10$.}
\label{tab_gamma}
\end{table}

As is often adopted in the literature, the desired enhancement of primordial curvature perturbations should exceed the CMB scales by the factor of $10^7$, in order to efficiently produce PBHs, although the authors of Ref.~\cite{Germani:2018jgr} argued by using peak theory that, given a {\it broad} peak, the required enhancement in the power spectrum drops by one order of the magnitude to $\sim 10^6$. Our numerical estimates with $\Delta N_2=10$ 
 (see Fig.~\ref{Pk_int_gamma}) show that the required enhancement of the power spectrum is not achieved. However, the enhancement grows as we increase $\Delta N_2$ (see below).

\begin{figure}
\centering
\begin{subfigure}{.49\textwidth}
  \centering
  \includegraphics[width=1\linewidth]{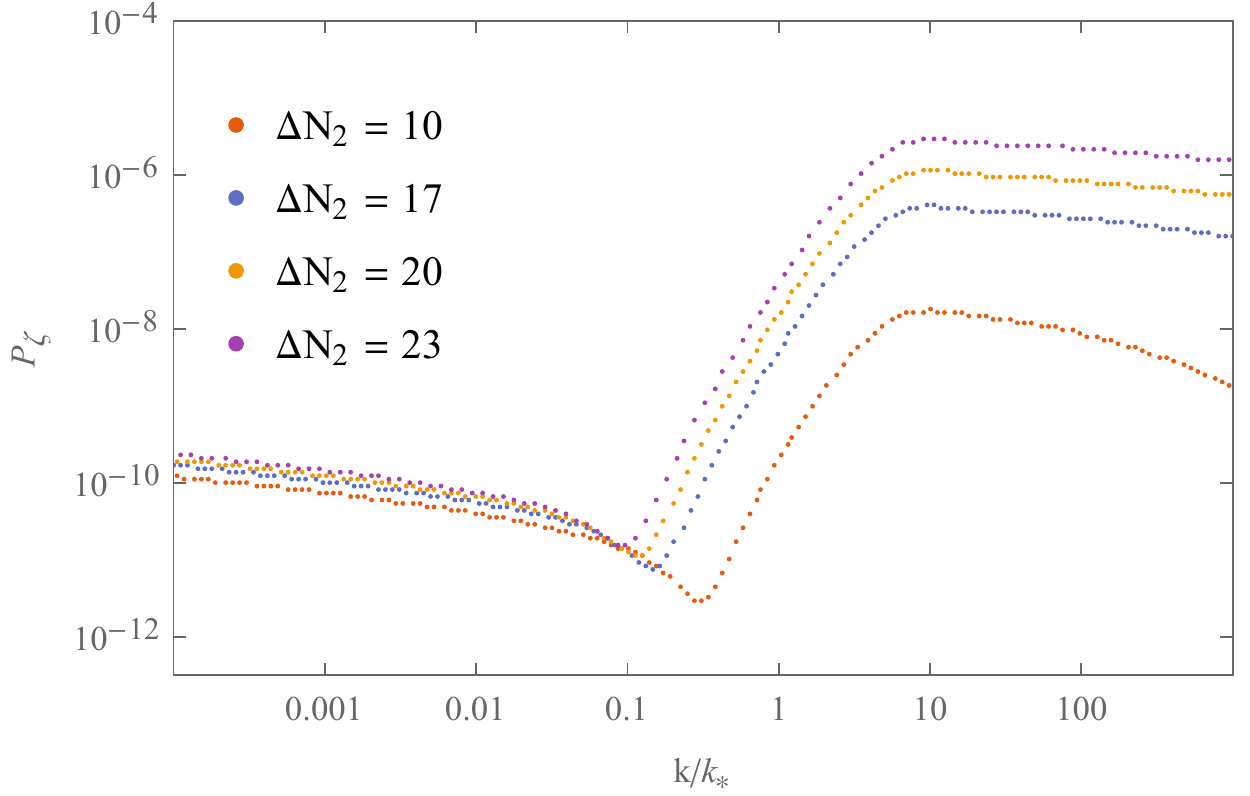}
\end{subfigure}
\begin{subfigure}{.49\textwidth}
  \centering
  \includegraphics[width=1\linewidth]{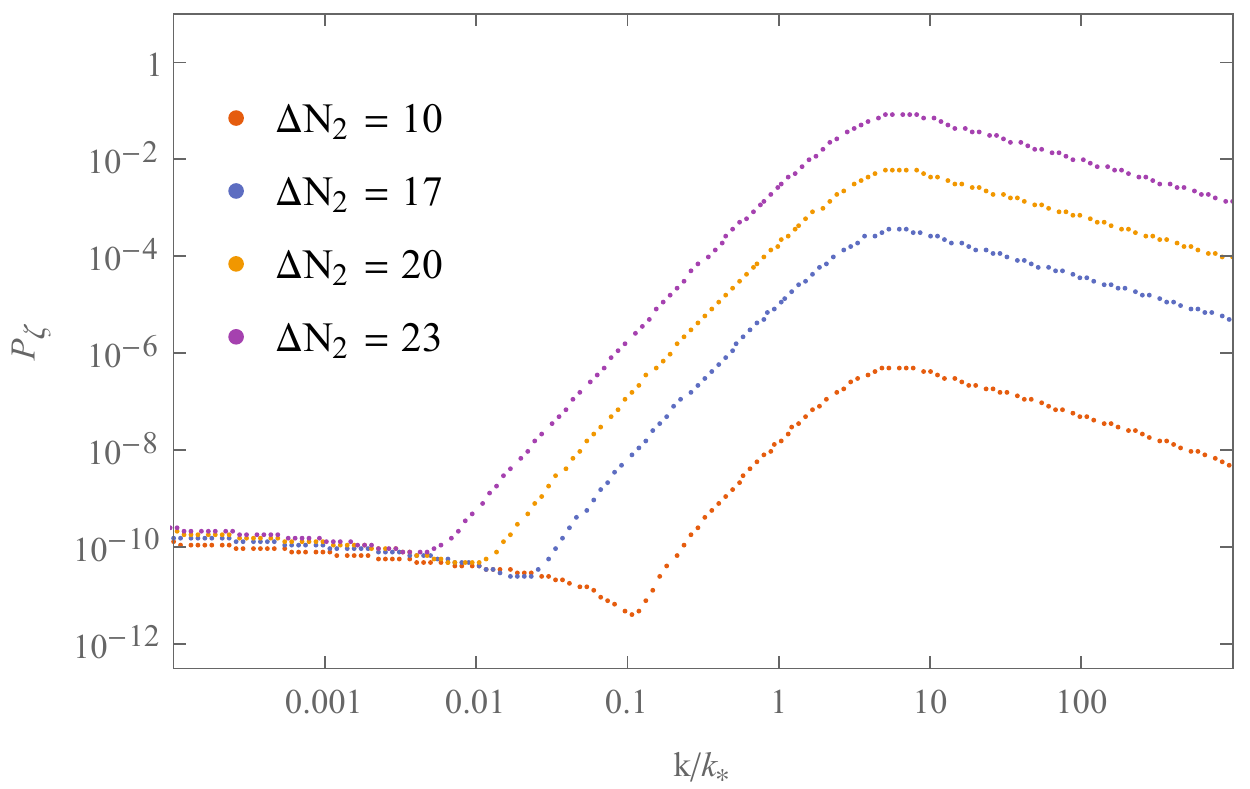}
\end{subfigure}
\captionsetup{width=.9\linewidth}
\caption{The power spectrum at $\gamma=0.1$ (on the left side) and $\gamma=1$ (on the right side) for $\Delta N_2=10,17,20,23$.}
\label{Pk_gamma_dn}
\end{figure}

\textbf{Changing $\Delta N_2$}. Let us examine how the power spectrum changes with the duration of the USR regime $\Delta N_2$. To demonstrate that dependence, we consider the power spectrum at $\gamma=0.1$ and $\gamma=1$  with various values of duration of the USR stage, $\Delta N_2=10,17,20,23$ for each $\gamma$. The results are collected in Fig.~\ref{Pk_gamma_dn}. The case of  $\gamma\lesssim 0.1$ can be excluded because the power spectrum peak is too small (technically, a larger enhancement is still possible but requires a very long USR stage that pushes the spectral index well outside of the $3\sigma$ (lower) limit of $n_s\approx 0.946$ ({\it cf.} Refs.~\cite{Jiang:2017nou,Pi:2017gih}). In the case of $\gamma=1$, the required enhancement is possible provided that $\Delta N_2\gtrsim 20$, and, therefore, the values of $\gamma\gtrsim 1$ are favored for efficient production of PBHs.

In Table~\ref{tab_gamma_dn} we collect the approximate values of $n_s$ and $r_{\rm max.}$ (at the CMB scales)  for the values of $\Delta N_2=10,17,20,23$, universally across the considered values of $\gamma=0.1,1,10,100,1000$. The tensor-to-scalar ratio $r$ is well within the observational limits in all those cases, but the scalar tilt $n_s$ is outside the $3\sigma$ limit when $\Delta N_2>17$, assuming the standard reheating scenario.

\begin{table}[ht]
\centering
\begin{tabular}{l r r r r}
\toprule
$\Delta N_2$ & $10$ & $17$ & $20$ & $23$ \\
\hline
$n_s$ & $0.955$ & $0.946$ & $0.942$ & $0.936$ \\
$r_{\rm max}$ & $0.006$ & $0.008$ & $0.009$ & $0.011$ \\\bottomrule
\hline
\end{tabular}
\captionsetup{width=.9\linewidth}
\caption{The approximate values of $n_s$ and $r$ for various choices of $\Delta N_2$, obtained by tuning 
the parameter $\zeta$ around its inflection point value.}
\label{tab_gamma_dn}
\end{table}

\textbf{PBH masses and their density fraction}. The mass of a PBH created by late-inflationary overdensities was estimated in Ref.~\cite{Pi:2017gih} as follows:
\begin{eqnarray}
    M_{\rm PBH}\simeq \fracmm{M^2_{\rm Pl}}{H(t_*)}\exp\left[2(N_{\rm end}-N_*)+\int^{t_{\rm exit}}_{t_*}\epsilon(t)H(t)dt\right]~,\label{MPBH}
\end{eqnarray}
where $t_*$ is the time when the first (slow-roll) stage ends, whereas $t_{\rm exit}$ is 
the time when the CMB pivot scale $k=0.05~{\rm Mpc}^{-1}$ exits the horizon. The formula is independent of the period between $t_*$ and the time of PBHs formation during the radiation-dominated era.

We estimate the values of $M_{\rm PBH}$ for various values of $\Delta N_2$ by using Eq.~\eqref{MPBH}. The results are shown in Table~\ref{tab_MPBH_gamma} 
together with the corresponding values of the spectral index. Our estimates are universal across the values of $\gamma=0.1,1,10,100,1000$. PBHs with masses smaller than $\sim 10^{16}{\rm g}$ would have already evaporated by now via Hawking radiation. Thus, on one hand, we need $\Delta N_2>17$. On the other hand, the lower $3\sigma$ limit on the spectral index, $n_s\approx 0.946$ \cite{Akrami:2018odb}, requires $\Delta N_2\leq 17$. Hence, the  
$\gamma$-extension alone is apparently ruled out as a model of PBH DM when we assume the standard reheating scenario and demand PBHs formation during the radiation era. Therefore, either we need yet another extension of our ansatz in modified supergravity or we have to assume some alternative (non-standard) cosmological scenarios.

\begin{table}[ht]
\centering
\begin{tabular}{l r r r r}
\toprule
$\Delta N_2$ & $10$ & $17$ & $20$ & $23$ \\
\hline
$M_{\rm PBH}$, g & $10^{9}$ & $10^{15}$ & $10^{17}$ & $10^{20}$ \\
$n_s$ & $0.955$ & $0.946$ & $0.942$ & $0.936$ \\\bottomrule
\hline
\end{tabular}
\captionsetup{width=.9\linewidth}
\caption{The PBH masses estimated by Eq.~\eqref{MPBH} for the $\gamma$-extension with the corresponding (approximate) values of the spectral index. In the Solar mass units, $1~{\rm g}\approx 5.03\times 10^{-34}~M_\odot$.}
\label{tab_MPBH_gamma}
\end{table}

As regards the constraints on $\gamma$, the power spectrum in Fig.~\ref{Pk_gamma_dn} tells us that it is sufficient to have
 $\gamma\geq{\cal O}(1)$ in order to produce the required enhancement in the spectrum.

We also estimate the PBHs density fraction by using {\it Press-Schechter} formalism \cite{Press:1973iz}. The useful formulae include the PBH mass $\tilde{M}_{\rm PBH}(k)$, the production rate $\beta_f(k)$, and the density contrast $\sigma(k)$ coarse-grained over $k$ as follows (see e.g., Refs.~\cite{Inomata:2017okj,Inomata:2017vxo} and references   therein):
\begin{gather}
    \tilde{M}_{\rm PBH}\simeq 10^{20}\left(\fracmm{7\times 10^{12}}{k~{\rm Mpc}}\right)^2{\rm g}~, \quad \beta_f(k)\simeq\fracmm{\sigma(k)}{\sqrt{2\pi}\delta_c}
    e^{-\fracmm{\delta^2_c}{2\sigma^2(k)}}~,\\
    \sigma^2(k)=\fracmm{16}{81}\int\fracmm{dq}{q}\left(\fracmm{q}{k}\right)^4e^{-q^2/k^2}P_\zeta(q)~,\label{PBH_productionE}
\end{gather}
respectively, where we have chosen the {\it Gaussian} window function for the density contrast and have introduced $\delta_c$ as a constant representing the density threshold for PBH formation, which is usually estimated as $\delta_c\approx 1/3$ \cite{Carr:1975qj} for simplicity (its more precise value depends upon details of the power spectrum). In terms of the above functions, the PBH-to-DM density fraction can be estimated as follows \cite{Inomata:2017okj,Inomata:2017vxo}:
\begin{eqnarray}
    \fracmm{\Omega_{\rm PBH}(k)}{\Omega_{\rm DM}}\equiv f(k)\simeq\fracmm{1.2\times 10^{24}\beta_f(k)}{\sqrt{\tilde{M}_{\rm PBH}(k){\rm g}^{-1}}}~~,\label{f_PBH}
\end{eqnarray}
where the numerical factor is computed for the Minimal Supersymmetric Standard Model degrees of freedom (in the case of the Standard Model it would be approximately $1.4\times 10^{24}$).

\begin{figure}
\centering
  \includegraphics[width=.7\linewidth]{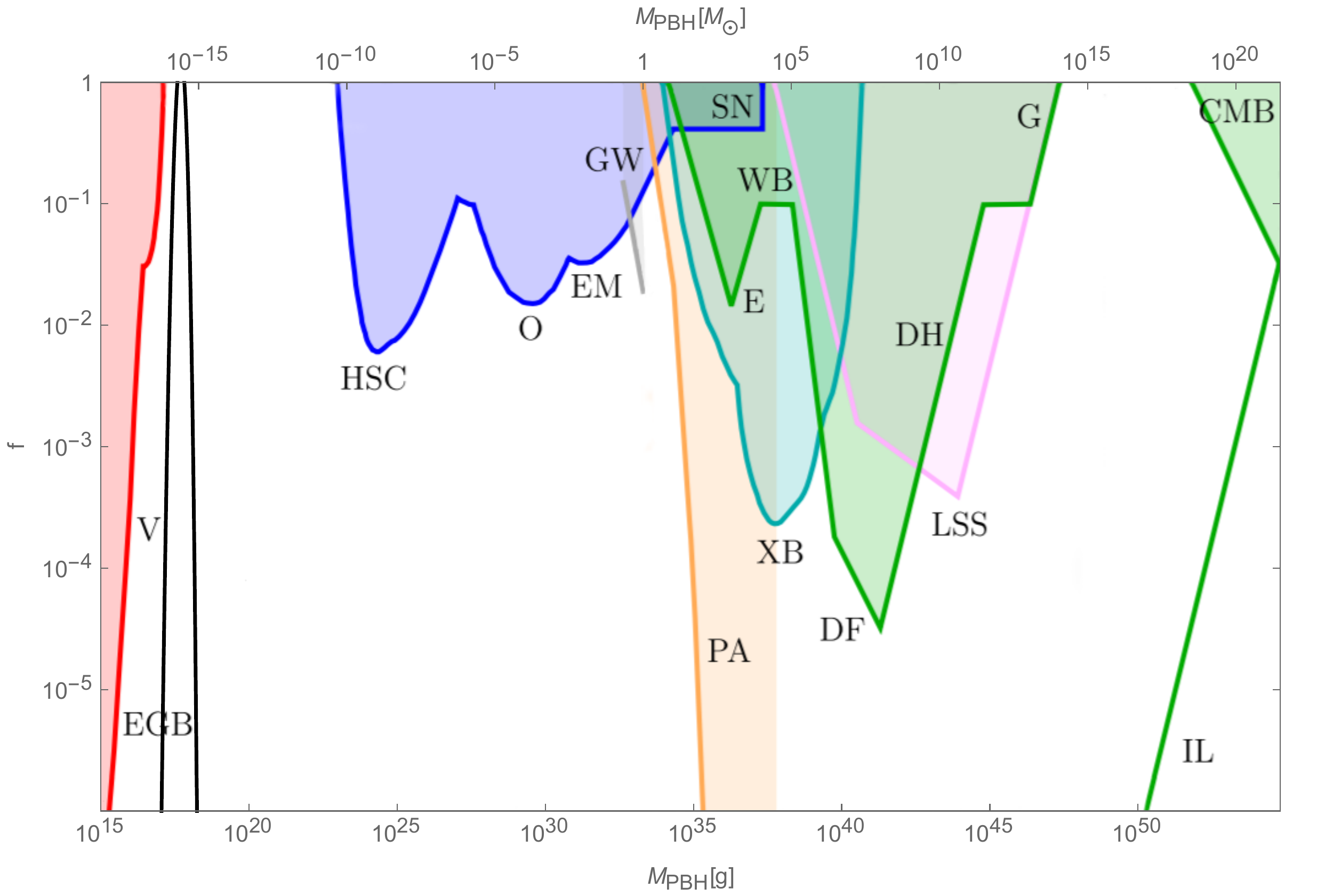}
\captionsetup{width=.9\linewidth}
\caption{The PBHs fraction obtained with the parameters $\gamma=1.5$, $\Delta N_2=20$, and $\delta_c=0.4$ (black curve). The shaded regions represent the  observational constraints of Ref.~\cite{Carr:2020xqk}: from evaporation (red), lensing (blue), gravitational waves (gray), various dynamical effects (green), accretion (light blue), large-scale structure (pink), and CMB distortions (orange).}
\label{Fig_f_gamma}
\end{figure}

In order to numerically evaluate the functions \eqref{PBH_productionE} and \eqref{f_PBH}, we need to normalize the values of $k$ in terms of the observable scales today. As we already mentioned above, we use the scale $k=0.05~{\rm Mpc}^{-1}$ leaving the horizon $54$ e-folds before the end of inflation. To find a specific example, we search for a set 
 of the parameters that can lead to a substantial PBH density with the critical density in the range $1/3\lesssim\delta_c\lesssim 2/3$. We get such an example with $\gamma=1.5$, $\Delta N_2=20$, and $\delta_c=0.4$ (this leads to $n_s\approx 0.942$ as can be seen in Table \ref{tab_MPBH_gamma}). The resulting PBH fraction is shown in Fig.~\ref{Fig_f_gamma} on the background of the observational constraints of Ref.~\cite{Carr:2020xqk} (see also Ref.~\cite{Carr:2020gox}). According to Fig.~\ref{Fig_f_gamma}, our peak is located at the edge of the lowest-mass window, between $10^{17}$ and $10^{18}$ grams. The constraints of Fig.~\ref{Fig_f_gamma} are imposed by assuming a {\it monochromatic} PBH mass spectrum. In our case the mass distribution is narrow, albeit is not strictly monochromatic.

The total PBH-to-DM density fraction, given by
\begin{eqnarray}
    f_{\rm tot}=\int d(\log\tilde{M}_{\rm PBH})f(\tilde{M}_{\rm PBH})~,
\end{eqnarray}
is estimated for Figure \ref{Fig_f_gamma} as $f_{\rm tot}\approx 1$, i.e. PBHs can constitute the whole DM in that case.

\subsection{The $\delta$-extension}

\begin{figure}
\centering
\begin{subfigure}{.4\textwidth}
  \centering
  \includegraphics[width=1\linewidth]{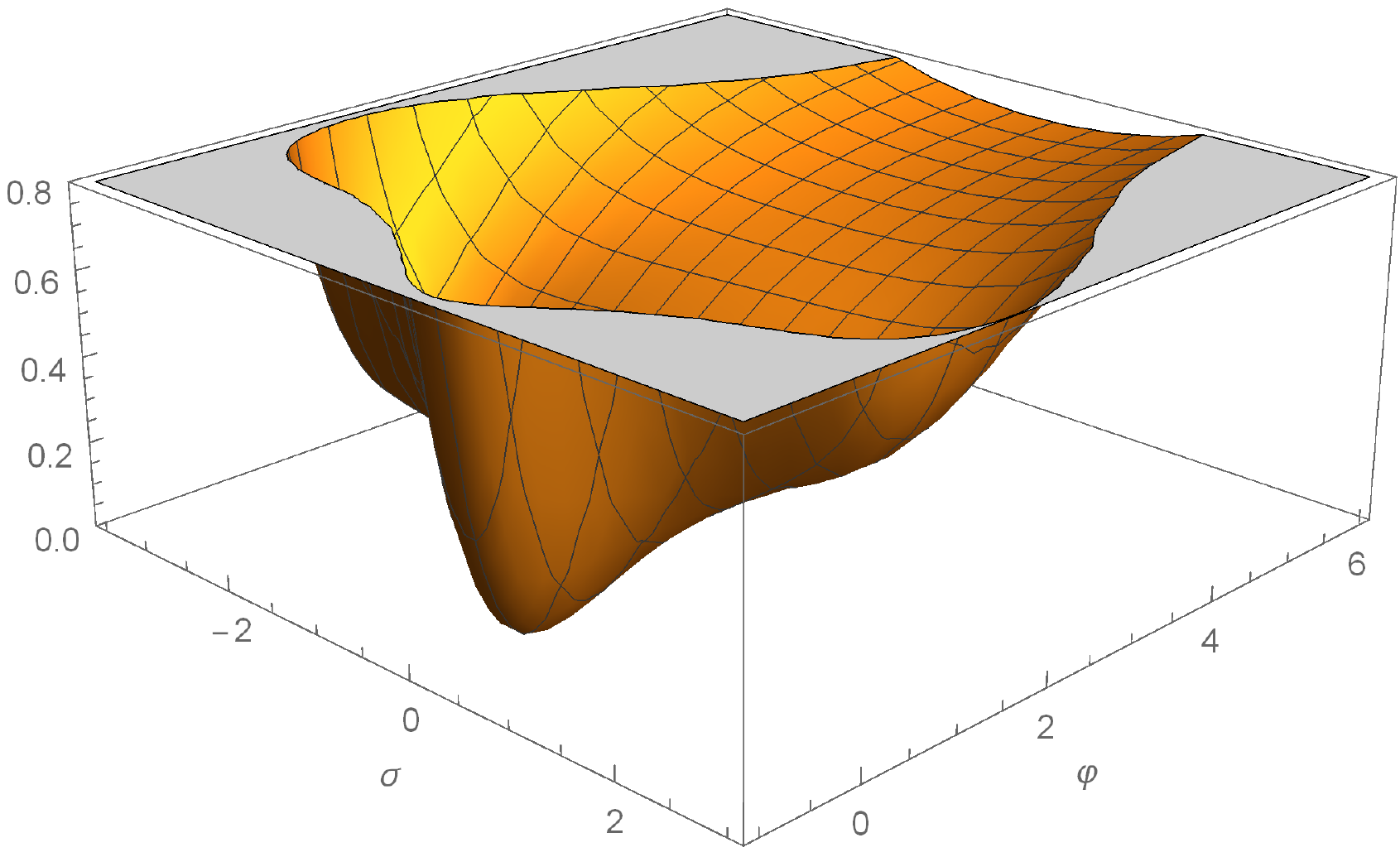}
  \label{V_3d1_delta}
\end{subfigure}
\begin{subfigure}{.49\textwidth}
  \centering
  \includegraphics[width=.85\linewidth]{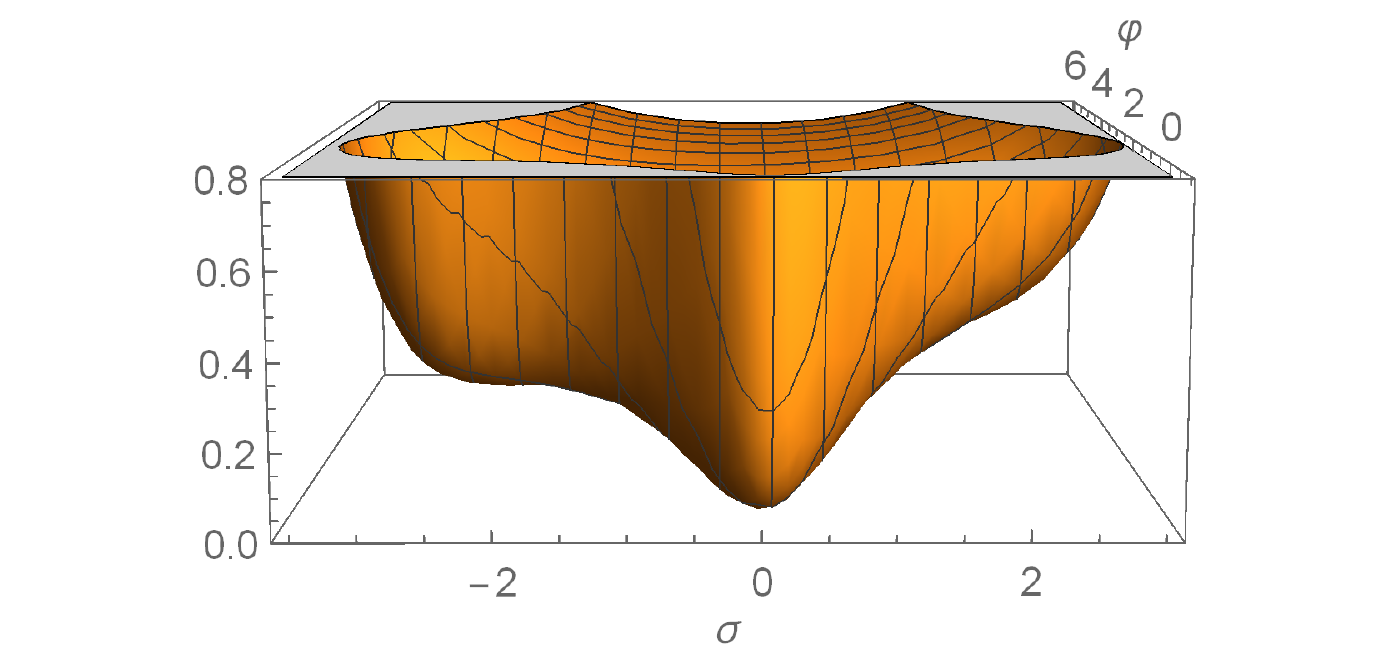}
  \label{V_3d2_delta}
\end{subfigure}
\captionsetup{width=.9\linewidth}
\caption{The scalar potential in the Lagrangian \eqref{L_varphi2} for $\gamma=0$, $\delta=0.1$ and $\zeta=0.033407$.}
\label{V_3d_delta}
\end{figure}

Having established that the PBH DM scenario in the $\gamma$-extension is in conflict with the CMB constraint, the next possibility is to study the $\delta$-extension. In this Subsection, we take $\gamma=0$ for simplicity and take $\delta\neq 0$ in Eqs.~(\ref{N_choice2}) and (\ref{N_F_choice2}).~\footnote{A model similar to our $\delta$-extension was considered in Ref.~\cite{Dalianis:2014aya} in relation to spontaneous supersymmetry breaking after inflation. The difference between our model and that of Ref.~\cite{Dalianis:2014aya} is in the scalar potential: the potential of 
Ref.~\cite{Dalianis:2014aya} has an additional (Minkowski) minimum (away from  $\sigma=0$) that breaks both  supersymmetry and R-symmetry. In our case, we have a single, SUSY-preserving Minkowski minimum (at $\sigma=0$) and an inflection point away from $\sigma=0$, in order to achieve an ultra-slow-roll stage.} It breaks the R-symmetry and the reflection symmetry $\sigma\rightarrow -\sigma$ of the potential (see Fig.~\ref{V_3d_delta}), and generates $\theta$-dependent terms in the potential (we remind that $\theta$ is the angular component of the complex scalar $X\equiv {\cal R}|$). Those terms can be obtained by replacing $\delta\rightarrow\delta\cos\theta$ in 
Eq.~\eqref{ABU_tilde}. We assume that the angular scalar $\theta$ is stabilized during inflation. Then its VEV 
$\langle\theta\rangle$ can be absorbed into a redefinition of $\delta$. According to Eq.~\eqref{ABU_tilde}, the 
$\theta$ terms are multiplied by the factors of $\sigma$, so that they vanish when $\sigma=0$, including the Minkowski minimum. A stabilization of $\theta$ requires additional tools that we leave to future studies.

As far as the shape of the potential is concerned, for any non-zero $\delta$ there is a value of $\zeta$ that leads to an inflection point: for a positive $\delta$ the inflection point is at $\sigma=-|\sigma_{\rm inflec}|$ (as in our example of Fig.~\ref{V_3d_delta}), and for a negative $\delta$ the inflection point is at $\sigma=+|\sigma_{\rm inflec}|$.

In contrast to the $\gamma$-extension, here we have a single valley for large positive $\varphi$ and $\sigma=0$, so that in this limit the model reduces to a single-field Starobinsky model. As one approaches $\varphi=0$, the potential inclines towards the (near-)inflection point which could, in principle, guide the inflationary trajectory towards passing through the (near-)inflection point before falling to the Minkowski minimum at $\varphi=\sigma=0$.

\begin{figure}
\centering
\begin{subfigure}{.49\textwidth}
  \centering
  \includegraphics[width=0.85\linewidth]{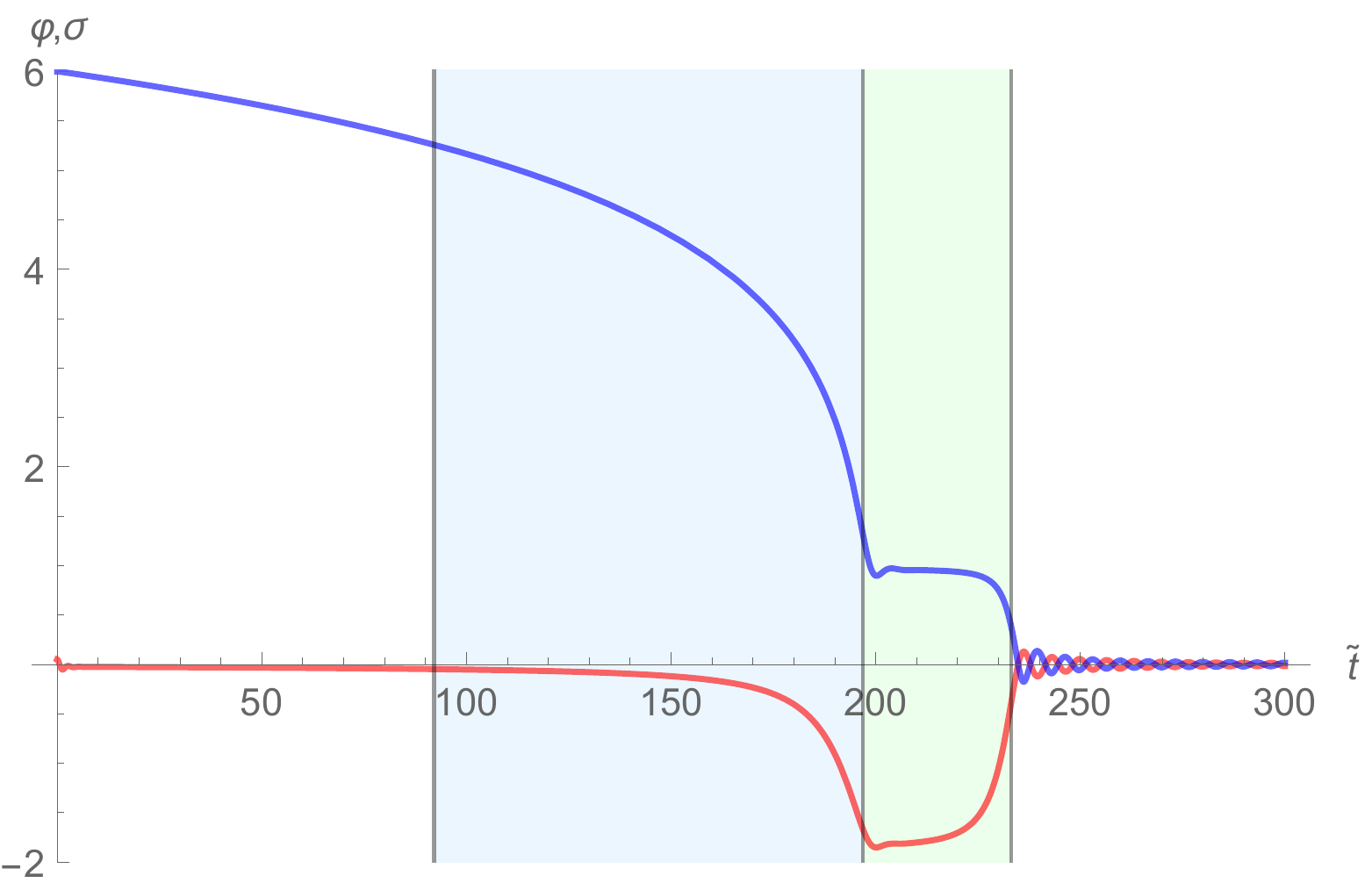}
  \caption{}
  \label{fs_sol_delta}
\end{subfigure}
\begin{subfigure}{.49\textwidth}
  \centering
  \includegraphics[width=.75\linewidth]{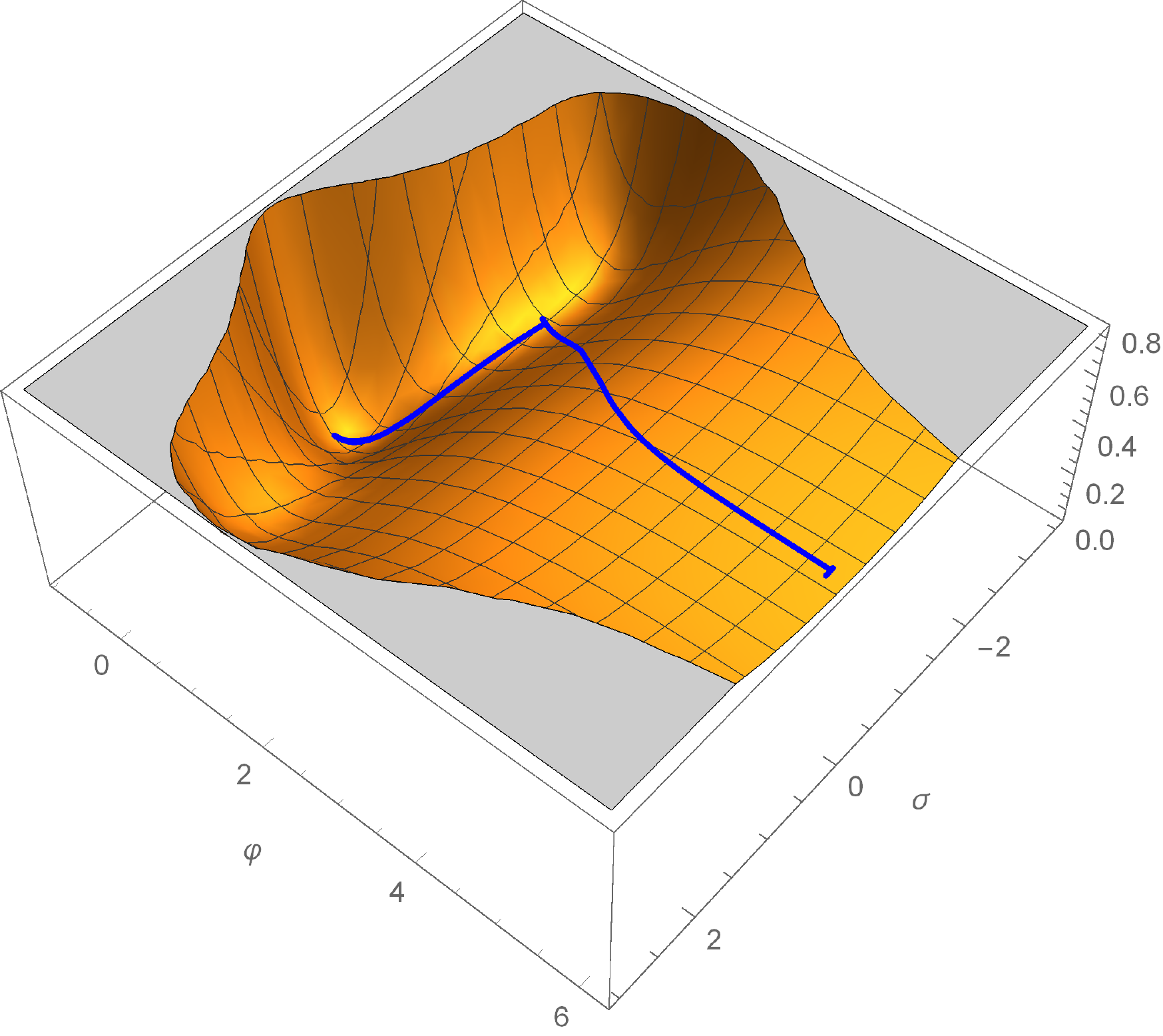}
  \caption{}
  \label{V_sol_delta}
\end{subfigure}
\begin{subfigure}{.32\textwidth}
  \centering
  \includegraphics[width=.9\linewidth]{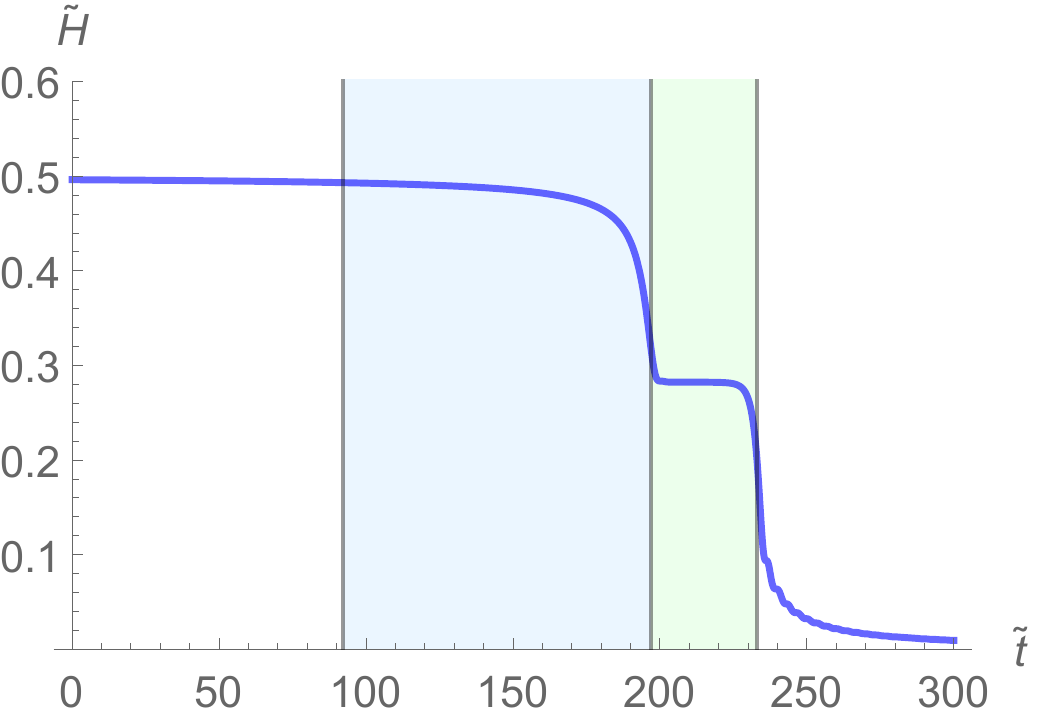}
  \caption{}
  \label{H_sol_delta}
\end{subfigure}
\begin{subfigure}{.32\textwidth}
  \centering
  \includegraphics[width=.9\linewidth]{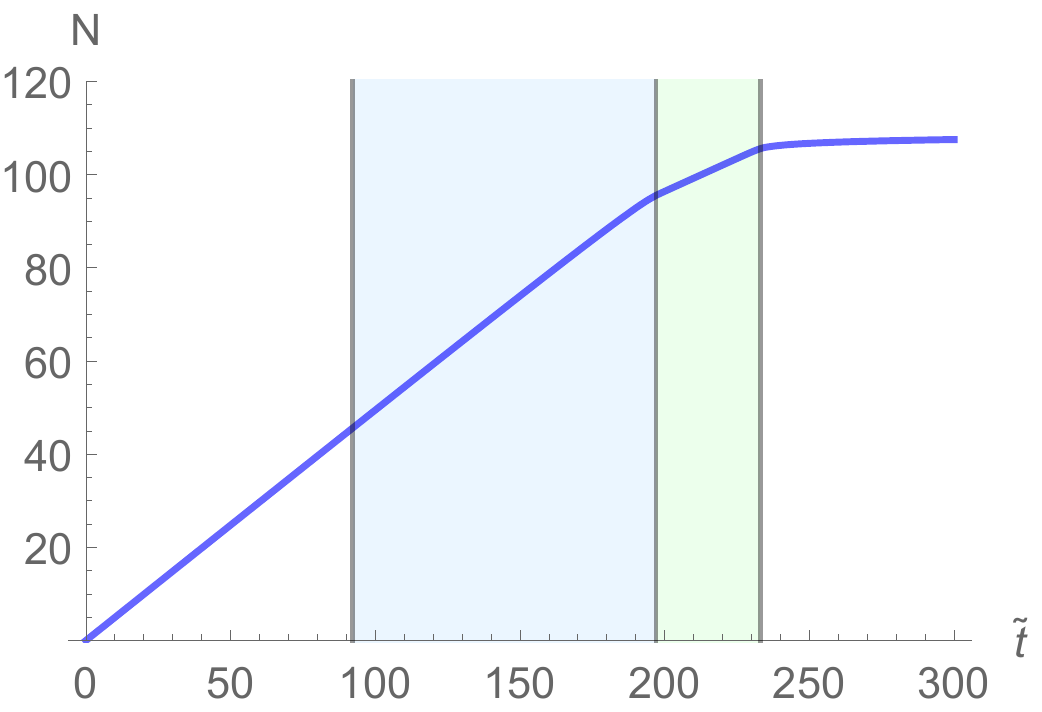}
  \caption{}
  \label{N_sol_delta}
\end{subfigure}
\begin{subfigure}{.32\textwidth}
\centering
\includegraphics[width=.95\linewidth]{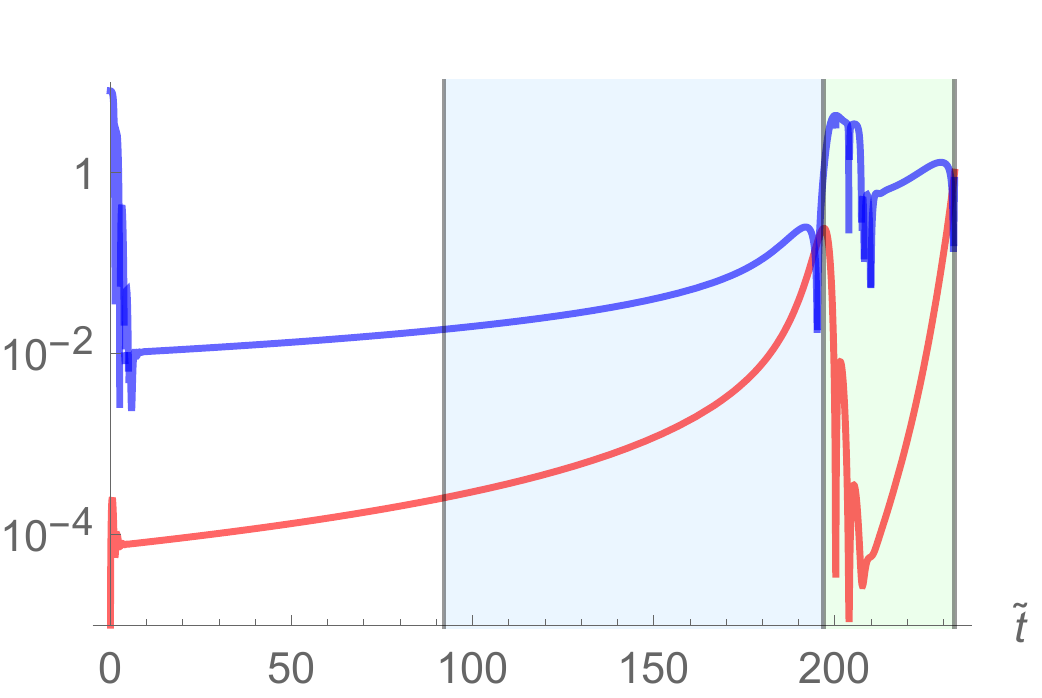}
\captionsetup{width=.9\linewidth}
\caption{}
\label{en_sol_delta}
\end{subfigure}
\captionsetup{width=.9\linewidth}
\caption{(a) The solution to the field equations \eqref{KG1_gamma} and \eqref{KG2_gamma} with the initial conditions 
$\varphi(0)=6$ and $ \sigma(0)=0.05$, the vanishing  initial velocities, and the parameter choice $\gamma=0$, $\delta=0.1$ and  $\zeta=0.033407$. (b) The trajectory of the solution ($\varphi$ -- blue, $\sigma$ -- red). (c) The corresponding Hubble function. (d) The number of e-folds. (e) The slow-roll parameters $\epsilon$ (red) and $\eta_{\Sigma\Sigma}$ (blue).}
\label{fsV_sol_delta}
\end{figure}

Let us consider, for example, the parameter values $\delta=0.1$ and $\zeta=0.033407$ ($\zeta$ is chosen to get $\Delta N_2=10$). After solving the corresponding field equations, we show the time dependence of $\varphi$, $\sigma$, $\tilde{H}$, $N$, $\epsilon$ and $\eta_{\Sigma\Sigma}$ in Fig.~\ref{fsV_sol_delta}. The near-inflection point divides inflation into two stages with $\Delta N_1=50$ (slow-roll) and $\Delta N_2=10$ (ultra-slow-roll). We set initial velocities to zero, with $\varphi(0)=6$ and $\sigma(0)=0.05$. Similarly to the $\gamma$-extension, the inflationary trajectory is stable against variations of the initial conditions, as long as they are not very large.

\begin{figure}
\centering
\begin{subfigure}{.49\textwidth}
  \centering
  \includegraphics[width=.9\linewidth]{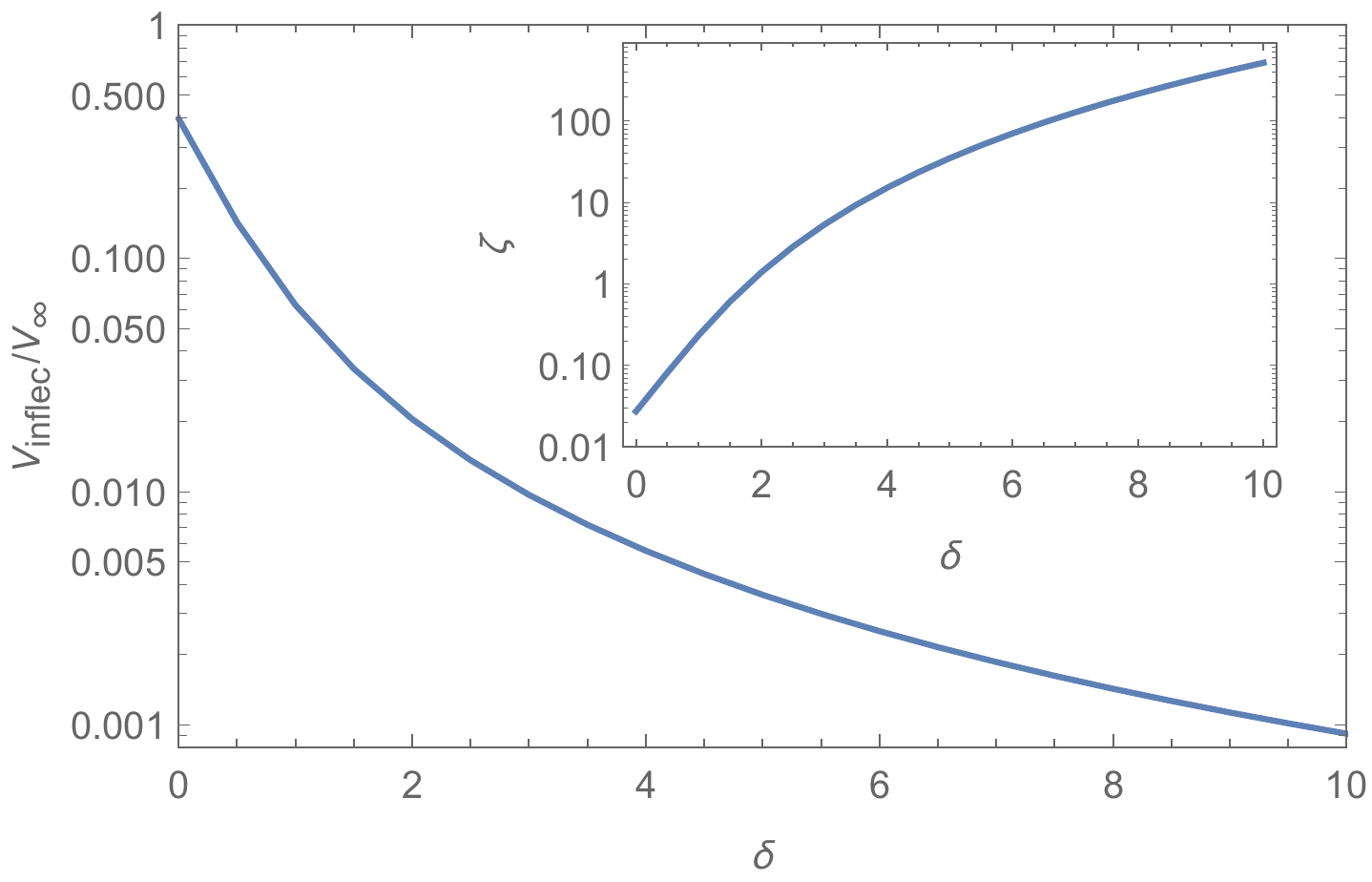}
\end{subfigure}
\begin{subfigure}{.49\textwidth}
  \centering
  \includegraphics[width=.87\linewidth]{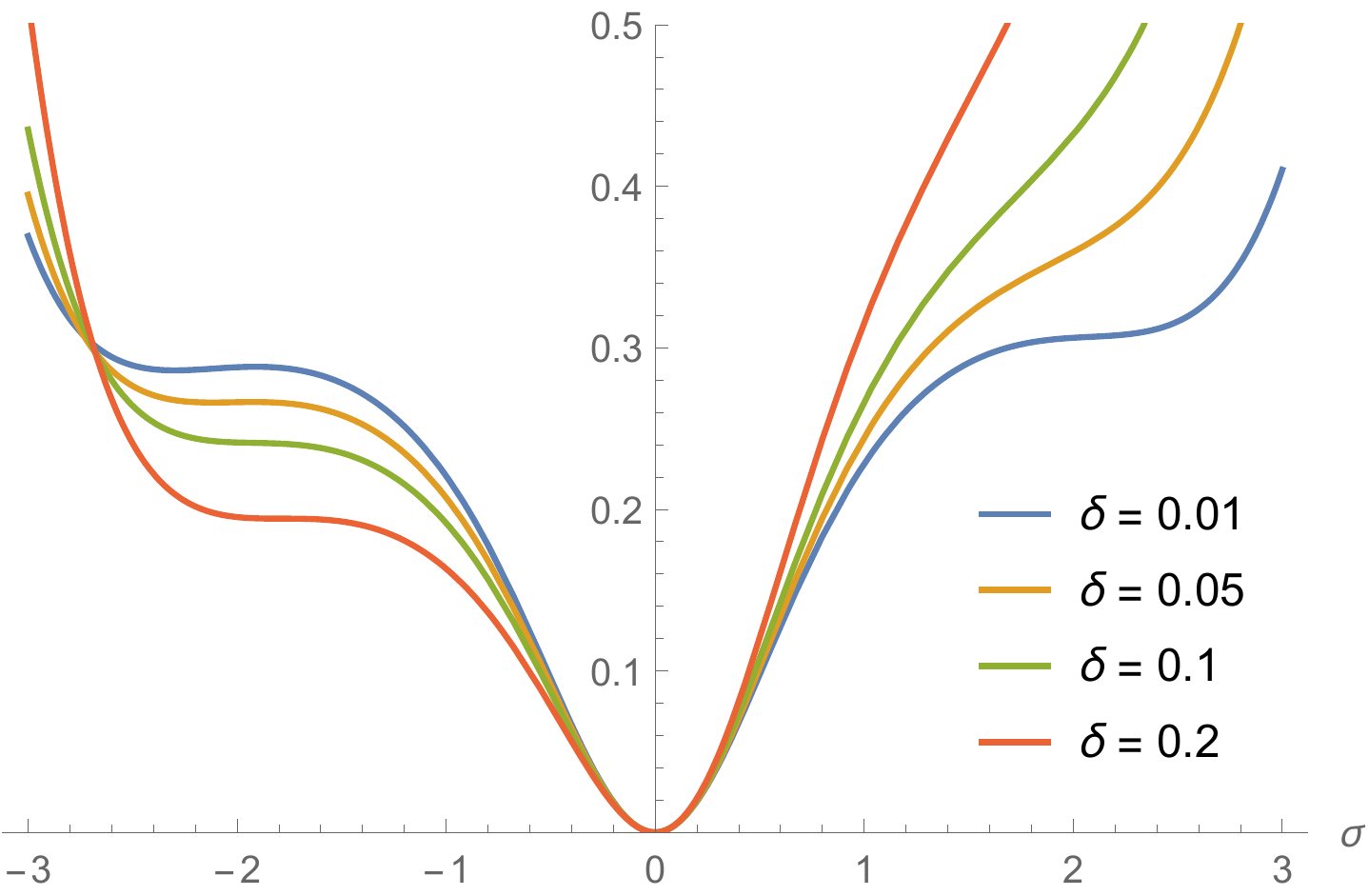}
\end{subfigure}
\captionsetup{width=.8\linewidth}
\caption{The left side: the ratio $V_{\rm inflec}/V_{\infty}$ and $\zeta$ as the functions of $\delta$ according to 
Eq.~\eqref{Hessian}. The right side: the profile of the potential $V/M^2$ with a single inflection point, when $\varphi$ is at its local minimum.}
\label{V_inflection_delta}
\end{figure}

\textbf{The parameter space}. When demanding the presence of a (near-)inflection point, the parameters must satisfy  Eq.~\eqref{Hessian}. The plot of $V_{\rm inflec.}/V_{\infty}$ ($V_\infty$ is taken for $\varphi\gg 1$ and $\sigma=0$) versus $\delta$, and the solution $\zeta(\delta)$ to Eq.~\eqref{Hessian}, are displayed on the left side of Fig.~\ref{V_inflection_delta}. On the right side of Fig.~\ref{V_inflection_delta} we show the profile of the potential with 
$\varphi$ at its local minimum satisfying $\partial_\varphi V=0$, for several  choices of $\delta$. In particular, our plot shows, when $\delta\rightarrow 0$, the inclination of the potential towards the inflection point becomes smaller until it vanishes when $\delta=0$ (in such case the potential coincides with the one in Fig.~\ref{V_3d2}). Therefore, when $\delta$ is very small, the inclination of the potential becomes insufficient for guiding the inflationary trajectory through the inflection point. Instead, the trajectory tends to the $\sigma=0$ path (when $\delta=0$, the trajectory exactly follows the $\sigma=0$ path).

\begin{figure}
\centering
  \includegraphics[width=.5\linewidth]{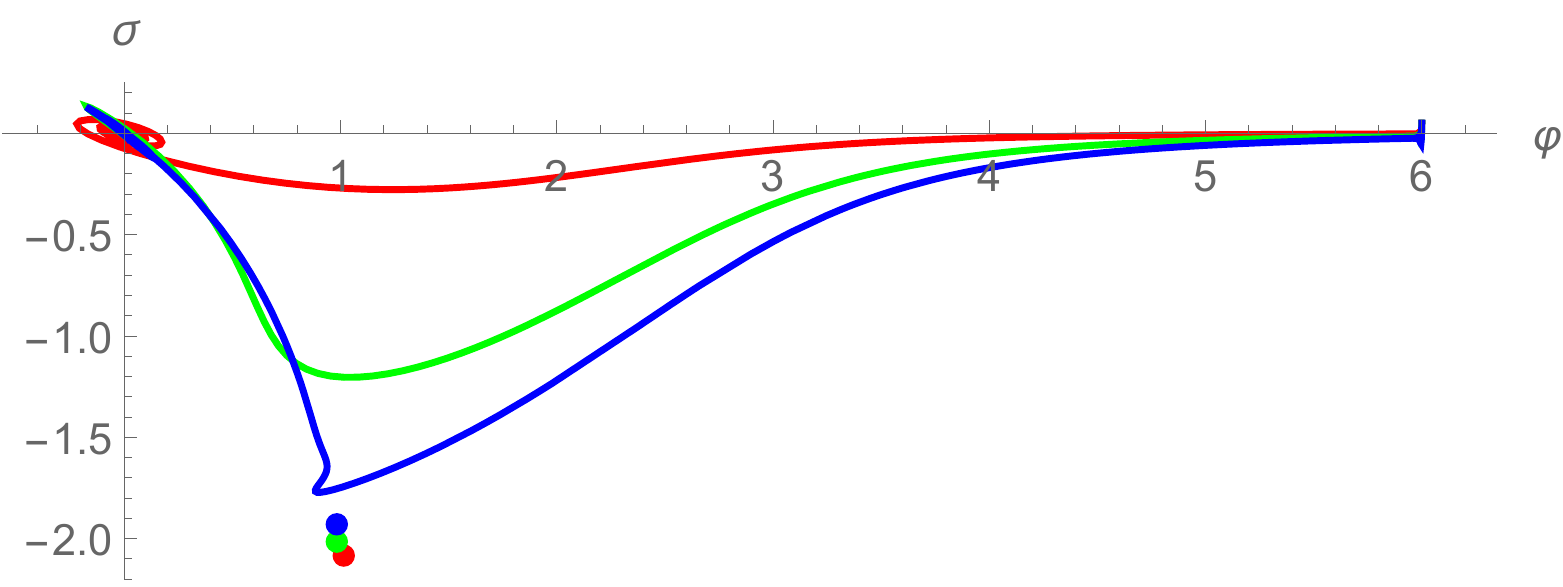}
\captionsetup{width=.9\linewidth}
\caption{The inflationary trajectories and the corresponding inflection points marked by the colored points. The parameters are $\delta=0.01$ (red), $\delta=0.05$ (green) and $\delta=0.1$ (blue).}
\label{Fig_fs_small_delta}
\end{figure}

We plot the inflationary trajectories in the $\varphi-\sigma$ plane for various choices of $\delta$ (with $\zeta$ being fixed by requiring the existence of an inflection point) in Fig.~\ref{Fig_fs_small_delta}. The colored spots represent the inflection points. As can be seen in Fig.~\ref{Fig_fs_small_delta} for $\delta=0.01$ and $\delta=0.05$, the trajectory misses the corresponding inflection point by a large margin and, therefore, avoids the USR regime. On the other hand, when $\delta=0.1$, the trajectory stops near the inflection point and then oscillates a few times before going to the minimum at 
$\varphi=\sigma=0$. This indicates the possibility of an USR stage, and it happens in Fig.~\ref{fsV_sol_delta} for this parameter choice indeed.

\begin{figure}
\centering
\begin{subfigure}{.49\textwidth}
  \centering
  \includegraphics[width=1\linewidth]{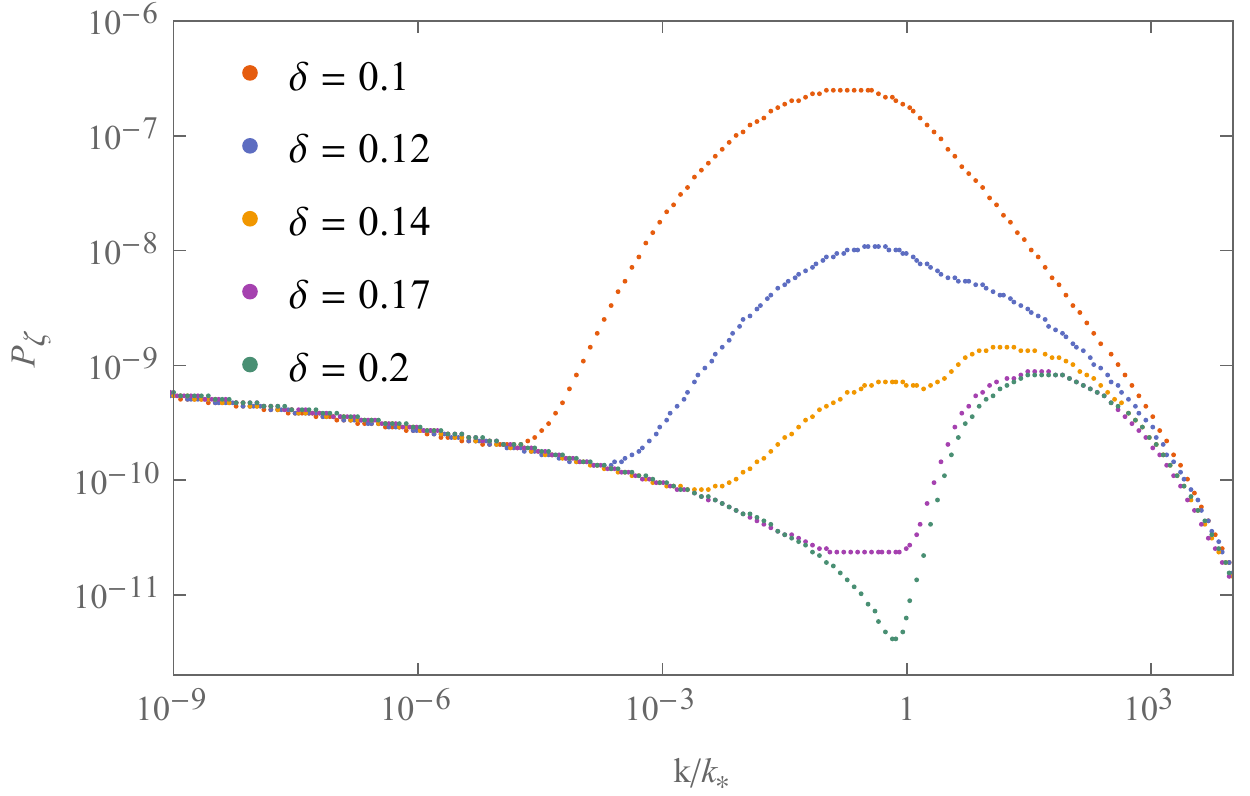}
\end{subfigure}
\begin{subfigure}{.49\textwidth}
  \centering
  \includegraphics[width=1\linewidth]{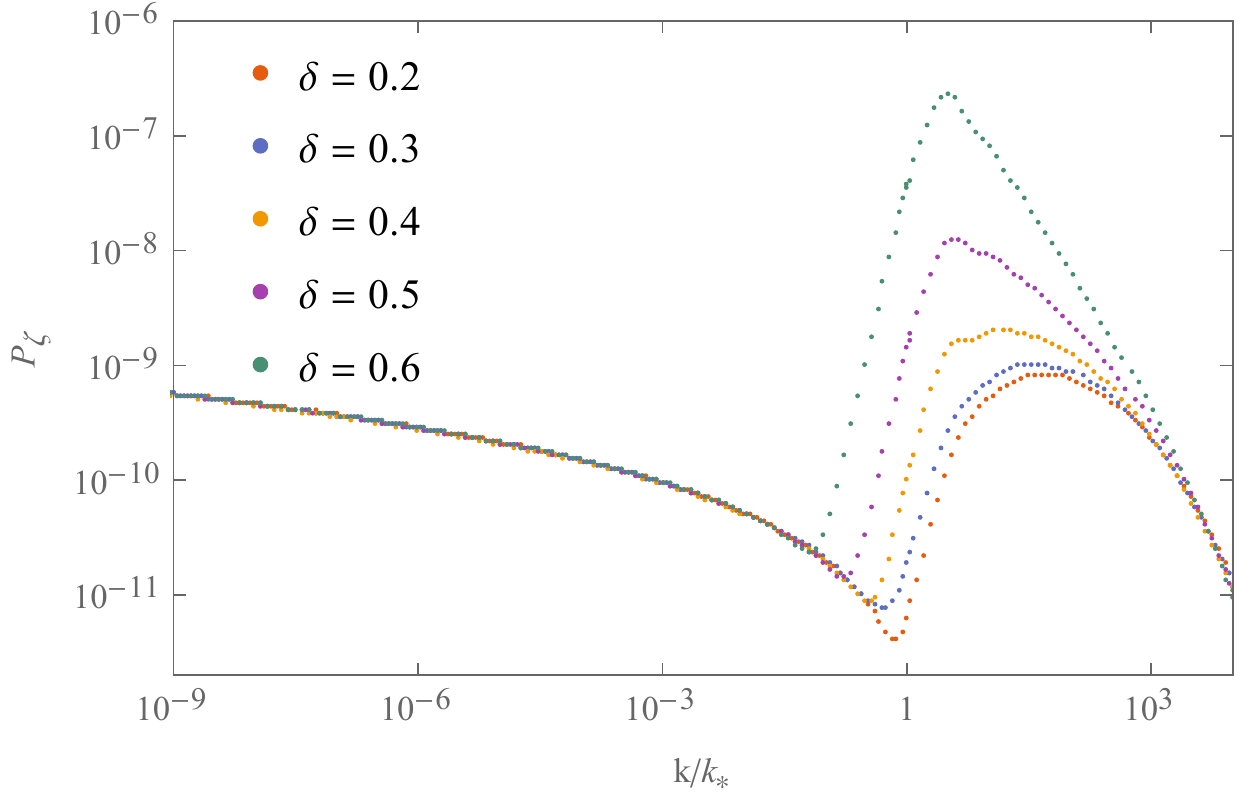}
\end{subfigure}
\captionsetup{width=.9\linewidth}
\caption{The power spectrum for various values of $\delta$ from $0.1$ to $0.2$ (on the left side), and from $0.2$ to $0.6$ (on the right side). The pivot scale is $k_*=k_{\Delta N_2}$ with $\Delta N_2=10$.}
\label{Pk_int_delta}
\end{figure}

\textbf{The scalar power spectrum at fixed $\Delta N_2$ and $\delta\gtrsim 0.1$}. Let us  fix $\Delta N_2=10$, and consider the power spectrum for several values of $\delta$. We find that the spectrum has a non-trivial dependence on $\delta$, see Fig.~\ref{Pk_int_delta}. In the left plot, $\delta$ is varied from $0.1$ to $0.2$, and we observe the spectrum enhancement to become smaller as $\delta$ grows. In the right plot, once $\delta$ reaches $0.2$, the enhancement starts growing with $\delta$ and develops a sharper peak.

As regards larger values of $\delta$, our numerical results show, when $\delta\gtrsim 0.6$, it becomes increasingly more difficult to maintain the USR stage and to achieve $\Delta N_2>10$, in particular. It may be due to the need of an extreme fine-tuning of the parameter $\zeta$ when $\delta$ is large.

\begin{figure}
\centering
\begin{subfigure}{.49\textwidth}
  \centering
  \includegraphics[width=1\linewidth]{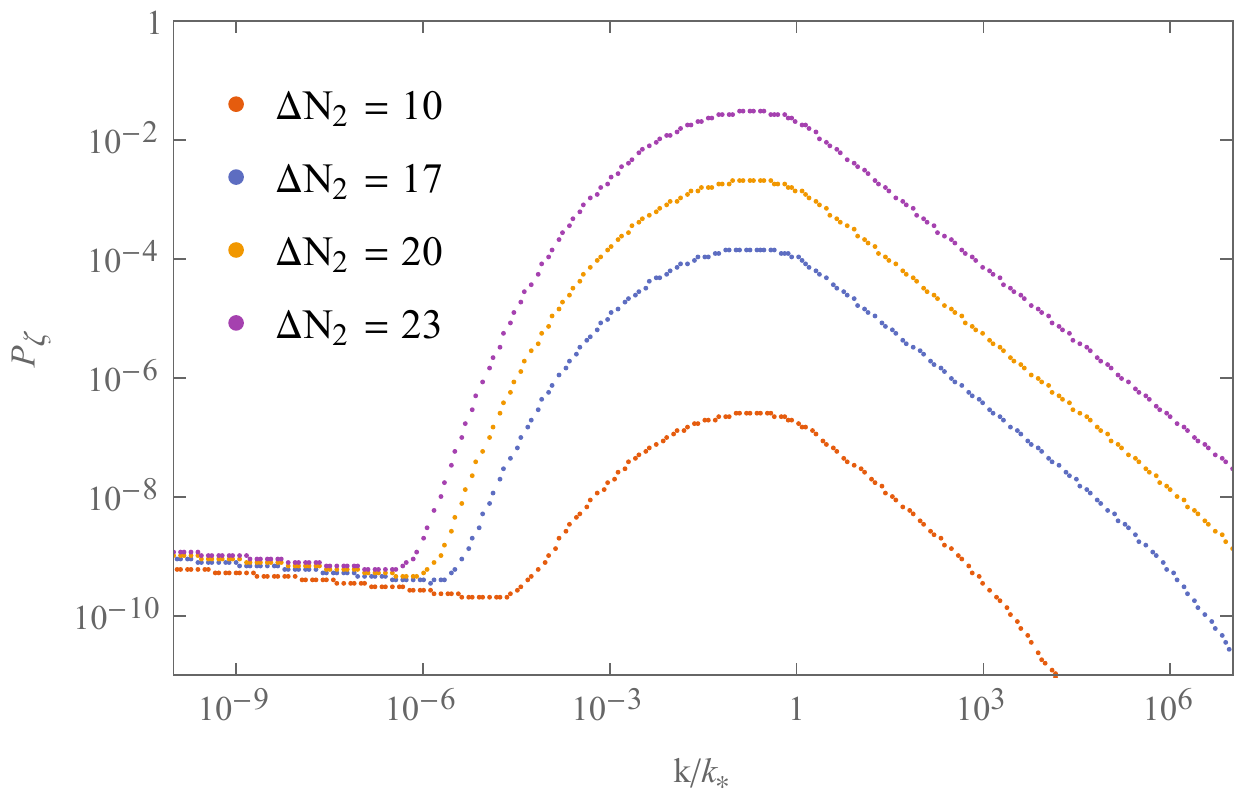}
\end{subfigure}
\begin{subfigure}{.49\textwidth}
  \centering
  \includegraphics[width=1\linewidth]{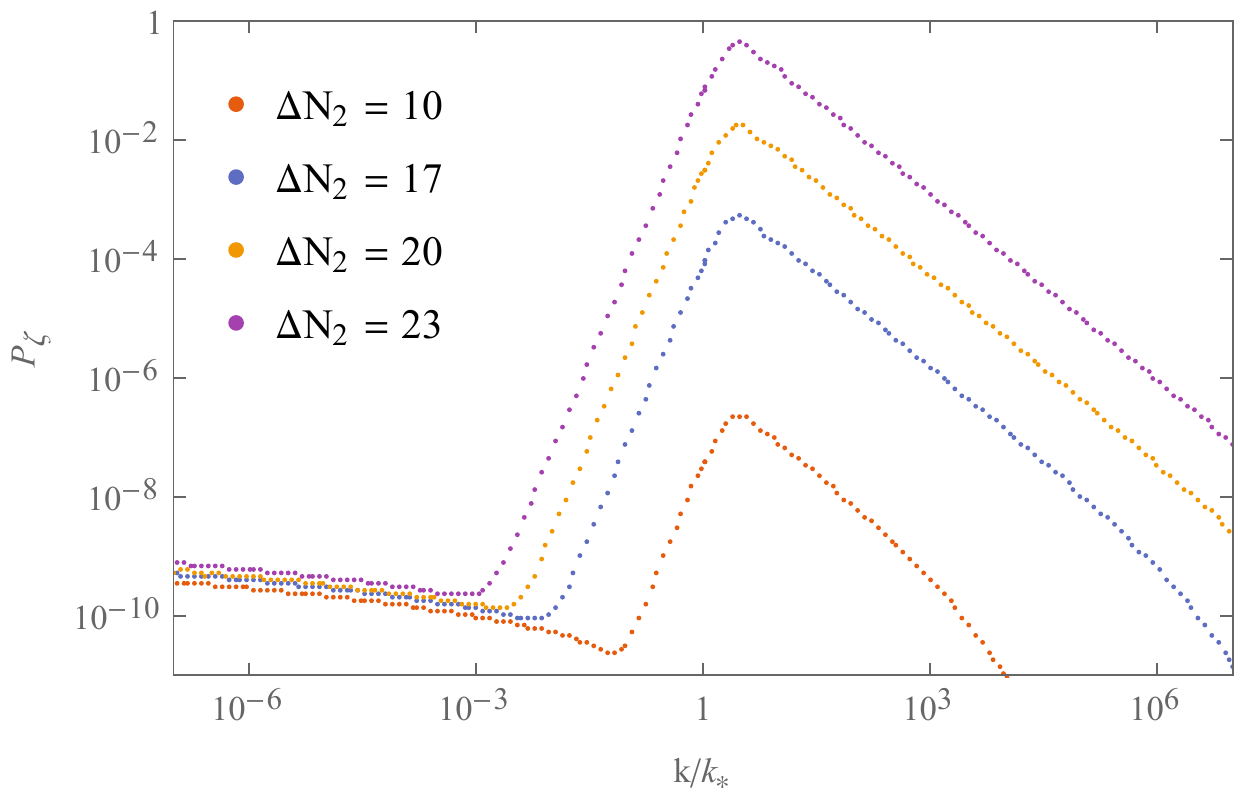}
\end{subfigure}
\captionsetup{width=.9\linewidth}
\caption{The power spectrum for $\delta=0.1$ (left) and $\delta=0.6$ (right) around the pivot scale $k_*=k_{\Delta N_2}$ when changing $\Delta N_2$.}
\label{Pk_delta_dn}
\end{figure}

\textbf{Changing $\Delta N_2$}. Amongst the values of $\delta$ studied above, let us pick up those with the highest power spectrum peaks, namely, $\delta=0.1$ and $\delta=0.6$, and plot the spectrum for $\Delta N_2=10,17,20,23$. The results are displayed in Fig.~\ref{Pk_delta_dn} with the plots on the left side and the right side corresponding to $\delta=0.1$ and $0.6$, respectively. As expected, the enhancement becomes larger with increasing $\Delta N_2$.

\textbf{PBH masses and their density fraction}. To be specific, let us consider two different examples: a {\it smooth} peak for $\delta\sim 0.1$, and a {\it sharp} peak for $\delta\sim 0.6$, in the power spectrum. Requiring the total PBH density fraction $f_{\rm tot}\approx 1$ and the corresponding density threshold in the region $1/3\leq\delta_c\leq 2/3$, we find $\delta=0.09$ and $\delta=0.61$ are suitable for efficient generation of PBHs.

\begin{figure}
\centering
  \includegraphics[width=.7\linewidth]{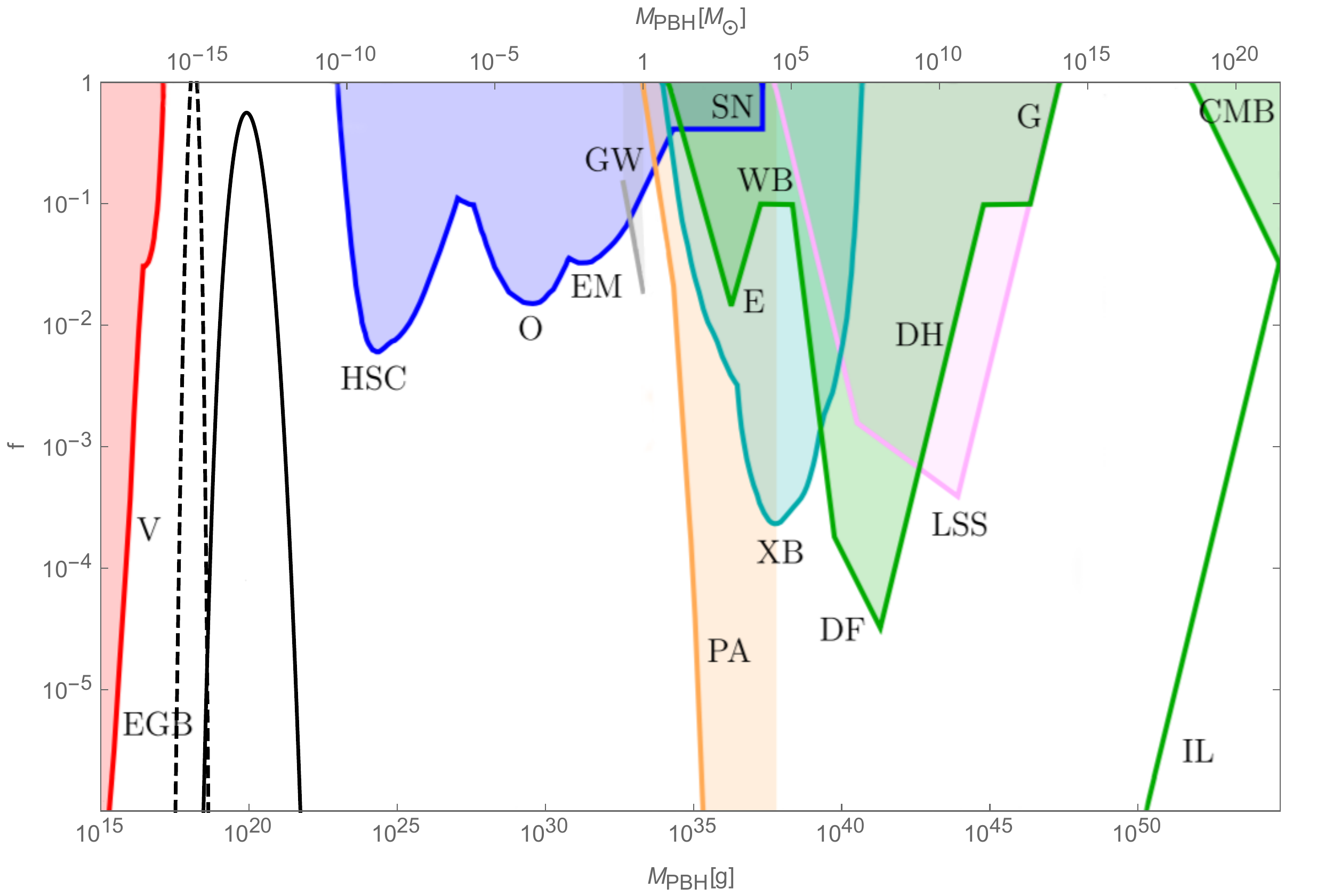}
\captionsetup{width=.9\linewidth}
\caption{The PBH fraction in the two working examples of the $\delta$-type extension, including $f_{\rm tot}\approx 1$. The solid black curve corresponds to  $\delta=0.09$, $\Delta N_2=19$, and $\delta_c=0.47$; the dashed black curve corresponds to $\delta=0.61$, $\Delta N_2=20$, and $\delta_c=0.4$. The  observational constraints are taken from Ref.~\cite{Carr:2020xqk}.}
\label{Fig_f_delta}
\end{figure}

We estimate the PBH masses by using Eq.~\eqref{MPBH} and summarize our results in Table \ref{tab_MPBH_delta}. We find when $\delta=0.09$, the $3\sigma$ value of $n_s\gtrsim 0.946$ requires $\Delta N_2\lesssim 19$. When $\delta=0.61$, it requires $\Delta N_2\lesssim 20$. In the examples of $\delta=0.09$ and $\delta=0.61$ we take the upper limits $\Delta N_2=19$ and $\Delta N_2=20$, respectively, and compute the PBH density fraction from  Eq.~\eqref{f_PBH}. The results are shown in Fig.~\ref{Fig_f_delta} where the observational constraints are included for the reference purposes only, as they assume the monochromatic PBH mass function. The density functions of Fig.~\ref{Fig_f_delta} peak at $M_{\rm PBH}\sim 10^{20}~{\rm g}$ and $M_{\rm PBH}\sim 10^{18}~{\rm g}$ when $\delta=0.09$ and $\delta=0.61$, respectively.~\footnote{The peak in the PBH density of the case $\delta=0.09$ can be seen to deviate from the prediction of Eq.~\eqref{MPBH} shown in Table \ref{tab_MPBH_delta}, likely due to the relatively broad nature of the peak.}

\begin{table}[ht]
\centering
\begin{tabular}{l r r r r r r r r}
\toprule
& \multicolumn{4}{c}{$\delta=0.09$} & \multicolumn{4}{c}{$\delta=0.61$}\\
\cmidrule(r){2-5} \cmidrule(l){6-9}
$\Delta N_2$ & $10$ & $17$ & $20$ & $23$ & $10$ & $17$ & $20$ & $23$ \\
\hline
$M_{\rm PBH}$, g & $10^{9}$ & $10^{15}$ & $10^{18}$ & $10^{20}$ & $10^{9}$ & $10^{15}$ & $10^{18}$ & $10^{20}$ \\
$n_s$ & $0.9566$ & $0.9486$ & $0.9443$ & $0.9390$ & $0.9581$ & $0.9504$ & $0.9461$ & $0.9409$\\
$r_{\rm max}$ & $0.005$ & $0.007$ & $0.008$ & $0.010$ & $0.004$ & $0.006$ & $0.007$ & $0.008$ \\\bottomrule
\hline
\end{tabular}
\captionsetup{width=.9\linewidth}
\caption{The PBH masses estimated from Eq.~\eqref{MPBH} for $\delta=0.09$ and $\delta=0.61$, with the corresponding values of $n_s$ and $r_{\rm max}$.}
\label{tab_MPBH_delta}
\end{table}

\section{Conclusions and comments}

In this paper, we analyze several supergravity extensions of the Starobinsky inflationary model.  We explore possibilities 
of PBHs genesis that could account for part of Cold Dark Matter.  We find that PBHs generation can be efficiently catalyzed by primordial perturbations sourced by the Starobinsky scalaron coupled to a new supersymmetric "modulus" (scalar) field.

Let us summarize our strategy. 

We rely on theoretical considerations {\it before} comparing them with cosmological observations, as a top-down approach. As our starting point, we adopt the Starobinsky inflationary model serving as the theoretical tool and pointing out the need of modified gravity in a more fundamental approach, i.e. beyond  considering the Starobinsky model as merely the best phenomenological fit to CMB observations. We extend the modified $(R+\zeta R^2)$ gravity to the modified supergravity, where the latter is considered as the candidate (or as the approximation) of a more fundamental theory of quantum gravity.  Amongst the theoretical advantages of modified supergravity are (i) the
use of supergravitational couplings only, (ii) predicted new physical degrees of freedom, and (iii) its formal equivalence to the standard (matter-coupled) supergravities. However,  unlike the standard supergravities coupled to matter, modified supergravity can be limited to the supergravity fields alone, where the new physical scalar naturally accompanies 
the inflaton (scalaron), together with metric and gravitino. It happens because modified supergravity is a higher-derivative theory, so that the "auxiliary" scalar of the standard (off-shell) supergravity multiplet becomes  {\it dynamical}. We find that modified supergravity naturally leads to the two-field inflationary models with restricted couplings and a small number of free parameters. Therefore, local supersymmetry has predictive power for phenomenology of the early universe cosmology via the double-inflation scenario. Indeed, the second field coupled to the Starobinsky scalaron is not introduced {\it ad hoc}  but is predicted by the supergravity extension of the Starobinsky model.  Our strategy is to use those models for a viable description of Starobinsky inflation together with the PBH production after inflation. Cosmological inflation and the PBH production can be considered as {\it probes} of supergravity for its use as a more fundamental approach, and {\it vice versa}: modified supergravity provides a theoretical input for the {\it discrimination} of phenomenological models of inflation and PBHs.

We summarize our main {\it results} as follows.

A generic modified supergravity Lagrangian in the {\it manifestly} supersymmetric form (with all couplings included) is given by Eq.~(\ref{L_master}). After (Taylor) expanding its potentials $N$ and 
${\cal F}$ in powers of the scalar curvature superfield ${\cal R}$ and keeping only the leading terms (needed for minimal embedding of $R+\zeta R^2$ gravity), we arrive at our basic model defined by Eq.~(\ref{N_F_choice}), whose relevant bosonic terms (in Jordan frame) are given by Eqs.~(\ref{L_master_comp}) and (\ref{U(X)}). As the next step, we perform the duality transformation of the derived bosonic terms to Einstein frame, and arrive at the two-scalar NLSM minimally coupled to gravity with the derived NLSM metric and the scalar potential, given by Eqs.~(\ref{L_varphi}) and (\ref{V_varphi}).  We also provide the manifestly supersymmetric (complete) duality transformation in terms of the superfields, and compute the corresponding K\"ahler potential and the superpotential, given by Eqs.~(\ref{Kael}) and (\ref{supV}) in the case of the basic model as an example. Then, we study the critical points (vacua) of the derived scalar potentials and the inflationary dynamics of two scalars in the context of two-field inflation, and we find consistency of the basic model with CMB observations. However, we also observe that such a scenario can work only with an extreme fine-tuning of initial conditions for efficient formation of PBHs. 

To overcome that problem, we add the next (subleading) terms to our basic model within the same
modified supergravity master Lagrangian (\ref{L_master}). There are two such terms, see Eqs.~(\ref{N_choice2}) and (\ref{N_F_choice2}), so we study them separately. We numerically compute the power spectra, estimate PBH masses and their density fraction, in both cases. We find that any of the extended models can {\it simultaneously} describe viable (Starobinsky-type) inflation and the PBH production after inflation, with limited fine-tuning of the parameters, exhibiting an attractor-type behavior. Actually, the PBH production is less sensitive to changes of the parameter $\gamma$ in the $\gamma$-extension of the $N$-potential. Next, we confront our theoretical predictions for PBHs (as part of DM) with current observations in Figs.~\ref{Fig_f_gamma} and \ref{Fig_f_delta}; in the cases of the $\gamma$- and $\delta$-extensions, respectively. When assuming the standard reheating temperature of $10^9$ GeV with PBHs  formation during the radiation era, the $\gamma$-model is apparently ruled out by the CMB constraints because it predicts $n_s$ outside the $3\sigma$ limit. It motivates us to consider the $\delta$-extension that predicts a larger $n_s$ within the CMB constraint. It is, therefore, quite possible that having both the $\gamma$- and $\delta$-terms (and, perhaps, even higher order terms) in the Lagrangian will render our supergravity model more flexible in accommodating the PBHs DM. 

Of course, modified supergravity  does not pretend on the status of an {\it ultimate} fundamental theory. However, there are  indications that it may be embedded into superstrings considered as an ultra-violet complete theory of quantum gravity.  Here it is worthwhile to mention that (i) modified supergravity always leads to the {\it no-scale} K\"ahler potential (\ref{Kael}) that often arises in superstring compactifications (see e.g., Ref.~\cite{Ellis:2013nka}), and (ii) there is a possibility of interpreting (some) modified supergravity theories as the D3-brane worldvolume theories in type II superstrings 
\cite{Binetruy:2004hh,Aldabergenov:2020pry}. Thus, the exploration of cosmological predictions from modified supergravity provides a remarkable bridge between quantum gravity on one side and phenomenology of inflation and PBHs on the other side.

PBH formation necessarily leads to Gravitational Waves (GWs) because large scalar overdensities act as a source for stochastic GWs background.  Frequencies of those GWs can be directly related to expected PBHs masses and duration of the second stage of inflation \cite{Bartolo:2018evs}. Those GWs may be detected in the future  ground-based experiments, such as the Einstein telescope \cite{ET} and the global network of GWs interferometers including advanced LIGO, Virgo and KAGRA \cite{ALAVK}, as well as in the space-based GWs interferometers such as LISA \cite{LISA}, TAIJI (old ALIA) \cite{TAIJI}, TianQin \cite{TQ} and DECIGO \cite{DEC}.

\section*{Acknowledgements}

The authors are grateful to Fotis Farakos, Anirudh Gundhi, Florian K\"uhnel, Hayato Motohashi, Misao Sasaki, Alexei Starobinsky, Christian Steinwachs, Spyros Sypsas and Ying-li Zhang for discussions and comments. Y.A. is supported by the CUniverse research promotion project of Chulalongkorn University in Bangkok, Thailand, under the grant reference CUAASC, and by the Ministry of Education and Science of the Republic of Kazakhstan under the grant reference BR05236730. S.V.K. is supported by Tokyo Metropolitan University, the World Premier International Research Center Initiative (WPI), MEXT, Japan, and the Competitiveness Enhancement Program of Tomsk Polytechnic University in Russia.

\begin{appendices}

\section{\hspace{-15pt}: supergravity in curved superspace}\label{App_superspace}

We follow the notation and conventions of Ref.~\cite{Wess:1992cp} with a few obvious modifications. A standard superspace Lagrangian of chiral superfields $\mathbf{\Phi}^i$ coupled to supergravity reads ($M_{\rm Pl}=1$)
\begin{equation}
    {\cal L}=\int d^2\Theta 2{\cal E}\left[\frac{3}{8}(\overbar{\cal D}^2-8{\cal R})e^{-K(\mathbf{\Phi}^i,\overbar{\mathbf{\Phi}}^i)/3}+W(\mathbf{\Phi}^i)\right]+{\rm h.c.}~,\label{App_L_super}
\end{equation}
where $\cal E$ is the chiral density superfield, $\cal R$ is the chiral curvature superfield, ${\cal D}_\alpha,\overbar{\cal D}_{\dot{\alpha}}$ are the superspace covariant derivatives with ${\cal D}^2\equiv{\cal D}^\alpha{\cal D}_\alpha$ and $\overbar{\cal D}^2\equiv\overbar{\cal D}_{\dot{\alpha}}\overbar{\cal D}^{\dot{\alpha}}$. A (real) K\"ahler potential $K$  and a  (holomorphic) superpotential $W$ are functions of the superfields, as indicated above.

A chiral superfield can be expanded in terms of its field components as 
\begin{equation}
    {\bf\Phi}=\Phi+\sqrt{2}\Theta\chi + \Theta^2F~.\label{Phi_expansion}
\end{equation}

The $\Theta$-expansion of $\cal E$ and $\cal R$ is given by
\begin{eqnarray}
    2{\cal E}&=&e\left[1+i\Theta\sigma^m\overbar{\psi}_m+\Theta^2(6\overbar{X}-\overbar{\psi}_m\overbar{\sigma}^{mn}\overbar{\psi}_n)\right]~,\\
    {\cal R}&=&X+\Theta\left(-\frac{1}{6}\sigma^m\overbar{\sigma}^n\psi_{mn}-i\sigma^m\overbar{\psi}_mX-\frac{i}{6}\psi_mb^m\right)+\nonumber\\
    &~&+\Theta^2\left(-\frac{1}{12}R-\frac{i}{6}\overbar{\psi}^m\overbar{\sigma}^n\psi_{mn}-4X\overbar{X}-\frac{1}{18}b_mb^m+\frac{i}{6}\nabla_mb^m+\right.\nonumber\\
    &~&+\left.\frac{1}{2}\overbar{\psi}_m\overbar{\psi}^mX+\frac{1}{12}\psi_m\sigma^m\overbar{\psi}_nb^n-\frac{1}{48}\varepsilon^{abcd}(\overbar{\psi}_a\overbar{\sigma}_b\psi_{cd}+\psi_a\sigma_b\overbar{\psi}_{cd})\right)~,\label{R_expansion}
\end{eqnarray}
where $e\equiv{\rm det}(e^a_m)$ and $\psi_{mn}\equiv\tilde{D}_m\psi_n-\tilde{D}_n\psi_m$ with the covariant derivative 
$\tilde{D}_m\psi_n\equiv\partial_m\psi_n+\psi_n\omega_m$. The vector $b_m$ and complex scalar $X$ are known in the literature as the  {\it old-minimal\/} set of supergravity auxiliary fields. In {\it modified} supergravity, those "auxiliary" fields become {\it dynamical} (or propagating) because of the presence of higher-derivatives in the Lagrangian (see e.g., Ref.~\cite{Ketov:2013dfa} for details). In our notation, the scalar curvature $R$ has the opposite sign compared to 
that in Wess--Bagger notation \cite{Wess:1992cp}.

In the standard supergravity, after eliminating the auxiliary fields and going to Einstein frame, the bosonic part of the Lagrangian of matter superfields  ${\bf\Phi}^i$ takes the form
\begin{equation}
    e^{-1}{\cal L}=\frac{1}{2}R-K_{i\bar{j}}\partial_m\Phi^i\partial^m\overbar{\Phi}^j-e^K\left(K^{i\bar{j}}D_iW D_{\bar{j}}\overbar{W}-3|W|^2\right)~,\label{App_L}
\end{equation}
where $K=K(\Phi^i,\overbar{\Phi}^i)$ is the K\"ahler potential, and $W=W(\Phi^i)$ is the superpotential, while the
same notation is used for the superfields and their leading field components, together with the standard definitions
\begin{gather}
    K_{i\bar{j}}\equiv\fracmm{\partial^2 K}{\partial\Phi^i\partial\overbar{\Phi}^j}~,\quad K^{i\bar{j}}\equiv K^{-1}_{i\bar{j}}~,\quad D_iW\equiv\fracmm{\partial W}{\partial\Phi^i}+W\fracmm{\partial K}{\partial\Phi^i}~~.
\end{gather}

\section{\hspace{-15pt}: estimating the transfer functions and the isocurvature fraction}\label{App_transfer}

\begin{figure}
\centering
  \includegraphics[width=.5\linewidth]{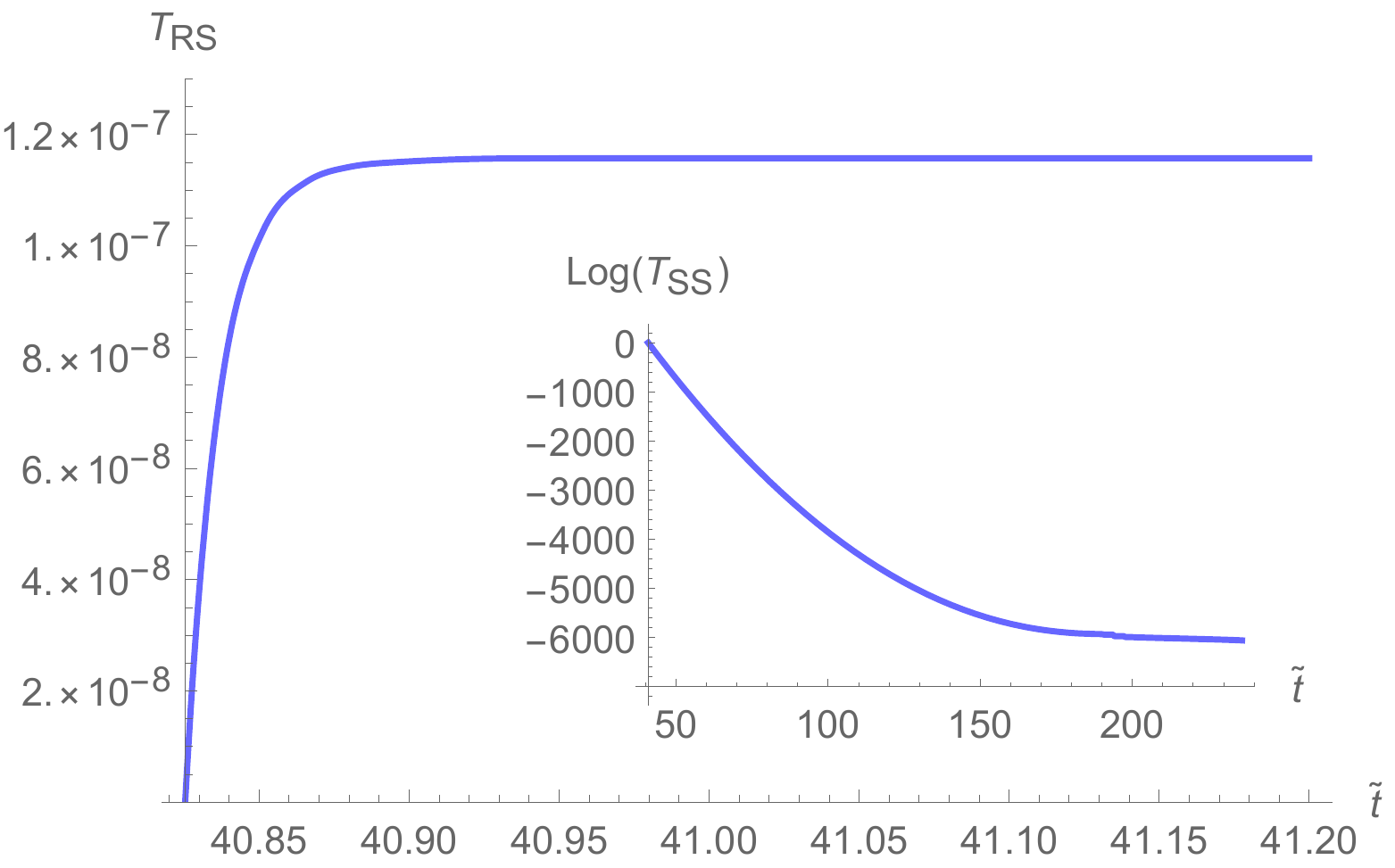}
\captionsetup{width=.9\linewidth}
\caption{The time dependence of the transfer functions.}
\label{Fig_transfer}
\end{figure}

Let us consider the case of $\gamma=1$ and $\Delta N_2=10$ as an example. After computing the transfer functions in Eq.~\eqref{Transfer_fns} as functions of $t_2$, with $t_1$ being fixed as the time corresponding to $60$ e-folds before the end of inflation (it corresponds to the horizon exit of the largest observable scale of around $k=10^{-4}~{\rm Mpc}^{-1}$), we find the result shown in Fig.~\ref{Fig_transfer}.

Having determined $T_{\rm RS}$ and $T_{\rm SS}$, we compute the isocurvature fraction at the end of inflation, i.e. with $t_2=t_{\rm end}$, and get
\begin{eqnarray}
    \beta_{\rm iso}=\fracmm{T^2_{\rm SS}}{1+T^2_{\rm SS}+T^2_{\rm RS}}={\cal O}(e^{-1200})~,
\end{eqnarray}
which is truly negligible.

\end{appendices}

\providecommand{\href}[2]{#2}\begingroup\raggedright\endgroup


\begin{thebibliography}{10}

\bibitem{Novikov:1967tw}
I.~Novikov and Y.~Zeldovic, ``{Cosmology},''
  \href{http://dx.doi.org/10.1146/annurev.aa.05.090167.003211}{{\em Ann. Rev.
  Astron. Astrophys.} {\bfseries 5} (1967) 627--649}.

\bibitem{Hawking:1971ei}
S.~Hawking, ``{Gravitationally collapsed objects of very low mass},'' {\em Mon.
  Not. Roy. Astron. Soc.} {\bfseries 152} (1971) 75.

\bibitem{M1}
M.~Khlopov, B.~Malomed, and I.~Zeldovich, ``{Gravitational instability of
  scalar fields and formation of primordial black holes},'' {\em Mon. Not. Roy.
  Astron. Soc.} {\bfseries 215} (1985) 575--589.

\bibitem{M4}
R.~Konoplich, S.~Rubin, A.~Sakharov, and M.~Khlopov, ``{Formation of black
  holes in first-order phase transitions as a cosmological test of
  symmetry-breaking mechanisms},'' {\em Phys. Atom. Nucl.} {\bfseries 62}
  (1999) 1593--1600.

\bibitem{M5}
M.~Khlopov, R.~Konoplich, S.~Rubin, and A.~Sakharov, ``{First-order phase
  transitions as a source of black holes in the early universe},'' {\em Grav.
  Cosmol.} {\bfseries 6} (2000) 153--156.

\bibitem{Addazi:2018nzm}
A.~Addazi, A.~Marcianò, and R.~Pasechnik, ``{Probing Trans-electroweak First
  Order Phase Transitions from Gravitational Waves},''
  \href{http://dx.doi.org/10.3390/physics1010010}{{\em MDPI Physics} {\bfseries
  1} no.~1, (2019) 92--102}, \href{http://arxiv.org/abs/1811.09074}{{\ttfamily
  arXiv:1811.09074 [hep-ph]}}.

\bibitem{Vilenkin:2018zol}
A.~Vilenkin, Y.~Levin, and A.~Gruzinov, ``{Cosmic strings and primordial black
  holes},'' \href{http://dx.doi.org/10.1088/1475-7516/2018/11/008}{{\em JCAP}
  {\bfseries 11} (2018) 008}, \href{http://arxiv.org/abs/1808.00670}{{\ttfamily
  arXiv:1808.00670 [astro-ph.CO]}}.

\bibitem{Belotsky:2018wph}
K.~M. Belotsky, V.~I. Dokuchaev, Y.~N. Eroshenko, E.~A. Esipova, M.~Y. Khlopov,
  L.~A. Khromykh, A.~A. Kirillov, V.~V. Nikulin, S.~G. Rubin, and I.~V.
  Svadkovsky, ``{Clusters of primordial black holes},''
  \href{http://dx.doi.org/10.1140/epjc/s10052-019-6741-4}{{\em Eur. Phys. J. C}
  {\bfseries 79} no.~3, (2019) 246},
  \href{http://arxiv.org/abs/1807.06590}{{\ttfamily arXiv:1807.06590
  [astro-ph.CO]}}.

\bibitem{Liu:2019lul}
J.~Liu, Z.-K. Guo, and R.-G. Cai, ``{Primordial Black Holes from Cosmic Domain
  Walls},'' \href{http://dx.doi.org/10.1103/PhysRevD.101.023513}{{\em Phys.
  Rev. D} {\bfseries 101} no.~2, (2020) 023513},
  \href{http://arxiv.org/abs/1908.02662}{{\ttfamily arXiv:1908.02662
  [astro-ph.CO]}}.

\bibitem{Barrow:1992hq}
J.~D. Barrow, E.~J. Copeland, and A.~R. Liddle, ``{The Cosmology of black hole
  relics},'' \href{http://dx.doi.org/10.1103/PhysRevD.46.645}{{\em Phys. Rev.
  D} {\bfseries 46} (1992) 645--657}.

\bibitem{Sasaki:2018dmp}
M.~Sasaki, T.~Suyama, T.~Tanaka, and S.~Yokoyama, ``{Primordial black
  holes---perspectives in gravitational wave astronomy},''
  \href{http://dx.doi.org/10.1088/1361-6382/aaa7b4}{{\em Class. Quant. Grav.}
  {\bfseries 35} no.~6, (2018) 063001},
  \href{http://arxiv.org/abs/1801.05235}{{\ttfamily arXiv:1801.05235
  [astro-ph.CO]}}.

\bibitem{Ketov:2019mfc}
S.~V. Ketov and M.~Y. Khlopov, ``{Cosmological Probes of Supersymmetric Field
  Theory Models at Superhigh Energy Scales},''
  \href{http://dx.doi.org/10.3390/sym11040511}{{\em Symmetry} {\bfseries 11}
  no.~4, (2019) 511}.

\bibitem{Carr:2003bj}
B.~J. Carr, ``{Primordial black holes as a probe of cosmology and high energy
  physics},'' \href{http://dx.doi.org/10.1007/978-3-540-45230-0\_7}{{\em Lect.
  Notes Phys.} {\bfseries 631} (2003) 301--321},
  \href{http://arxiv.org/abs/astro-ph/0310838}{{\ttfamily
  arXiv:astro-ph/0310838}}.

\bibitem{Pi:2017gih}
S.~Pi, Y.-l. Zhang, Q.-G. Huang, and M.~Sasaki, ``{Scalaron from $R^2$-gravity
  as a heavy field},''
  \href{http://dx.doi.org/10.1088/1475-7516/2018/05/042}{{\em JCAP} {\bfseries
  05} (2018) 042}, \href{http://arxiv.org/abs/1712.09896}{{\ttfamily
  arXiv:1712.09896 [astro-ph.CO]}}.

\bibitem{Germani:2018jgr}
C.~Germani and I.~Musco, ``{Abundance of Primordial Black Holes Depends on the
  Shape of the Inflationary Power Spectrum},''
  \href{http://dx.doi.org/10.1103/PhysRevLett.122.141302}{{\em Phys. Rev.
  Lett.} {\bfseries 122} no.~14, (2019) 141302},
  \href{http://arxiv.org/abs/1805.04087}{{\ttfamily arXiv:1805.04087
  [astro-ph.CO]}}.

\bibitem{Fumagalli:2020adf}
J.~Fumagalli, S.~Renaux-Petel, J.~W. Ronayne, and L.~T. Witkowski, ``{Turning
  in the landscape: a new mechanism for generating Primordial Black Holes},''
  \href{http://arxiv.org/abs/2004.08369}{{\ttfamily arXiv:2004.08369
  [hep-th]}}.

\bibitem{Palma:2020ejf}
G.~A. Palma, S.~Sypsas, and C.~Zenteno, ``{Seeding primordial black holes in
  multi-field inflation},'' \href{http://arxiv.org/abs/2004.06106}{{\ttfamily
  arXiv:2004.06106 [astro-ph.CO]}}.

\bibitem{Cai:2019bmk}
R.-G. Cai, Z.-K. Guo, J.~Liu, L.~Liu, and X.-Y. Yang, ``{Primordial black holes
  and gravitational waves from parametric amplification of curvature
  perturbations},'' \href{http://dx.doi.org/10.1017/S174392131900869X}{{\em
  JCAP} {\bfseries 06} (2020) 013},
  \href{http://arxiv.org/abs/1912.10437}{{\ttfamily arXiv:1912.10437
  [astro-ph.CO]}}.

\bibitem{Cai:2018dig}
R.-G. Cai, S.~Pi, and M.~Sasaki, ``{Gravitational Waves Induced by non-Gaussian
  Scalar Perturbations},''
  \href{http://dx.doi.org/10.1103/PhysRevLett.122.201101}{{\em Phys. Rev.
  Lett.} {\bfseries 122} no.~20, (2019) 201101},
  \href{http://arxiv.org/abs/1810.11000}{{\ttfamily arXiv:1810.11000
  [astro-ph.CO]}}.

\bibitem{Deng:2018wmy}
C.-M. Deng, Y.~Cai, X.-F. Wu, and E.-W. Liang, ``{Fast Radio Bursts From
  Primordial Black Hole Binaries Coalescence},''
  \href{http://dx.doi.org/10.1103/PhysRevD.98.123016}{{\em Phys. Rev. D}
  {\bfseries 98} no.~12, (2018) 123016},
  \href{http://arxiv.org/abs/1812.00113}{{\ttfamily arXiv:1812.00113
  [astro-ph.HE]}}.

\bibitem{Akrami:2018odb}
{\bfseries Planck} Collaboration, Y.~Akrami {\em et~al.}, ``{Planck 2018
  results. X. Constraints on inflation},''
  \href{http://arxiv.org/abs/1807.06211}{{\ttfamily arXiv:1807.06211
  [astro-ph.CO]}}.

\bibitem{Starobinsky:1980te}
A.~A. Starobinsky, ``A new type of isotropic cosmological models without
  singularity,''
  \href{http://dx.doi.org/https://doi.org/10.1016/0370-2693(80)90670-X}{{\em
  Phys. Lett. B} {\bfseries 91} no.~1, (1980) 99 -- 102}.

\bibitem{Garcia-Bellido:2017mdw}
J.~Garcia-Bellido and E.~Ruiz~Morales, ``{Primordial black holes from single
  field models of inflation},''
  \href{http://dx.doi.org/10.1016/j.dark.2017.09.007}{{\em Phys. Dark Univ.}
  {\bfseries 18} (2017) 47--54},
  \href{http://arxiv.org/abs/1702.03901}{{\ttfamily arXiv:1702.03901
  [astro-ph.CO]}}.

\bibitem{Motohashi:2017kbs}
H.~Motohashi and W.~Hu, ``{Primordial Black Holes and Slow-Roll Violation},''
  \href{http://dx.doi.org/10.1103/PhysRevD.96.063503}{{\em Phys. Rev. D}
  {\bfseries 96} no.~6, (2017) 063503},
  \href{http://arxiv.org/abs/1706.06784}{{\ttfamily arXiv:1706.06784
  [astro-ph.CO]}}.

\bibitem{Passaglia:2018ixg}
S.~Passaglia, W.~Hu, and H.~Motohashi, ``{Primordial black holes and local
  non-Gaussianity in canonical inflation},''
  \href{http://dx.doi.org/10.1103/PhysRevD.99.043536}{{\em Phys. Rev. D}
  {\bfseries 99} no.~4, (2019) 043536},
  \href{http://arxiv.org/abs/1812.08243}{{\ttfamily arXiv:1812.08243
  [astro-ph.CO]}}.

\bibitem{Farakos:2013cqa}
F.~Farakos, A.~Kehagias, and A.~Riotto, ``{On the Starobinsky Model of
  Inflation from Supergravity},''
  \href{http://dx.doi.org/10.1016/j.nuclphysb.2013.08.005}{{\em Nucl. Phys. B}
  {\bfseries 876} (2013) 187--200},
  \href{http://arxiv.org/abs/1307.1137}{{\ttfamily arXiv:1307.1137 [hep-th]}}.

\bibitem{Ferrara:2013rsa}
S.~Ferrara, R.~Kallosh, A.~Linde, and M.~Porrati, ``{Minimal Supergravity
  Models of Inflation},''
  \href{http://dx.doi.org/10.1103/PhysRevD.88.085038}{{\em Phys. Rev. D}
  {\bfseries 88} no.~8, (2013) 085038},
  \href{http://arxiv.org/abs/1307.7696}{{\ttfamily arXiv:1307.7696 [hep-th]}}.

\bibitem{Aldabergenov:2016dcu}
Y.~Aldabergenov and S.~V. Ketov, ``{SUSY breaking after inflation in
  supergravity with inflaton in a massive vector supermultiplet},''
  \href{http://dx.doi.org/10.1016/j.physletb.2016.08.016}{{\em Phys. Lett. B}
  {\bfseries 761} (2016) 115--118},
  \href{http://arxiv.org/abs/1607.05366}{{\ttfamily arXiv:1607.05366
  [hep-th]}}.

\bibitem{Aldabergenov:2017bjt}
Y.~Aldabergenov and S.~V. Ketov, ``{Higgs mechanism and cosmological constant
  in $N=1$ supergravity with inflaton in a vector multiplet},''
  \href{http://dx.doi.org/10.1140/epjc/s10052-017-4807-8}{{\em Eur. Phys. J. C}
  {\bfseries 77} no.~4, (2017) 233},
  \href{http://arxiv.org/abs/1701.08240}{{\ttfamily arXiv:1701.08240
  [hep-th]}}.

\bibitem{Addazi:2018pbg}
A.~Addazi, A.~Marciano, S.~V. Ketov, and M.~Y. Khlopov, ``{Physics of
  superheavy dark matter in supergravity},''
  \href{http://dx.doi.org/10.1142/S0218271818410110}{{\em Int. J. Mod. Phys. D}
  {\bfseries 27} no.~06, (2018) 1841011}.

\bibitem{GarciaBellido:1996qt}
J.~Garcia-Bellido, A.~D. Linde, and D.~Wands, ``{Density perturbations and
  black hole formation in hybrid inflation},''
  \href{http://dx.doi.org/10.1103/PhysRevD.54.6040}{{\em Phys. Rev. D}
  {\bfseries 54} (1996) 6040--6058},
  \href{http://arxiv.org/abs/astro-ph/9605094}{{\ttfamily
  arXiv:astro-ph/9605094}}.

\bibitem{Kawasaki:2015ppx}
M.~Kawasaki and Y.~Tada, ``{Can massive primordial black holes be produced in
  mild waterfall hybrid inflation?},''
  \href{http://dx.doi.org/10.1088/1475-7516/2016/08/041}{{\em JCAP} {\bfseries
  08} (2016) 041}, \href{http://arxiv.org/abs/1512.03515}{{\ttfamily
  arXiv:1512.03515 [astro-ph.CO]}}.

\bibitem{Braglia:2020eai}
M.~Braglia, D.~K. Hazra, F.~Finelli, G.~F. Smoot, and A.~A. Starobinsky,
  ``{Generating PBHs and small-scale GWs in two-field models of inflation},''
  \href{http://arxiv.org/abs/2005.02895}{{\ttfamily arXiv:2005.02895
  [astro-ph.CO]}}.

\bibitem{Ketov:2019toi}
S.~V. Ketov, ``{On the equivalence of Starobinsky and Higgs inflationary models
  in gravity and supergravity},''
  \href{http://dx.doi.org/10.1088/1751-8121/ab6a33}{{\em J. Phys. A} {\bfseries
  53} no.~8, (2020) 084001}, \href{http://arxiv.org/abs/1911.01008}{{\ttfamily
  arXiv:1911.01008 [hep-th]}}.

\bibitem{Ketov:2012jt}
S.~V. Ketov and A.~A. Starobinsky, ``{Inflation and non-minimal
  scalar-curvature coupling in gravity and supergravity},''
  \href{http://dx.doi.org/10.1088/1475-7516/2012/08/022}{{\em JCAP} {\bfseries
  08} (2012) 022}, \href{http://arxiv.org/abs/1203.0805}{{\ttfamily
  arXiv:1203.0805 [hep-th]}}.

\bibitem{Ketov:2013dfa}
S.~V. Ketov and T.~Terada, ``{Old-minimal supergravity models of inflation},''
  \href{http://dx.doi.org/10.1007/JHEP12(2013)040}{{\em JHEP} {\bfseries 12}
  (2013) 040},
\href{http://arxiv.org/abs/1309.7494}{{\ttfamily arXiv:1309.7494 [hep-th]}}.

\bibitem{Addazi:2017rkc}
A.~Addazi and S.~V. Ketov, ``{Energy conditions in Starobinsky supergravity},''
  \href{http://dx.doi.org/10.1088/1475-7516/2017/03/061}{{\em JCAP} {\bfseries
  03} (2017) 061}, \href{http://arxiv.org/abs/1701.02450}{{\ttfamily
  arXiv:1701.02450 [hep-th]}}.

\bibitem{Cecotti:1987sa}
S.~Cecotti, ``{Higher derivative supergravity is equivalent to standard
  supergravity coupled to matter. 1.},''
\href{http://dx.doi.org/10.1016/0370-2693(87)90844-6}{{\em Phys. Lett.}
  {\bfseries B190} (1987) 86--92}.

\bibitem{Wess:1992cp}
J.~Wess and J.~Bagger, {\em {Supersymmetry and supergravity}}.
\newblock Princeton University Press, Princeton, NJ, USA,
1992.
\newblock

\bibitem{Kallosh:2013xya}
R.~Kallosh and A.~Linde, ``{Superconformal generalizations of the Starobinsky
  model},'' \href{http://dx.doi.org/10.1088/1475-7516/2013/06/028}{{\em JCAP}
  {\bfseries 1306} (2013) 028},
\href{http://arxiv.org/abs/1306.3214}{{\ttfamily arXiv:1306.3214 [hep-th]}}.

\bibitem{Gates:2009hu}
J.~Gates, S.James and S.~V. Ketov, ``{Superstring-inspired supergravity as the
  universal source of inflation and quintessence},''
  \href{http://dx.doi.org/10.1016/j.physletb.2009.03.005}{{\em Phys. Lett. B}
  {\bfseries 674} (2009) 59--63},
  \href{http://arxiv.org/abs/0901.2467}{{\ttfamily arXiv:0901.2467 [hep-th]}}.

\bibitem{Schutz:2013fua}
K.~Schutz, E.~I. Sfakianakis, and D.~I. Kaiser, ``{Multifield Inflation after
  Planck: Isocurvature Modes from Nonminimal Couplings},''
  \href{http://dx.doi.org/10.1103/PhysRevD.89.064044}{{\em Phys. Rev. D}
  {\bfseries 89} no.~6, (2014) 064044},
  \href{http://arxiv.org/abs/1310.8285}{{\ttfamily arXiv:1310.8285
  [astro-ph.CO]}}.

\bibitem{Gundhi:2018wyz}
A.~Gundhi and C.~F. Steinwachs, ``{Scalaron-Higgs inflation},''
  \href{http://dx.doi.org/10.1016/j.nuclphysb.2020.114989}{{\em Nucl. Phys.}
  {\bfseries B954} (2020) 114989},
\href{http://arxiv.org/abs/1810.10546}{{\ttfamily arXiv:1810.10546 [hep-th]}}.

\bibitem{Canko:2019mud}
D.~D. Canko, I.~D. Gialamas, and G.~P. Kodaxis, ``{A simple $F({\cal R},\phi)$
  deformation of Starobinsky inflationary model},''
  \href{http://dx.doi.org/10.1140/epjc/s10052-020-8025-4}{{\em Eur. Phys. J. C}
  {\bfseries 80} no.~5, (2020) 458},
  \href{http://arxiv.org/abs/1901.06296}{{\ttfamily arXiv:1901.06296
  [hep-th]}}.

\bibitem{Liddle:1994dx}
A.~R. Liddle, P.~Parsons, and J.~D. Barrow, ``{Formalizing the slow roll
  approximation in inflation},''
  \href{http://dx.doi.org/10.1103/PhysRevD.50.7222}{{\em Phys. Rev. D}
  {\bfseries 50} (1994) 7222--7232},
  \href{http://arxiv.org/abs/astro-ph/9408015}{{\ttfamily
  arXiv:astro-ph/9408015}}.

\bibitem{Terada:2014uia}
T.~Terada, Y.~Watanabe, Y.~Yamada, and J.~Yokoyama, ``{Reheating processes
  after Starobinsky inflation in old-minimal supergravity},''
  \href{http://dx.doi.org/10.1007/JHEP02(2015)105}{{\em JHEP} {\bfseries 02}
  (2015) 105}, \href{http://arxiv.org/abs/1411.6746}{{\ttfamily arXiv:1411.6746
  [hep-ph]}}.

\bibitem{Motohashi:2014ppa}
H.~Motohashi, A.~A. Starobinsky, and J.~Yokoyama, ``{Inflation with a constant
  rate of roll},'' \href{http://dx.doi.org/10.1088/1475-7516/2015/09/018}{{\em
  JCAP} {\bfseries 09} (2015) 018},
  \href{http://arxiv.org/abs/1411.5021}{{\ttfamily arXiv:1411.5021
  [astro-ph.CO]}}.

\bibitem{Mulryne:2009kh}
D.~J. Mulryne, D.~Seery, and D.~Wesley, ``{Moment transport equations for
  non-Gaussianity},''
  \href{http://dx.doi.org/10.1088/1475-7516/2010/01/024}{{\em JCAP} {\bfseries
  01} (2010) 024}, \href{http://arxiv.org/abs/0909.2256}{{\ttfamily
  arXiv:0909.2256 [astro-ph.CO]}}.

\bibitem{Mulryne:2010rp}
D.~J. Mulryne, D.~Seery, and D.~Wesley, ``{Moment transport equations for the
  primordial curvature perturbation},''
  \href{http://dx.doi.org/10.1088/1475-7516/2011/04/030}{{\em JCAP} {\bfseries
  04} (2011) 030}, \href{http://arxiv.org/abs/1008.3159}{{\ttfamily
  arXiv:1008.3159 [astro-ph.CO]}}.

\bibitem{Dias:2015rca}
M.~Dias, J.~Frazer, and D.~Seery, ``{Computing observables in curved multifield
  models of inflation---A guide (with code) to the transport method},''
  \href{http://dx.doi.org/10.1088/1475-7516/2015/12/030}{{\em JCAP} {\bfseries
  12} (2015) 030}, \href{http://arxiv.org/abs/1502.03125}{{\ttfamily
  arXiv:1502.03125 [astro-ph.CO]}}.

\bibitem{Jiang:2017nou}
H.~Jiang and Y.~Wang, ``{Massive Fields as Systematics for Single Field
  Inflation},'' \href{http://dx.doi.org/10.1088/1475-7516/2017/06/038}{{\em
  JCAP} {\bfseries 06} (2017) 038},
  \href{http://arxiv.org/abs/1703.04477}{{\ttfamily arXiv:1703.04477
  [astro-ph.CO]}}.

\bibitem{Press:1973iz}
W.~H. Press and P.~Schechter, ``{Formation of galaxies and clusters of galaxies
  by selfsimilar gravitational condensation},''
  \href{http://dx.doi.org/10.1086/152650}{{\em Astrophys. J.} {\bfseries 187}
  (1974) 425--438}.

\bibitem{Inomata:2017okj}
K.~Inomata, M.~Kawasaki, K.~Mukaida, Y.~Tada, and T.~T. Yanagida,
  ``{Inflationary Primordial Black Holes as All Dark Matter},''
  \href{http://dx.doi.org/10.1103/PhysRevD.96.043504}{{\em Phys. Rev. D}
  {\bfseries 96} no.~4, (2017) 043504},
  \href{http://arxiv.org/abs/1701.02544}{{\ttfamily arXiv:1701.02544
  [astro-ph.CO]}}.

\bibitem{Inomata:2017vxo}
K.~Inomata, M.~Kawasaki, K.~Mukaida, and T.~T. Yanagida, ``{Double inflation as
  a single origin of primordial black holes for all dark matter and LIGO
  observations},'' \href{http://dx.doi.org/10.1103/PhysRevD.97.043514}{{\em
  Phys. Rev. D} {\bfseries 97} no.~4, (2018) 043514},
  \href{http://arxiv.org/abs/1711.06129}{{\ttfamily arXiv:1711.06129
  [astro-ph.CO]}}.

\bibitem{Carr:1975qj}
B.~J. Carr, ``{The Primordial black hole mass spectrum},''
  \href{http://dx.doi.org/10.1086/153853}{{\em Astrophys. J.} {\bfseries 201}
  (1975) 1--19}.

\bibitem{Carr:2020xqk}
B.~Carr and F.~Kuhnel, ``{Primordial Black Holes as Dark Matter: Recent
  Developments},'' \href{http://arxiv.org/abs/2006.02838}{{\ttfamily
  arXiv:2006.02838 [astro-ph.CO]}}.

\bibitem{Carr:2020gox}
B.~Carr, K.~Kohri, Y.~Sendouda, and J.~Yokoyama, ``{Constraints on Primordial
  Black Holes},'' \href{http://arxiv.org/abs/2002.12778}{{\ttfamily
  arXiv:2002.12778 [astro-ph.CO]}}.

\bibitem{Dalianis:2014aya}
I.~Dalianis, F.~Farakos, A.~Kehagias, A.~Riotto, and R.~von Unge,
  ``{Supersymmetry Breaking and Inflation from Higher Curvature
  Supergravity},'' \href{http://dx.doi.org/10.1007/JHEP01(2015)043}{{\em JHEP}
  {\bfseries 01} (2015) 043}, \href{http://arxiv.org/abs/1409.8299}{{\ttfamily
  arXiv:1409.8299 [hep-th]}}.

\bibitem{Ellis:2013nka}
J.~Ellis, D.~V. Nanopoulos, and K.~A. Olive, ``{A no-scale supergravity
  framework for sub-Planckian physics},''
  \href{http://dx.doi.org/10.1103/PhysRevD.89.043502}{{\em Phys. Rev. D}
  {\bfseries 89} no.~4, (2014) 043502},
  \href{http://arxiv.org/abs/1310.4770}{{\ttfamily arXiv:1310.4770 [hep-ph]}}.

\bibitem{Binetruy:2004hh}
P.~Binetruy, G.~Dvali, R.~Kallosh, and A.~Van~Proeyen, ``{Fayet-Iliopoulos
  terms in supergravity and cosmology},''
  \href{http://dx.doi.org/10.1088/0264-9381/21/13/005}{{\em Class. Quant.
  Grav.} {\bfseries 21} (2004) 3137--3170},
  \href{http://arxiv.org/abs/hep-th/0402046}{{\ttfamily arXiv:hep-th/0402046}}.

\bibitem{Aldabergenov:2020pry}
Y.~Aldabergenov, S.~Aoki, and S.~V. Ketov, ``{Minimal Starobinsky supergravity
  coupled to a dilaton-axion superfield},''
  \href{http://dx.doi.org/10.1103/PhysRevD.101.075012}{{\em Phys. Rev. D}
  {\bfseries 101} no.~7, (2020) 075012},
  \href{http://arxiv.org/abs/2001.09574}{{\ttfamily arXiv:2001.09574
  [hep-th]}}.

\bibitem{Bartolo:2018evs}
N.~Bartolo, V.~De~Luca, G.~Franciolini, A.~Lewis, M.~Peloso, and A.~Riotto,
  ``{Primordial Black Hole Dark Matter: LISA Serendipity},''
  \href{http://dx.doi.org/10.1103/PhysRevLett.122.211301}{{\em Phys. Rev.
  Lett.} {\bfseries 122} no.~21, (2019) 211301},
  \href{http://arxiv.org/abs/1810.12218}{{\ttfamily arXiv:1810.12218
  [astro-ph.CO]}}.

\bibitem{ET}
{\bfseries ET} Collaboration, M.~Punturo {\em et~al.}, ``{The Einstein
  Telescope: a third-generation gravitational wave observatory},''
  \href{http://dx.doi.org/10.1088/0264-9381/27/19/194002}{{\em Class. Quant.
  Grav.} {\bfseries 27} no.~19, (9, 2010) 194002}.

\bibitem{ALAVK}
{\bfseries KAGRA, LIGO Scientific, VIRGO} Collaboration, B.~Abbott {\em
  et~al.}, ``{Prospects for Observing and Localizing Gravitational-Wave
  Transients with Advanced LIGO, Advanced Virgo and KAGRA},''
  \href{http://dx.doi.org/10.1007/s41114-018-0012-9}{{\em Living Rev. Rel.}
  {\bfseries 21} no.~1, (2018) 3},
  \href{http://arxiv.org/abs/1304.0670}{{\ttfamily arXiv:1304.0670 [gr-qc]}}.

\bibitem{LISA}
{\bfseries LISA} Collaboration, P.~Amaro-Seoane {\em et~al.}, ``{Laser
  Interferometer Space Antenna},''
  \href{http://arxiv.org/abs/1702.00786}{{\ttfamily arXiv:1702.00786
  [astro-ph.IM]}}.

\bibitem{TAIJI}
X.~Gong {\em et~al.}, ``{Descope of the ALIA mission},''
  \href{http://dx.doi.org/10.1088/1742-6596/610/1/012011}{{\em J. Phys. Conf.
  Ser.} {\bfseries 610} no.~1, (2015) 012011},
  \href{http://arxiv.org/abs/1410.7296}{{\ttfamily arXiv:1410.7296 [gr-qc]}}.

\bibitem{TQ}
{\bfseries TianQin} Collaboration, J.~Luo {\em et~al.}, ``{TianQin: a
  space-borne gravitational wave detector},''
  \href{http://dx.doi.org/10.1088/0264-9381/33/3/035010}{{\em Class. Quant.
  Grav.} {\bfseries 33} no.~3, (2016) 035010},
  \href{http://arxiv.org/abs/1512.02076}{{\ttfamily arXiv:1512.02076
  [astro-ph.IM]}}.

\bibitem{DEC}
H.~Kudoh, A.~Taruya, T.~Hiramatsu, and Y.~Himemoto, ``{Detecting a
  gravitational-wave background with next-generation space interferometers},''
  \href{http://dx.doi.org/10.1103/PhysRevD.73.064006}{{\em Phys. Rev. D}
  {\bfseries 73} (2006) 064006},
  \href{http://arxiv.org/abs/gr-qc/0511145}{{\ttfamily arXiv:gr-qc/0511145}}.

\end{thebibliography}
\end{document}